\documentclass[journa]{IEEEtran}
\usepackage{booktabs} 
\usepackage{array}    
\usepackage{mathtools, cuted}
\usepackage{lipsum}
\usepackage{graphicx}
\usepackage{booktabs}                                                       
\usepackage{amssymb}
\usepackage{amsmath}
\usepackage{amsthm}
\usepackage{diagbox}
\usepackage{multirow}
\usepackage{lineno}
\usepackage{bm}
\usepackage{cite}
\usepackage{float}
\usepackage{subfig}
\usepackage[ruled,linesnumbered]{algorithm2e}
\usepackage{color}
\usepackage{ragged2e}
\usepackage{lscape}
\usepackage{threeparttable} 
\usepackage{acronym}
\usepackage{glossaries}
\newtheorem{theorem}{Theorem}

\newtheorem{lemma}{Lemma}
\newtheorem{proposition}{Proposition}
\usepackage[justification=centering]{caption}
\newtheoremstyle{noparens}%
{}{}%
{\itshape}{}%
{\bfseries}{.}%
{ }%
{}

\theoremstyle{noparens}
\newtheorem{mylemma}[lemma]{Lemma}
\newtheorem{mytheorem}[theorem]{Theorem}

\newtheorem{myproposition}[proposition]{Proposition}

\newtheoremstyle{boldtitle}
{}{}
{\normalfont} 
{0pt}
{\bfseries} 
{.}{ }{}
\theoremstyle{boldtitle}
\newtheorem{myremark}{Remark}

\makeglossaries
\setglossarystyle{alttree}
\glssetwidest{NNNNNNNNNNNNN} 
\newacronym[sort={1}]{AMF}{AMF}{adaptive matched filter}
\newacronym[sort={2}]{ACE}{ACE}{adaptive coherent estimator}
\newacronym[sort={3}]{CFAR}{CFAR}{constant false alarm rate}
\newacronym[sort={4}]{DL}{DL}{diagonally loaded}
\newacronym[sort={5}]{EL}{EL}{expected likelihood}
\newacronym[sort={6}]{GLRT}{GLRT}{generalized likelihood ratio test}
\newacronym[sort={7}]{IID}{i.i.d.}{independent and identically distributed}
\newacronym[sort={8}]{LDR}{LDR}{large dimensional regime}
\newacronym[sort={9}]{NP}{NP}{Neyman-Pearson}
\newacronym[sort={10}]{PDF}{PDF}{probability density function}
\newacronym[sort={11}]{RMT}{RMT}{random matrix theory}
\newacronym[sort={12}]{ROC}{ROC}{receiver operating characteristic}
\newacronym[sort={13}]{SCM}{SCM}{sample covariance matrix}
\newacronym[sort={14}]{SCNR}{SCNR}{signal-to-clutter-plus-noise ratio}

\newacronym[sort={15}]{DL-AMF}{DL-AMF}{diaginally loaded AMF}
\newacronym[sort={16}]{EL-AMF}{EL-AMF}{EL based AMF}
\newacronym[sort={17}]{EL-GLRT}{EL-GLRT}{EL based GLRT}
\newacronym[sort={18}]{CFAR-DL-AMF}{CFAR-DL-AMF}{CFAR DL adaptive matched filter}
\newacronym[sort={19}]{CFAR-DL-SCMF}{CFAR-DL-SCMF}{CFAR DL semi-clairvoyant matched filter}
\newacronym[sort={20}]{CFAR-EL-AMF}{CFAR-EL-AMF}{CFAR version of EL-AMF}
\newacronym[sort={21}]{opt-CFAR-DL-AMF}{opt-CFAR-DL-AMF}{optimal CFAR-DL-AMF}
\newacronym[sort={22}]{opt-CFAR-DL-SCMF}{opt-CFAR-DL-SCMF}{optimal CFAR-DL-SCMF}

\begin{document}
\captionsetup[figure]{name={Fig.},labelsep=period,font=small,justification=justified} 
\captionsetup[table]{name={TABLE},labelsep=newline,font=small} 
\title{On the Asymptotic Performance of Diagonally Loaded Detectors for Large Arrays: To Achieve CFAR and Optimality}

\author{Jie~Zhou, ~Junhao~Xie,~\IEEEmembership{Senior Member,~IEEE} 
\thanks{ This work was supported in part by the National Natural Science Foundation of China under Project 62371162. \textit{(Corresponding author: Junhao Xie.)}}
\thanks{The authors are with Department of Electronic Engineering,
Harbin Institute of Technology, Harbin 150001, China (e-mail: 22b905037@stu.hit.edu.cn; xj@hit.edu.cn). }
}

%

%
\markboth{Submitted to IEEE Transactions on Aerospace and Electronic Systems}{SHORT ARTICLE TITLE}
\maketitle
\begin{abstract}
This paper addresses two critical limitations in  diagonally loaded (DL) adaptive matched filter (AMF) detector: (1) the lack of constant false alarm rate (CFAR) property with respect to arbitrary covariance matrices, and (2) the absence of selection criteria for optimal loading factor from the perspective of maximizing the detection probability ($P_{\rm d}$). We provide solutions to both challenges through a comprehensive analysis for the asymptotic performance of DL-AMF under large dimensional regime (LDR) where the dimension $N$ and sample size $K$ tend to infinity whereas their ratio $N/K$ converges to a constant $c\in(0,1)$. The analytical results show that any DL detectors constructed by normalizing the random variable $|\hat{\alpha}_{\rm DL}(\lambda)|^2=|\mathbf{s}^H(\hat{\mathbf{R}}+\lambda\mathbf{I}_N)^{-1}\mathbf{y}_0|^2$ with a deterministic quantity or a random variable that converges almost surely to a deterministic value will exhibit equivalent performance under LDR. Hence, following this idea, we derive two CFAR DL detectors: CFAR DL semi-clairvoyant matched filter (CFAR-DL-SCMF) detector and CFAR DL adaptive matched filter (CFAR-DL-AMF) detector, by normalizing $|\hat{\alpha}_{\rm DL}(\lambda)|^2$ with an appropriate deterministic quantity and its consistent estimate, respectively. The theoretical analysis and simulations show that both CFAR-DL-SCMF and CFAR-DL-AMF achieve CFAR with respect to covariance matrix, target steering vector and loading factor. Furthermore, we derive the asymptotically optimal loading factor $\lambda_{\rm opt}$ by maximizing the explicit expression of asymptotic $P_{\rm d}$. For practical implementation, we provide a consistent estimator for $\lambda_{\rm opt}$ under LDR. Based on $\lambda_{\rm opt}$ and its consistent estimate, we establish  the optimal CFAR-DL-SCMF (opt-CFAR-DL-SCMF) and the optimal CFAR-DL-AMF (opt-CFAR-DL-AMF). Numerical examples validate our theoretical results and demonstrate that the proposed opt-CFAR-DL-SCMF and opt-CFAR-DL-AMF consistently outperform expected likelihood based AMF and persymmetric AMF in both full-rank and low-rank clutter plus noise environments.
\end{abstract}

\printglossary[type=\acronymtype, nonumberlist,title={}]
\section{Introduction}
\subsection{Background and Related Works}
Detecting signals embedded in clutter-plus-noise with unknown correlation properties has gained considerable attention in many fields, including radar, sonar, and communications. Particularly, in the context of radar detection, if the information of both target and clutter-plus-noise environment is exactly known, an optimal detector is easily designed via the \gls{NP} criterion. Unfortunately, in practical applications, such information is typically unavailable and must be estimated from finite observations. This requisite leads to the development of adaptive detection techniques, such as Kelly's \gls{GLRT} \cite{Kelly1986An}, the \gls{AMF} test \cite{Robey1992A}, the \gls{ACE} test \cite{Kraut2005The} and Rao test \cite{Maio2007Rao}. 

In these detectors, a standard routine involves estimating the unknown correlation properties of the background by employing the \gls{SCM} obtained from the secondary data.  Typically, these secondary data are subject to certain standard statistical assumptions, such as being \gls{IID} and containing no contributions of targets, so that the SCM possesses some favorable properties. One of the most crucial properties of SCM is its consistency in the large sample size regime, where the sample size (denoted by $K$) tends to infinity while the dimension (denoted by $N$) remains fixed. This consistency ensures the excellent performance of SCM-based detectors in the case where the sample size $K$ significantly exceeds the dimension $N$.

In practice, however, due to the non-stationary nature of background, as well as the potential presence of active or passive interference from external sources, the available data is always limited. We frequently encounter the insufficient sample scenarios where the sample size $K$ is comparable in magnitude to (sometimes even smaller than) the dimension $N$. In this case, the SCM fails to provide a reliable estimate of the population covariance matrix, thereby leading to a significant degradation in the detection performance of adaptive detectors. For instance, Robey, Fuhrmann, Kelly and Nitzberg (RFKN) \cite{Robey1992A} demonstrated that the SCM-based AMF suffers approximately 3 dB detection loss compared to the optimal detector in the case $K=2N$. 
Similarly, Kelly's GLRT also incurs nearly identical performance loss under the same situation \cite{Kelly1986An}.

Over the past decades, numerous strategies have been proposed to address this issue. The general idea of these methods is to effectively employ the priori information, such as a known array geometry structure or some knowledge about the background. A typical example of the specific structure is the persymmetry of the covariance matrix \cite{Pailloux2011Persymmetric, Gao2014Persymmetric, Liu2015On}, which arises in active radar systems using symmetrically spaced linear arrays for spatial processing or symmetrically spaced pulses train for temporal processing. Other commonly used structures include Toeplitz \cite{Fuhrmann1991Application, Du2020Toeplitz} and Kronecker \cite{Raghavan2017CFAR}. Both the theoretical analysis and experiments have demonstrated that such structure-based methods enhance the detection performance in insufficient sample scenario. However, there is no evidence that these detectors achieve optimality. Moreover, their performance heavily relies on the assumed structural properties. The deviations between the assumed and true covariance matrices may lead to performance degradation.

An alternative strategy to mitigate performance decline in insufficient sample scenarios is to leverage prior knowledge about the background. Such knowledge is typically obtained through extensive and comprehensive analysis of empirical radar data. For example, several studies (see e.g., \cite{Abramovich1981Controlled, Klemm1983Adaptive, Kirsteins1994Adaptive, Jain2023Radar}) have reported that the clutter acquired by airborne radar systems is composed of a high power low-rank clutter component plus a white Gaussian noise component. Consequently, the SCM of such clutter exhibits the so-called low-rank structure, whose covariance matrix can be mathematically expressed as the sum of a low-rank matrix and an identity matrix scaled by a loading factor. 
Based on the low-rank structure, researchers have developed \gls{DL} detectors to enhance the detection performance in insufficient sample scenarios. In fact, the concept of DL detectors originates from diagonal loading techniques used for adaptive filtering and/or adaptive beamformer. Since the 1980s, researchers \cite{Abramovich1981Controlled, Carlson1988Covariance} have observed that incorporating a diagonal matrix into the SCM enhances the interference suppression performance of adaptive filters/beamformers, particularly in the insufficient sample scenario.  A key challenge in DL techniques is the determination of the loading factor. Most of the existing work has addressed this problem from an estimation perspective, such as by minimizing the mean square error \cite{Maio2019Loading, Pajovic2019Analysis}. Furthermore, some studies have explored the selection of loading factor by maximizing the output \gls{SCNR} of the beamformer \cite{Kim1998Optimal,Mestre2006Finite}. 

A major breakthrough in the development of DL detectors was achieved by Abramovich, Spencer and Gorokhov (ASG) \cite{Abramovich2007Modified}. They proposed the \gls{EL} approach as an alternative to the maximum likelihood method for estimating the loading factor.  The core idea of the EL approach is to adjust the loading factor such that the likelihood ratio derived from the observations aligns as closely as possible with the true one. Based on the EL estimation, ASG \cite{Abramovich2007Modified} developed adaptive detectors, including \gls{EL-AMF} and \gls{EL-GLRT}. While simulation experiments demonstrate that these detectors exhibit near-optimal performance, they face two critical limitations. First, their near optimality lacks rigorous theoretical proof and we don't know whether the EL-based detectors reach the optimality. Second, as shown in \cite{Abramovich2007Modified}, EL-based detectors only maintain the \gls{CFAR} property for low-rank covariance matrices and fail to achieve CFAR for full-rank covariance matrices. As a result, ASG's EL-AMF and EL-GLRT can not guarantee CFAR for arbitrary covariance matrices, severely restricting their practical applicability. 

This paper addresses the above two limitations of EL-based detectors. We aim to design an adaptive detector that achieves the following two goals: (1) to maintain the CFAR property with respect to arbitrary covariance matrix and (2) to achieve the optimal detection performance in the sense of NP criterion. To accomplish these two goals, it is necessary to conduct a comprehensive theoretical analysis of false alarm and detection performance of DL detectors.  Unlike traditional approaches that focus on the large sample size regime, our work considers the \gls{LDR}, a standard asymptotic framework in \gls{RMT}.  In the LDR, both the dimension $N$ and the sample size $K$ are assumed to tend to infinity, while their ratio $N/K$ converges to a constant $c\in(0,1)$.  It should be noted that this new asymptotic setting is not an abstract or unrealistic assumption; rather, it is particularly relevant in the context of modern radar systems where the widespread adoption of large-scale antenna arrays, such as those used in multiple-input multiple-output and phased array radars, has significantly increased the dimensionality of the data. Meanwhile, the demand for high-resolution detection and/or tracking in complex environments requires the processing of larger sample sizes. Consequently, both the dimension and the sample size of the data to be processed are growing in practical applications, making the LDR a natural and realistic framework for advanced radar signal processing. Furthermore, we highlight that the traditional large sample size regime can be regarded as a special case of the LDR when $c\to0$. Thus the results in this paper are also applicable to scenarios with large samples, effectively bridging the gap between different theoretical frameworks.
\subsection{Overview of Main Ideas and Contributions}

\subsubsection{Main idea for achieving CFAR}
We emphasize that the CFAR property is crucial for practical adaptive detectors because it ensures that the detection threshold remains invariant to unknown disturbance parameters. This invariance allows for straightforward determination of threshold without prior knowledge of the environment. In contrast, the non-CFAR detectors require their thresholds to be determined as functions of unknown disturbance parameters. This typically necessitates computationally intensive Monte Carlo simulations for each new operational scenario.

The main idea for achieving CFAR in this paper is inspired by the asymptotic property of the test statistic of AMF \cite{Zhou2024On}. 
As suggested by \cite{Zhou2024On}, the decision rule of AMF can be expressed by
\begin{equation}\label{key}
	\hat{\eta}_{\rm AMF} = \frac{|\hat{\alpha}|^2}{\hat{\beta}} \mathop\gtrless\limits_{H_0}^{H_1}\tau_{\rm AMF}
\end{equation}
where $H_1$ and $H_0$ represent the hypotheses of target presence and absence, respectively; $\tau_{\rm AMF}$ is the threshold; $\hat{\alpha}$ and $\hat{\beta}$ are two bilinear forms, defined by $
	\hat{\alpha} = \mathbf{s}^H{\hat{\mathbf{R}}}^{-1}\mathbf{y}_0$
and $
	\hat{\beta} =\ \mathbf{s}^H\hat{\mathbf{R}}^{-1}\mathbf{s}$. Here, $\mathbf{s}\in\mathbb{C}^{N\times 1}$ represents the known steering vector; $\hat{\mathbf{R}}\in\mathbb{C}^{N\times N}$ is the SCM and $\mathbf{y}_0\in\mathbb{C}^{N\times 1}$ is the primary data under test.  Under $H_0$, $\mathbf{y}_0$ is assumed to follow an $N$-dimensional complex Gaussian distribution with mean $\mathbf{0}$ and covariance matrix $\mathbf{R}\in\mathbb{C}^{N\times N}$.  It is shown in \cite{Zhou2024On} that, under $H_0$ and LDR (i.e., as $N,K\to\infty, N/K\to c\in(0,1)$), the following asymptotic properties hold:
\begin{itemize}
	\item  $|\hat{\alpha}|^2$ converges in distribution to an Exponential distribution with scale parameter $\mu = \frac{\mathbf{s}^H\mathbf{R}^{-1}\mathbf{s}}{(1-c)^3}$;
	\item $\hat{\beta}$ almost surely converges to a deterministic quantity $\widetilde{\beta} = \frac{\mathbf{s}^H\mathbf{R}^{-1}\mathbf{s}}{1-c}$.
\end{itemize} 
These two convergence results imply that $\hat{\eta}_{\rm AMF}$ converges in distribution to an Exponential distribution with scale parameter ${\mu}/{\widetilde{\beta}} = \frac{1}{(1-c)^2}$ (from Slutsky’s Theorem). It is observed that the term $\mathbf{s}^H\mathbf{R}^{-1}\mathbf{s}$ cancels out in the limiting distribution, thereby resulting in the CFAR property with respect to an arbitrary covariance matrix $\mathbf{R}$. 

The above analysis implies that $\hat{\beta}$ plays a crucial role in maintaining the CFAR property by eliminating the influence of the term associated with the population covariance matrix $\mathbf{R}$. Moreover, one can see that $\hat{\beta}$ remains invariant under both $H_0$ and $H_1$, implying that $\hat{\beta}$ has no impact on the detection performance. In summary, we conclude that the random variables $\hat{\alpha}$ and $\hat{\beta}$ play distinct roles in the AMF:
\begin{itemize}
	\item $\hat{\alpha}$ determines the detection performance;
	\item  $\hat{\beta}$ guarantees the CFAR property.
\end{itemize}
This conclusion offers a crucial insight: to enhance the detection performance of AMF, it suffices to perform diagonal loading solely on $\hat{\alpha}$, rather than on both $\hat{\alpha}$ and $\hat{\beta}$. In fact, as we will see in Section II, the detectors established by scaling the diagonally loaded $\hat{\alpha}$ with any deterministic quantity—or by a random variable that converges almost surely to a deterministic quantity—will tend to exhibit equivalent detection performance under LDR. This equivalence means that they share a unified expression for the \gls{ROC}. For example, if we define the diagonally loaded $\hat{\alpha}$ and $\hat{\beta}$ respectively by
\begin{equation}\label{4}
	 \hat{\alpha}_{ \rm DL}(\lambda) = \mathbf{s}^H\left(\hat{\mathbf{R}}+\lambda\mathbf{I}_N\right)^{-1}\mathbf{y}_0
\end{equation}
and 
\begin{equation}\label{5}
	\hat{\beta}_{\rm DL}(\lambda) = \mathbf{s}^H\left(\hat{\mathbf{R}}+\lambda\mathbf{I}_N\right)^{-1}\mathbf{s}
\end{equation}
where $\lambda$ is the loading factor and $\mathbf{I}_N$ is an $N\times N$ identity matrix, then the detection performance of the following three detectors tend to be equivalent under LDR (as we will see in Section II):
\begin{equation}\label{8}
	\frac{|\hat{\alpha}_{\rm DL}(\lambda)|^2}{\hat{\beta}_{\rm DL}(\lambda)}\mathop\gtrless\limits_{H_0}^{H_1}\tau_1,
\end{equation}
\begin{equation}\label{7}
	\frac{|\hat{\alpha}_{\rm DL}(\lambda)|^2}{\hat{\beta}}\mathop\gtrless\limits_{H_0}^{H_1}\tau_2,
\end{equation}
and
\begin{equation}\label{6}
	|\hat{\alpha}_{\rm DL}(\lambda)|^2\mathop\gtrless\limits_{H_0}^{H_1}\tau_3
\end{equation}
where $\tau_1$, $\tau_2$ and $\tau_3$ are thresholds.  Notably, the detector \eqref{8},  referred to as the DL-AMF, is the traditional DL detector that has been widely studied in existing adaptive radar studies \cite{Abramovich2007Modified,Maio2019Loading}.

Unfortunately, none of the above DL detectors are able to maintain the CFAR property with respect to $\mathbf{R}$. To achieve CFAR, we need to identify a random variable that plays a role analogous to that of $\hat{\beta}$ in the AMF.  To this end, we need to explore the asymptotic distribution of $|\hat{\alpha}_{\rm DL}(\lambda)|^2$ under $H_0$ within the LDR setting. As we will see in Section II, this distribution is an Exponential distribution whose scale parameter is a function of $\mathbf{R}$, $\mathbf{s}$ and $\lambda$, denoted as $\mu_0(\mathbf{R},\mathbf{s},\lambda)$. Then in order to achieve CFAR, a natural idea is to normalize $|\hat{\alpha}_{\rm DL}(\lambda)|^2$ with $\mu_0(\mathbf{R},\mathbf{s},\mathbf{\lambda})$, yielding the following detector
\begin{equation}\label{9}
		\frac{|\hat{\alpha}_{\rm DL}(\lambda)|^2}{\mu_0(\mathbf{R},\mathbf{s},\lambda)}\mathop\gtrless\limits_{H_0}^{H_1}\tau.
\end{equation}
Since $|\hat{\alpha}_{\rm DL}(\lambda)|^2$ is asymptotically Exponentially distributed with scale parameter $\mu_0(\mathbf{R},\mathbf{s},\lambda)$, it is straightforward to show that the test statistic of detector \eqref{9} asymptotically follows a standard Exponential distribution (with scale parameter $1$). Obviously, this distribution is invariant to $\mathbf{R}$, demonstrating that the detector \eqref{9} achieves CFAR property. We refer to the detector in \eqref{9} as the \textit{\gls{CFAR-DL-SCMF} detector} because it retains a form similar to the matched filter detector whereas relying on the unknown covariance matrix $\mathbf{R}$. Owing to this reliance on the unknown parameter, it actually cannot be implemented in practice. Therefore, for practicability, it is necessary to derive a consistent estimate for $\mu_0(\mathbf{R},\mathbf{s},\lambda)$. Importantly, we must ensure consistency of this estimate in LDR rather than the traditional large sample size regime. Denoting this consistent estimate as $\widehat{\mu_0(\mathbf{R},\mathbf{s},\lambda)}$ and replacing the original $\mu_0(\mathbf{R},\mathbf{s},\lambda)$, we readily obtain an adaptive version of CFAR-DL-SCMF, given by
\begin{equation}\label{10}
	\frac{|\hat{\alpha}_{\rm DL}(\lambda)|^2}{\widehat{\mu_0(\mathbf{R},\mathbf{s},\lambda)}}\mathop\gtrless\limits_{H_0}^{H_1}\tau.
\end{equation}
The detector \eqref{10} is termed the \textit{\gls{CFAR-DL-AMF} detector}. Because $\widehat{\mu_0(\mathbf{R},\mathbf{s},\lambda)}$ almost surely converges to $\mu_0(\mathbf{R},\mathbf{s},\lambda)$ under LDR, the test statistic of CFAR-DL-AMF shares the same asymptotic distribution as that of detector \eqref{9}. Consequently, the CFAR-DL-AMF achieves the CFAR property. 

\subsubsection{Main idea for achieving optimality}
 Now the key issue we aim to address is: what value of $\lambda$ will enable the DL detector to achieve optimality?  As we have introduced in previous subsection, the existing studies focus on determining $\lambda$ from an estimation perspective. Although the DL detectors using these estimated $\lambda$ exhibit significant performance improvement compared to the SCM-based AMF, neither theoretic analysis nor experiment results demonstrate that they have reached optimal detection performance. 

From the perspective of detection theory (rather than estimation theory), an optimal detector should be designed under the NP criterion, i.e., by maximizing the  detection probability ($P_{\rm d}$) for a given false alarm probability ($P_{\rm fa}$). Hence, in order to construct an optimal DL detector,  it is necessary to derive the expression for $P_{\rm d}$, which is expected to depend on $\lambda$. Subsequently, the optimal value of $\lambda$ should be determined by maximizing $P_{\rm d}$. Nevertheless, deriving an exact expression for $P_{\rm d}$ currently poses a major challenge due to the difficulty in determining the exact statistical distribution of the test statistic for the DL detector under the $H_1$ hypothesis. 

An alternative solution is to derive the asymptotic distribution of the test statistic. Using tools from RMT, we establish the asymptotic distribution for the test statistics of CFAR-DL-SCMF and CFAR-DL-AMF under $H_1$, with the target following Swerling 0 and Swerling I models.  From this asymptotic distribution, the analytical expression for asymptotic $P_{\rm d}$ is easily derived as a function of $\lambda$, denoted by $\widetilde{P}_{\rm d}(\lambda)$. The optimal $\lambda$, denoted by $\lambda_{\rm opt}$,  is therefore obtained by 
\begin{equation}\label{key}
	\lambda_{\rm opt} = \arg\max_{\lambda} \widetilde{P}_{\rm d}(\lambda).
\end{equation}
Unfortunately, as will be shown in Section IV, $\lambda_{\rm opt}$ depends on the true covariance matrix $\mathbf{R}$, making it impractical for real-world implementation. To overcome this limitation, we further derive a consistent estimate for $\lambda_{\rm opt}$, denoted by $\hat{\lambda}_{\rm opt}$. Importantly, this consistency is established within the LDR framework.
Thanks to the consistency, the DL detector employing $\hat{\lambda}_{\rm opt}$ performs equivalently to the detector based on $\lambda_{\rm opt}$ under LDR, thereby achieving optimality.

\subsubsection{Contributions}
The main contributions of this work are as follows:
\begin{itemize}
	\item \textbf{Explicit expressions for asymptotic $P_{\rm fa}$ and $P_{\rm d}$ of DL-AMF under LDR:} A rigorous theoretical analysis of the performance of DL-AMF detector is fundamental to understanding its behavior. Unfortunately, to our best knowledge, an analytical analysis of the performance of DL-AMF has remained a significant challenge. In this work, we overcome this limitation through the tools from RMT. We successfully derive the explicit expressions for asymptotic $P_{\rm fa}$ and $P_{\rm d}$ of  DL-AMF under LDR. These theoretical results yields several important insights. First, it is shown that the asymptotic $P_{\rm fa}$ depends explicitly on $\mathbf{R}$, $\mathbf{s}$, and $\lambda$, which demonstrates that DL-AMF inherently fails to maintain CFAR with respect to these parameters. Furthermore, the analysis of the asymptotic $P_{\rm d}$ indicates that the performance advantage of DL-AMF over SCM-based AMF is not guaranteed for arbitrary loading factors. In some cases (specifically when $\frac{1}{(\mathbf{s}^H\mathbf{R}^{-1}\mathbf{s})(\mathbf{s}^H\mathbf{R}\mathbf{s})}<1-c$), an excessive loading factor may actually degrade DL-AMF's detection performance. Most significantly, we prove that the detection performance of a DL-AMF under LDR depends only on $|\hat{\alpha}_{\rm DL}(\lambda)|^2$. This critical discovery serves as the theoretical foundation for the development of CFAR version of DL detectors.
	\item \textbf{Breakthrough in achieving CFAR property for DL detectors:} For decades, the absence of CFAR property of DL detectors has significantly limited their applicability in real-world scenarios. This work presents a complete solution to this longstanding problem. We prove that the detection performance of DL detectors depends solely on the random variable $|\hat{\alpha}_{\rm DL}(\lambda)|^2$, which asymptotically follows an Exponential distribution  with scale parameter $\mu_0(\mathbf{R},\mathbf{s},\lambda)$ under $H_0$. This finding motivates us to construct the CFAR-DL-SCMF detector by normalizing $|\hat{\alpha}_{\rm DL}(\lambda)|^2$ with $\mu_0(\mathbf{R},\mathbf{s},\lambda)$. We further develop a consistent estimator for $\mu_0(\mathbf{R},\mathbf{s},\lambda)$ under LDR, denoted by $\widehat{\mu_0(\mathbf{R},\mathbf{s},\lambda)}$. Substituting this estimator yields the adaptive version of CFAR-DL-SCMF, i.e., CFAR-DL-AMF detector. Both theoretical analysis and numerical simulations demonstrate that CFAR-DL-SCMF and CFAR-DL-AMF maintain CFAR properties with respect to $\mathbf{R}$, $\mathbf{s}$ and $\lambda$ under LDR. Particularly, the significance of CFAR-DL-AMF framework lies in its capability that enables any existing DL detectors to achieve CFAR property without compromising their original detection performance. As a concrete example, we apply this framework to ASG's EL-AMF \cite{Abramovich2007Modified} and obtain a \gls{CFAR-EL-AMF}, which simultaneously attains CFAR property and preserves EL-AMF's superior detection performance.
	\item \textbf{Determination of optimal loading factor by maximizing the detection probability:} Previous approaches to determining optimal loading factors have been primarily estimation-driven, inherently incapable of developing detectors with optimal performance. In this work, we fundamentally reformulate this problem by seeking the optimal loading factor under the NP criterion, i.e., by maximizing $P_{\rm d}$ for a given $P_{\rm fa}$. This breakthrough is enabled by the explicit expression for the asymptotic $P_{\rm d}$ of a DL detector under LDR.	By maximizing the asymptotic $P_{\rm d}$, we obtain an optimal loading factor $\lambda_{\rm opt}$. Incorporating $\lambda_{\rm opt}$ into CFAR-DL-SCMF yields the \gls{opt-CFAR-DL-SCMF} detector. For practical implementation, we further develop a consistent estimator for $\lambda_{\rm opt}$ under LDR, based on which the \gls{opt-CFAR-DL-AMF} is established. Experimental results demonstrate that both detectors consistently outperform EL-AMF and persymmetric AMF, two famous detectors for their excellent performance in insufficient sample scenarios, in both full-rank clutter and low-rank clutter plus noise background. 
\end{itemize}
To provide readers with a clear and intuitive comparison of the key differences between the proposed detectors and the competitors, we systematically summarize the core characteristics of various detectors in Table \ref{tab:stap_comparison} and offer a quick-reference overview of the advantages of the proposed detectors.

\vspace{-1em}
\begin{table*}[h]
	\centering
	\caption{Summary of the Characteristics of Different Detectors}
	\label{tab:stap_comparison}
	\begin{tabular}{>{\raggedright\arraybackslash}p{3cm}ccccc}
		\toprule
		\textbf{Detector Type} & \textbf{Test statistic}&\textbf{CFAR} & \textbf{Optimality} & \textbf{Practicality} & \textbf{Performance in insufficient sample scenarios} \\
		\midrule
		SCM-based AMF \cite{Robey1992A}&See \cite[Eq. (8) ]{Robey1992A}& $\checkmark$ & $\times$& $\checkmark$ & Poor \\
		\addlinespace
		
		Persymmetric AMF \cite{Pailloux2011Persymmetric} &See \cite[Eq. (25)]{Pailloux2011Persymmetric}& $\checkmark$& $\times$ & $\checkmark$ & Moderate \\
		\addlinespace
		
		ASG's EL-AMF \cite{Abramovich2007Modified} & See \cite[Eq. (110)]{Abramovich2007Modified}&$\times$ & $\times$ (near-optimal) & $\checkmark$ & Good \\
		\addlinespace
		
		CFAR-EL-AMF &See \eqref{70-1} & $\checkmark$& $\times$(near-optimal) & $\checkmark$ & Good \\
		\addlinespace
		
		CFAR-DL-SCMF &See \eqref{55}& $\checkmark$ & $\times$ & $\times$ &  Good \\
		\addlinespace
		
		CFAR-DL-AMF &See \eqref{66-1}& $\checkmark$ & $\times$ & $\checkmark$ &  Good  \\
		\addlinespace
		
		opt-CFAR-DL-SCMF&See \eqref{76-1} & $\checkmark$ & $\checkmark$ & $\times$ &  Good  \\
		\addlinespace
		
		opt-CFAR-DL-AMF &See \eqref{81}& $\checkmark$ & $\checkmark$ & $\checkmark$ &   Good \\
		\bottomrule
	\end{tabular}
	\vspace{0.2cm}
	\footnotesize
	\begin{tabular}{p{\textwidth}}
		$\checkmark$ = Fully satisfies, $\times$\ = Does not satisfy 
	\end{tabular}
\vspace{-1em}
\end{table*}
\subsection{Organization and Notations}
\noindent\textit{Organization:} The rest of this paper is organized as follows. Section II presents a rigorous asymptotic analysis for the performance of DL-AMF under LDR. Building on these theoretical foundations, Section III introduces the proposed CFAR-DL-SCMF and CFAR-DL-AMF. Section IV investigates the determination of optimal loading factor with its consistent estimator under LDR. Section V conducts a comparative study on the detection performance of the proposed opt-CFAR-DL-SCMF and opt-CFAR-DL-AMF against some competitors, including SCM-based AMF, CFAR version of ASG's EL-AMF and persymmetric AMF. Finally, Section VI concludes the paper and discusses promising directions for future research.

~\\
\noindent\textit{Notations:} Throughout the paper, the scalars, vectors and matrices are denoted by  lowercase, boldface lowercase and boldface uppercase, respectively. $(\cdot)^T$ and $(\cdot)^H$ represent transpose and conjugate transpose, respectively. $\mathbb{E}(\cdot)$, ${\rm Var}(\cdot)$ and ${\rm tr}(\cdot)$ denotes expectation, variance and trace, respectively. We denote convergence in distribution by $\overset{d}{\longrightarrow}$ and almost convergence by $\overset{a.s.}{\longrightarrow}$.  ${\rm Re}(z)$, ${\rm Im}(z)$ and $\bar{z}$ represent the real part, imaginary part and conjugate of the complex number $z$, respectively. $\mathsf{i}$ denotes the imag  inary unit.  $\mathbb{R}$ and $\mathbb{C}$ respectively denote the real and complex
fields with dimension specified by a superscript, and $\mathbb{C}^+ \equiv\{
z\in \mathbb{C}:\rm{Im}(z)>0\}$. The notation ${\rm eig}(\mathbf{R})$ denotes the set of eigenvectors of $\mathbf{R}$. Moreover, the following conventions are used:
\begin{itemize}
		\item$\mathbf{x}\sim\mathcal{CN}(\mathbf{0},\mathbf{R})$: the vector $\mathbf{x}$ follows an $N$-dimensional
	complex Gaussian distribution with mean $\mathbf{0}$ and covariance matrix $\mathbf{R}$.	
	\item $x\sim \mathcal{N}(\mu,\sigma^2)$:  $x$ follows a real Gaussian distribution with mean $\mu$ and variance $\sigma^2$.
	\item $x\sim{\rm Exp}(\mu)$: $x$ follows an Exponential distribution with scale parameter $\mu$, whose \gls{PDF} is given by $f(x) = \frac{1}{\mu}\exp\left(-\frac{x}{\mu}\right)$.
	\item $x\sim\chi^2(\nu,\sigma^2)$: $x$ follows a non-central chi-squared distribution of degree $2$ with parameters $\nu$ and $\sigma^2$, whose PDF and distribution function are respectively given by $f(x;\nu,\sigma^2)  =\frac{1}{2\sigma^2}\exp\left(-\frac{x+\nu}{2\sigma^2}\right)I_0\left(\frac{\sqrt{\nu x}}{\sigma^2}\right)$ and $F(x;\nu,\sigma^2) = 1-Q\left(\frac{\sqrt{\nu}}{\sigma},\frac{\sqrt{x}}{\sigma}\right)$. $I_0(x)$ denotes the modified Bessel function of the first kind of order 0 and $Q(a,b)$ denotes the Marcum Q-function.

\end{itemize}

\section{Asymptotic Performance of DL-AMF under LDR}
We start our analysis by examining the asymptotic performance (including $P_{\rm fa}$ and $P_{\rm d}$) of the DL-AMF (see \eqref{8}) under LDR. For clarity, we denote the test statistic of DL-AMF by 
\begin{equation}\label{12}
	\hat{\eta}_{_{\rm DL}} = \frac{|\hat{\alpha}_{\rm DL}(\lambda)|^2}{\hat{\beta}_{\rm DL}(\lambda)}.
\end{equation} 
Apparently, $\hat{\eta}_{_{\rm DL}}$ is a function of the random variables $\hat{\alpha}_{\rm DL}(\lambda)$ and $\hat{\beta}_{\rm DL}(\lambda)$. Hence, to derive the explicit expressions for asymptotic $P_{\rm fa}$ and $P_{\rm d}$ of DL-AMF, it is necessary to investigate the asymptotic behaviors of $\hat{\alpha}_{\rm DL}(\lambda)$ and $\hat{\beta}_{\rm DL}(\lambda)$ under the LDR framework.
%

Before proceeding, we clarify some standard assumptions used in this paper. We consider the LDR framework, which is mathematically represented by the following assumption:
\begin{itemize}
	\item (A1) As $N,K\to\infty$, $N/K\to c\in (0,1)$.
\end{itemize}

Regarding the secondary data $\mathbf{y}_1,...,\mathbf{y}_K$, we assume:
\begin{itemize}
	\item (A2) $\mathbf{y}_k=\mathbf{c}_k$ ($k=1,...,K$) follows independent and identically distributed (i.i.d.) $N$-dimensional complex circular Gaussian distribution with mean $\mathbf{0}$ and covariance matrix $\mathbf{R}\in\mathbb{C}^{N\times N}$. Furthermore, $\mathbf{R}$ is an Hermitian matrix with $N$ eigenvalues $\rho_1\leqslant\rho_2\leqslant\cdots\leqslant\rho_N$. There exists $C_{\rm min}$ and $C_{\rm max}$ such that $0<C_{\rm min}<\inf_N(\rho_1)\leqslant\sup_N(\rho_N)<C_{\rm max}<\infty$ for all large $N$.
\end{itemize}

Regarding the primary data
\begin{equation}\label{key}
	\mathbf{y}_0=\begin{cases}
		\mathbf{c}_0,& \text{under $H_0$}\\
		b\mathbf{s}+\mathbf{c}_0,&\text{under $H_1$}
	\end{cases},
\end{equation}
 we assume that:
\begin{itemize}
	\item (A3) $\mathbf{c}_0$ and $\mathbf{c}_k$ ($k=1,2...,K$) are i.i.d. $\mathbf{s}$ is an $N$ dimensional steering vector and satisfies $\mathbf{s}^H\mathbf{s}=1$;
	\item (A4) For the Swerling 0 target, $b$ is a deterministic complex number. For the Swerling I target, $b$ is a complex random variable with i.i.d. real and imaginary parts, satisfying ${\rm Re}(b)\sim\mathcal{N}(0,\frac{\sigma_t^2}{2})$.
\end{itemize}

\subsection{Asymptotic $P_{\rm fa}$ of DL-AMF under LDR}
In this subsection, we provide the asymptotic distribution of $\hat{\eta}_{_{\rm DL}}$ under $H_0$ and derive the asymptotic $P_{\rm fa}$ of DL-AMF under LDR.
	
We first present the asymptotic distribution of $\hat{\alpha}_{\rm DL}(\lambda)$ under $H_0$ in the following theorem.

\begin{mytheorem}\label{Theorem asymptotic distribution of alpha}
		Let assumptions (A1)-(A4) hold true. Then under $H_0$ and as $N,K\to\infty$, it holds that 
		\begin{equation}\label{key}
			\frac{{\rm Re}(\hat{\alpha}_{\rm DL}(\lambda))-\mu_{\hat{\alpha}_{\rm DL},H_0}}{\sigma_{\hat{\alpha}_{\rm DL},H_0}}\overset{d}{\longrightarrow}\mathcal{N}(0,1)
		\end{equation}
		where $\mu_{\hat{\alpha}_{\rm DL},H_0} = 0$ and $\sigma^2_{\hat{\alpha}_{\rm DL},H_0} = \frac{\mathbf{u}^H\mathbf{E}(\lambda)^2\mathbf{u}}{{2(1-\gamma(\lambda))}}$ with $\mathbf{u} = \mathbf{R}^{-1/2}\mathbf{s}$, $\mathbf{E}(\lambda) = \left(\delta\mathbf{I}_N+\lambda\mathbf{R}^{-1}\right)^{-1}$, $\gamma(\lambda) = \frac{\delta^2}{K}{\rm tr}\mathbf{E}(\lambda)^2$ and $\delta$ being the unique solution to the equation
		\begin{equation}\label{14}
			\delta\left[1+\frac{1}{K}{\rm tr}\left(\delta\mathbf{I}_N+\lambda\mathbf{R}^{-1}\right)^{-1}\right]=1.
		\end{equation}
	
	In addition, ${\rm Im}(\hat{\alpha}_{\rm DL}(\lambda))$ also converges in distribution to the same Gaussian distribution as that of ${\rm Re}(\hat{\alpha}_{\rm DL}(\lambda))$. Furthermore, ${\rm Im}(\hat{\alpha}_{\rm DL}(\lambda))$ and ${\rm Re}(\hat{\alpha}_{\rm DL}(\lambda))$ are uncorrelated.
\end{mytheorem}
\noindent\textit{Proof:} See Appendix A.\qed

To get the asymptotic distribution of the test statistic $\hat{\eta}_{_{\rm DL}}$, it remains to check the asymptotic behavior of $\hat{\beta}_{\rm DL}(\lambda)$ under LDR, which is provided in the following theorem.

\begin{mytheorem}\label{Theorem Almost Convergence of Beta}
	Let assumptions (A1)-(A3) hold true. Then as $N,K\to\infty$, it holds that
	\begin{equation}\label{key}
		\hat{\beta}_{\rm DL}(\lambda)\overset{a.s.}{\longrightarrow}\widetilde{\beta}_{\rm DL}
	\end{equation}
	where $\widetilde{\beta}_{\rm DL} = \mathbf{u}^H\mathbf{E}(\lambda)\mathbf{u}$ with $\mathbf{u}$ and $\mathbf{E}(\lambda)$ defined in the same way as in Theorem \ref{Theorem asymptotic distribution of alpha}.
\end{mytheorem}
\noindent\textit{Proof:} See Appendix B.\qed

We now use Theorems \ref{Theorem asymptotic distribution of alpha} and \ref{Theorem Almost Convergence of Beta} to derive the asymptotic $P_{\rm fa}$ of the DL-AMF.
It is well-known that if $x,y\sim\mathcal{N}(0,\frac{\sigma^2}{2})$ and $x,y$ are independent, then $x^2+y^2\sim{\rm Exp}(\sigma^2)$. Combining this result with Theorem \ref{Theorem asymptotic distribution of alpha},  we obtain 
\begin{equation}\label{17-1}
	|\hat{\alpha}_{\rm DL}(\lambda)|^2\overset{d}{\longrightarrow}{\rm Exp}\left[\mu_0(\mathbf{R},\mathbf{s},\lambda)\right]
\end{equation}
where
\begin{equation}\label{17}
	\mu_0(\mathbf{R},\mathbf{s},\lambda) = 2\sigma^2_{\hat{\alpha}_{\rm DL},H_0} = \frac{\mathbf{u}^H\mathbf{E}(\lambda)^2\mathbf{u}}{1-\gamma(\lambda)}.
\end{equation}

Furthermore, the strong convergence established in Theorem \ref{Theorem Almost Convergence of Beta} allows us to treat the random variable $\hat{\beta}_{\rm DL}(\lambda)$ as a deterministic quantity $\widetilde{\beta}_{\rm DL}$. Consequently, it follows from Slutsky’s theorem that
\begin{equation}\label{19-1}
	\hat{\eta}_{_{\rm DL}}\overset{d}{\longrightarrow}{\rm Exp}\left[\frac{\mu_0(\mathbf{R},\mathbf{s},\lambda)}{\widetilde{\beta}_{\rm DL}}\right].
\end{equation} 
Therefore, the asymptotic PDF of $\hat{\eta}_{_{\rm DL}}$ under $H_0$ can be written as
\begin{equation}\label{key}
	\begin{aligned}
	f_{{\rm DL}}(x|H_0) = \frac{\widetilde{\beta}_{\rm DL}}{\mu_0(\mathbf{R},\mathbf{s},\lambda)}\exp\left[-\frac{\widetilde{\beta}_{\rm DL}}{\mu_0(\mathbf{R},\mathbf{s},\lambda)}x\right]
\end{aligned}
\end{equation}
from which we can easily calculate the asymptotic $P_{\rm fa}$ of the DL-AMF, given as a function of $\lambda$ :
\begin{align}
	\widetilde{P}_{\rm fa,DL}(\lambda)&= \int_{\tau_1}^{\infty} f_{\rm DL}(x|H_0)dx \nonumber
	\\&= \exp\left[-\frac{\widetilde{\beta}_{\rm DL}}{\mu_0(\mathbf{R},\mathbf{s},\lambda)}\tau_1\right]\label{18}
	\\& =  \exp\left[-(1-\gamma(\lambda))\frac{\mathbf{u}^H\mathbf{E}(\lambda)\mathbf{u}}{\mathbf{u}^H\mathbf{E}(\lambda)^2\mathbf{u}}\tau_1\right]\label{19}
\end{align}
where $\tau_1$ is the threshold. 
\begin{myremark} As evident from \eqref{19}, the asymptotic $P_{\rm fa}$ of DL-AMF depends on the quantities $\mathbf{u}$, $\mathbf{E}(\lambda)$ and $\gamma(\lambda)$, all of which are functions of the covariance matrix $\mathbf{R}$. This dependence explicitly shows that the DL-AMF can not achieve the CFAR property with respect to arbitrary covariance matrix $\mathbf{R}$. Furthermore, the asymptotic $P_{\rm fa}$ also varies with the steering vector $\mathbf{s}$ and the loading factor $\lambda$, implying that the DL-AMF also fails to achieve the CFAR property with respect to $\mathbf{s}$ and $\lambda$.
\end{myremark}
\begin{myremark}Following the same analytical framework, we can also derive the asymptotic $P_{\rm fa}$ for the detectors given in \eqref{7} and \eqref{6}. In  \cite{Zhou2024On}, Zhou \textit{et al.} proved that $\hat{\beta}\overset{a.s.}{\longrightarrow}\widetilde{\beta}$ with $\widetilde{\beta} = \frac{\mathbf{s}^H\mathbf{R}^{-1}\mathbf{s}}{1-c}$ under LDR. Hence, similar to \eqref{18}, the asymptotic $P_{\rm fa}$ for detector \eqref{7} is calculated by
\begin{equation}\label{22}
	\begin{aligned}
	\tilde{P}_{\rm fa, \eqref{7}}(\lambda) &= \exp\left[-\frac{\widetilde{\beta}}{\mu_0(\mathbf{R},\mathbf{s},\lambda)}\tau_2\right]
\\&	= \exp\left[-\frac{(1-\gamma(\lambda))\mathbf{s}^H\mathbf{R}^{-1}\mathbf{s}}{(1-c)\mathbf{u}^H\mathbf{E}(\lambda)^2\mathbf{u}}\tau_2\right].
\end{aligned}
\end{equation}
Likewise, the asymptotic $P_{\rm fa}$ of the detector in \eqref{6} is given by
\begin{equation}\label{23}
\small	\begin{aligned}
		\tilde{P}_{\rm fa, \eqref{6}}(\lambda) = \exp\left[-\frac{1}{\mu_0(\mathbf{R},\mathbf{s},\lambda)}\tau_3\right]
		= \exp\left[-\frac{(1-\gamma(\lambda))}{\mathbf{u}^H\mathbf{E}(\lambda)^2\mathbf{u}}\tau_3\right].
	\end{aligned}
\end{equation}
It can be seen from \eqref{22} and \eqref{23} that neither detector \eqref{7} nor \eqref{6} maintains the CFAR property with respect to  $\mathbf{R}$,  $\mathbf{s}$ and $\lambda$.
\end{myremark}

\subsection{Asymptotic $P_{\rm d}$ of DL-AMF under LDR}
Here we derive the asymptotic $P_{\rm d}$ of the DL-AMF under LDR. This is equivalent to deriving the asymptotic distribution of $\hat{\eta}_{_{\rm DL}}$ under $H_1$. Notably,  since $\hat{\beta}_{\rm DL}(\lambda)$ has not changed under $H_1$, the convergence in Theorem \ref{Theorem Almost Convergence of Beta} remains valid under $H_1$.  Thus, our focus reduces to characterizing the asymptotic behavior of $\hat{\alpha}_{\rm DL}(\lambda)$. We consider two target models: Swerling 0 and Swerling I.

\subsubsection{Asymptotic $P_{\rm d}$ for Swerling 0 target}
In the following theorem, we provide the asymptotic distribution of $\hat{\alpha}_{\rm DL}(\lambda)$ under $H_1$ when the target follows Swerling 0 model.
\begin{mytheorem}\label{Theorem Asymptotic distribution of alpha under H1 Swerling 0}
	Let assumptions (A1)-(A4) hold true. If the target follows Swerling 0 model, then under $H_1$ and as $N,K\to \infty$, it holds that
\begin{equation}\label{20}
	\frac{{\rm Re}(\hat{\alpha})-\mu^{\rm real}_{\hat{\alpha}_{\rm DL},Swer0}}{\sigma_{\hat{\alpha}_{\rm DL},Swer0}}\overset{d}{\longrightarrow}\mathcal{N}(0,1)
\end{equation}
where $\mu_{\hat{\alpha}_{\rm DL},Swer0}^{\rm real} = {\rm Re}(b)\mathbf{u}^H\mathbf{E}(\lambda)\mathbf{u}$ and $\sigma^2_{\hat{\alpha}_{\rm DL},Swer0} = \frac{\mathbf{u}^H\mathbf{E}(\lambda)^2\mathbf{u}}{2(1-\gamma(\lambda))}$ with $\mathbf{E}(\lambda)$, $\mathbf{u}$ and $\gamma(\lambda)$ defined in the same manner as above.

 In addition, ${\rm Im}(\hat{\alpha}_{\rm DL}(\lambda))$ converges in distribution to a Gaussian distribution with mean $\mu_{\hat{\alpha}_{\rm DL},Swer0}^{\rm imag} = {\rm Im}(b)\mathbf{u}^H\mathbf{E}(\lambda)\mathbf{u}$ and the same variance as ${\rm Re}(\hat{\alpha}_{\rm DL}(\lambda))$. Moreover, ${\rm Im}(\hat{\alpha}_{\rm DL}(\lambda))$ and ${\rm Re}(\hat{\alpha}_{\rm DL}(\lambda))$ are uncorrelated.
\end{mytheorem}
\noindent\textit{Proof:} Under $H_1$, we have $\mathbf{\hat{\alpha}_{\rm DL}(\lambda)} = \mathbf{s}^H(\hat{\mathbf{R}}+\lambda\mathbf{I}_N)^{-1}(b\mathbf{s}+\mathbf{c}_0)$. Hence, ${\rm Re}(\hat{\alpha}_{\rm DL}(\lambda)) = {\rm Re}(b)\mathbf{s}(\hat{\mathbf{R}}+\lambda\mathbf{I}_N)^{-1}\mathbf{s}+{\rm Re}[\mathbf{s}^H(\hat{\mathbf{R}}+\lambda\mathbf{I}_N)^{-1}\mathbf{c}_0]$. It follows from Theorem \ref{Theorem Almost Convergence of Beta} that ${\rm Re}(b)\mathbf{s}(\hat{\mathbf{R}}+\lambda\mathbf{I}_N)^{-1}\mathbf{s}\overset{a.s.}{\longrightarrow}{\rm Re}(b)\mathbf{u}^H\mathbf{E}(\lambda)\mathbf{u}$. Likewise, by using Theorem \ref{Theorem asymptotic distribution of alpha}, we have ${\rm Re}[\mathbf{s}^H(\hat{\mathbf{R}}+\lambda\mathbf{I}_N)^{-1}\mathbf{c}_0]\overset{d}{\longrightarrow}\mathcal{N}(0,\sigma^2_{\hat{\alpha}_{\rm DL},H_0})$. Gathering the above two convergences and using the Slutsky’s Theorem, we finally get \eqref{20}. Using the same procedures, it is easy to get the asymptotic distribution of ${\rm Im}(\hat{\alpha}_{\rm DL}(\lambda))$ under $H_1$ when a Swerling 0 target exists. In addition, the fact that ${\rm Im}(\hat{\alpha}_{\rm DL}(\lambda))$ and ${\rm Re}(\hat{\alpha}_{\rm DL}(\lambda))$ are uncorrelated directly follows from Theorem \ref{Theorem asymptotic distribution of alpha}.\qed

Now  we derive the asymptotic distribution of $\hat{\eta}_{_{\rm DL}}$ under $H_1$ when a Swerling 0 target exists in the primary data. Firstly,  Theorems \ref{Theorem Almost Convergence of Beta} and \ref{Theorem Asymptotic distribution of alpha under H1 Swerling 0} together with Slutsky's theorem directly yield $\frac{{\rm Re}(\hat{\alpha}_{\rm DL}(\lambda))}{\sqrt{\hat{\beta}_{\rm DL}(\lambda)}}\overset{d}{\longrightarrow}\mathcal{N}\left(\frac{\mu_{\hat{\alpha}_{\rm DL},Swer0}^{\rm real}}{\sqrt{\widetilde{\beta}_{\rm DL}}}, \frac{\sigma^2_{\hat{\alpha}_{\rm DL},Swer0} }{\widetilde{\beta}_{\rm DL}}\right)$ and $\frac{{\rm Im}(\hat{\alpha}_{\rm DL}(\lambda))}{\sqrt{\hat{\beta}_{\rm DL}(\lambda)}}\overset{d}{\longrightarrow}\mathcal{N}\left(\frac{\mu_{\hat{\alpha}_{\rm DL},Swer0}^{\rm imag}}{\sqrt{\widetilde{\beta}_{\rm DL}}}, \frac{\sigma^2_{\hat{\alpha}_{\rm DL},Swer0} }{\widetilde{\beta}_{\rm DL}}\right)$.  The exact distribution theory asserts that if $x\sim\mathcal{N}(\mu_0,\sigma^2),y\sim\mathcal{N}(\mu_1,\sigma^2)$ and $x$ is independent on $y$, then it holds that $x^2+y^2\sim\chi^2(\mu_0^2+\mu_1^2,\sigma^2)$. Using this statement, we get 
\begin{equation}\label{key}
	\hat{\eta}_{_{\rm DL}}\overset{d}{\longrightarrow}\chi^2(\nu_{0},\sigma^2_{0})
\end{equation}with $\nu_0 = \frac{1}{\widetilde{\beta}_{\rm DL}}\left[{\left(\mu_{\hat{\alpha}_{\rm DL},Swer0}^{\rm real}\right)^2+\left(\mu_{\hat{\alpha}_{\rm DL},Swer0}^{\rm imag}\right)^2}\right]$ and $\sigma^2_0 =\frac{\sigma^2_{\hat{\alpha}_{\rm DL},Swer0} }{\widetilde{\beta}_{\rm DL}}$, which implies that the asymptotic PDF of $\hat{\eta}_{_{\rm DL}}$ in the presence of a Swerling 0 target is given by
\begin{equation}\label{key}
	f_{\rm DL}^{Swer0}(x|H_1) = \frac{1}{2\sigma_0^2}\exp\left(-\frac{x+\nu_0}{2\sigma^2_0}\right)I_0\left(\frac{\sqrt{x\nu_0}}{\sigma^2_0}\right)
\end{equation}
from which we can calculate the asymptotic $P_{\rm d}$ of DL-AMF for the Swerling 0 target, given by
\begin{equation}\label{key}
 \begin{aligned}
	\widetilde{P}_{\rm d,DL}^{Swer0}(\lambda)& = \int_{\tau_1}^{\infty}	f_{\rm DL}^{Swer0}(x|H_1)dx
	 \\&= Q\left(\frac{\sqrt{\nu_0}}{\sigma_0},\frac{\sqrt{\tau_1}}{\sigma_0}\right).
\end{aligned}
\end{equation}

Substituting the expressions of $\nu_0$ and $\sigma^2_0$, we have
\begin{equation}\label{25}
	\small\begin{aligned}
 &\widetilde{P}_{\rm d,DL}^{Swer0} (\lambda)
	\\&= Q\left(\frac{\sqrt{{\left(\mu_{\hat{\alpha}_{\rm DL},Swer0}^{\rm real}\right)^2+\left(\mu_{\hat{\alpha}_{\rm DL},Swer0}^{\rm imag}\right)^2}}}{\sigma_{\hat{\alpha}_{\rm DL},Swer0}},\frac{\sqrt{\widetilde{\beta}_{\rm DL}\tau_1}}{\sigma_{\hat{\alpha}_{\rm DL},Swer0}} \right).
\end{aligned}
\end{equation}

Now we denote the preassigned $P_{\rm fa}$ by $P_{\rm fa,pre}$. Then from \eqref{18}, it is easy to get the threshold associated with $P_{\rm fa,pre}$, given by  $\tau_1=-\frac{\mu_0(\mathbf{R},\mathbf{s},\lambda)}{\widetilde{\beta}_{\rm DL}}\log{P}_{\rm fa,pre}$. Substituting $\tau_1$ into \eqref{25} and noticing that $\mu_0(\mathbf{R},\mathbf{s},\lambda) = 2\sigma^2_{\hat{\alpha}_{\rm DL},Swer0}$, we finally get the ROC of DL-AMF for Swerling 0 target:
\begin{small}
\begin{align}
 &	\widetilde{P}_{\rm d,DL}^{Swer0}(\lambda) \nonumber
		\\&=   Q\left(\frac{\sqrt{{\left(\mu_{\hat{\alpha}_{\rm DL},Swer0}^{\rm real}\right)^2+\left(\mu_{\hat{\alpha}_{\rm DL},Swer0}^{\rm imag}\right)^2}}}{\sigma_{\hat{\alpha}_{\rm DL},Swer0}},\sqrt{-2\log{P}_{\rm fa,pre}} \right)\label{26}
		\\& = 	Q\left(\sqrt{2S_0\frac{(1-\gamma(\lambda))(\mathbf{u}^H\mathbf{E}(\lambda)\mathbf{u})^2}{(\mathbf{s}^H\mathbf{R}^{-1}\mathbf{s})(\mathbf{u}^H\mathbf{E}(\lambda)^2\mathbf{u})}},\sqrt{-2\log {P}_{\rm fa,pre}}\right)\label{27}
\end{align}
\end{small}where $S_0 = |b|^2\mathbf{s}^H\mathbf{R}^{-1}\mathbf{s}$ denotes the SCNR in the Swerling 0 case.

\begin{myremark}As shown in \eqref{26}, for a given preassigned $P_{\rm fa}$, the asymptotic $P_{\rm d}$ of DL-AMF is determined solely by the asymptotic means of ${\rm Re}(\hat{\alpha}_{\rm DL}(\lambda))$ and ${\rm Im}(\hat{\alpha}_{\rm DL}(\lambda))$ (i.e., $\mu_{\hat{\alpha}_{\rm DL},Swer0}^{\rm real}$ and $\mu_{\hat{\alpha}_{\rm DL},Swer0}^{\rm imag}$) and the asymptotic variance of ${\rm Re}(\hat{\alpha}_{\rm DL}(\lambda))$ or ${\rm Im}(\hat{\alpha}_{\rm DL}(\lambda))$ (i.e., $\sigma^2_{\hat{\alpha}_{\rm DL},Swer0}$). This means that the asymptotic detection performance of DL-AMF for detecting a Swerling 0 target only depends on the random variable $\hat{\alpha}_{\rm DL}(\lambda)$ or equivalently $|\hat{\alpha}_{\rm DL}(\lambda)|^2$. In other words, the random variable $\hat{\beta}_{\rm DL}(\lambda)$ actually has no effect on the detection performance.  Hence we conclude that any DL detector constructed by scaling $|\hat{\alpha}_{\rm DL}(\lambda)|^2$ with a deterministic quantity or a random variable that converges almost surely to a deterministic quantity will achieve identical detection performance under LDR. Therefore, the ROCs of detectors given in \eqref{7} and \eqref{6} can also be characterized by \eqref{27}.
\end{myremark} 
%

\subsubsection{Asymptotic $P_{\rm d}$ for Swerling I target}
In this part, we establish the asymptotic $P_{\rm d}$ of DL-AMF for Swerling I target. 

Firstly, as a preliminary step, the following theorem provides the asymptotic distribution of $\hat{\alpha}_{\rm DL}(\lambda)$ in the presence of a Swerling I target.
\begin{mytheorem}\label{Theorem Asymptotic distribution of alpha under H1 Swerling I}
	Let assumptions (A1)-(A4) hold true. If the target follows Swerling I model, then under $H_1$ and as $N,K\to \infty$, it holds that
	\begin{equation}\label{21}
		\frac{{\rm Re}(\hat{\alpha})-\mu_{\hat{\alpha},Swer1}}{\sigma_{\hat{\alpha},Swer1}}\overset{d}{\longrightarrow}\mathcal{N}(0,1)
	\end{equation}
	where $\mu_{\hat{\alpha}_{\rm DL},Swer1} = 0$ and $\sigma^2_{\hat{\alpha}_{\rm DL},Swer1} = \frac{\sigma_t^2}{2}(\mathbf{u}^H\mathbf{E}(\lambda)\mathbf{u})^2+ \frac{\mathbf{u}^H\mathbf{E}(\lambda)^2\mathbf{u}}{2(1-\gamma(\lambda))}$  with $\mathbf{E}(\lambda)$, $\mathbf{u}$ and $\gamma(\lambda)$ defined in the same manner as above.
	
	In addition,  ${\rm Im}(\hat{\alpha}_{\rm DL}(\lambda))$ has the same asymptotic distribution as ${\rm Re}(\hat{\alpha}_{\rm DL}(\lambda))$, and ${\rm Im}(\hat{\alpha}_{\rm DL}(\lambda))$ and ${\rm Re}(\hat{\alpha}_{\rm DL}(\lambda))$ are uncorrelated.
\end{mytheorem} 
\noindent\textit{Proof:} Reviewing the assumption (A4), we have ${\rm Re}(b)\sim\mathcal{N}(0,\frac{\sigma_t^2}{2})$. Then using the Slutsky’s Theorem and Theorem \ref{Theorem Almost Convergence of Beta}, we get ${\rm Re}(b)\mathbf{s}^H(\hat{\mathbf{R}}+\lambda\mathbf{I}_N)^{-1}\mathbf{s}\overset{d}{\longrightarrow}\mathcal{N}(0,\frac{\sigma_t^2}{2}(\mathbf{u}^H\mathbf{E}(\lambda)\mathbf{u})^2)$.  Because the sum of Gaussian random variables is still Gaussian, we finally get ${\rm Re}(\hat{\alpha}_{\rm DL}(\lambda)) = {\rm Re}(b)\mathbf{s}(\hat{\mathbf{R}}+\lambda\mathbf{I}_N)^{-1}\mathbf{s}+{\rm Re}[\mathbf{s}^H(\hat{\mathbf{R}}+\lambda\mathbf{I}_N)^{-1}\mathbf{c}_0]\overset{d}{\longrightarrow}\mathcal{N}\left(0, \frac{\sigma_t^2}{2}(\mathbf{u}^H\mathbf{E}(\lambda)\mathbf{u})^2+ \frac{\mathbf{u}^H\mathbf{E}(\lambda)^2\mathbf{u}}{2(1-\gamma(\lambda))}\right)$, which completes the proof of \eqref{21}. The asymptotic distribution of ${\rm Im}(\hat{\alpha}_{\rm DL}(\lambda))$ can be derived in the same manner. Furthermore, the uncorrelated nature of ${\rm Im}(\hat{\alpha}_{\rm DL}(\lambda))$ and ${\rm Re}(\hat{\alpha}_{\rm DL}(\lambda))$ is a direct consequence of Theorem \ref{Theorem asymptotic distribution of alpha} and assumption (A4).\qed

We are now in position to derive the asymptotic $P_{\rm d}$ of DL-AMF for Swerling I target. It follows from Theorem \ref{Theorem Asymptotic distribution of alpha under H1 Swerling I} that 
 \begin{equation}\label{key}
 	|\hat{\alpha}_{\rm DL}(\lambda)|^2\overset{d}{\longrightarrow}{\rm Exp}\left[\mu_1(\mathbf{R},\mathbf{s},\lambda)\right]
 \end{equation}
where
\begin{equation}\label{key}
	\mu_1(\mathbf{R},\mathbf{s},\lambda) = 2\sigma^2_{\hat{\alpha}_{\rm DL},Swer1} = \sigma_t^2(\mathbf{u}^H\mathbf{E}(\lambda)\mathbf{u})^2+\frac{\mathbf{u}^H\mathbf{E}(\lambda)^2\mathbf{u}}{1-\gamma(\lambda)}.
\end{equation}  Then applying Theorem \ref{Theorem Almost Convergence of Beta} and the Slutsky's theorem, we have 
\begin{equation}\label{key}
	\hat{\eta}_{_{\rm DL}} \overset{d}{\longrightarrow}{\rm Exp}\left[\frac{\mu_1(\mathbf{R},\mathbf{s},\lambda)}{\widetilde{\beta}_{\rm DL}}\right].
\end{equation}

The asymptotic PDF of $\hat{\eta}_{_{\rm DL}}$ in the presence of a Swerling I target is thus given by
\begin{equation}\label{key}
	f_{\rm DL}^{Swer1}(x|H_1) = \frac{\widetilde{\beta}_{\rm DL}}{\mu_1(\mathbf{R},\mathbf{s},\lambda)}\exp\left[-\frac{\widetilde{\beta}_{\rm DL}}{\mu_1(\mathbf{R},\mathbf{s},\lambda)}x\right]
\end{equation}
from which we obtain the asymptotic $P_{\rm d}$ of DL-AMF for Swerling I target, given by
\begin{align}
		\widetilde{P}_{\rm d,DL}^{Swer1}(\lambda) = \int_{\tau_1}^{\infty}f_{\rm DL}^{Swer1}(x|H_1) dx
		= \exp\left[-\frac{\widetilde{\beta}_{\rm DL}}{\mu_1(\mathbf{R},\mathbf{s},\lambda)}\tau_1\right]\label{32}.
\end{align}

Substituting $\tau_1=-\frac{\mu_0(\mathbf{R},\mathbf{s},\lambda)}{\widetilde{\beta}_{\rm DL}}\log{P}_{\rm fa,pre}$ into \eqref{32}, we get the ROC of DL-AMF for Swerling I target:
\begin{align}
	\widetilde{P}_{\rm d,DL}^{Swer1}(\lambda)& =\left( {P}_{\rm fa,pre}\right)^{\frac{\mu_0(\mathbf{R},\mathbf{s},\lambda)}{\mu_1(\mathbf{R},\mathbf{s},\lambda)}}\label{33}
	\\& = \left( {P}_{\rm fa,pre}\right)^{\frac{1}{1+S_1\frac{(1-\gamma(\lambda))(\mathbf{u}^H\mathbf{E}(\lambda)\mathbf{u})^2}{(\mathbf{s}^H\mathbf{R}^{-1}\mathbf{s})( \mathbf{u}^H\mathbf{E}(\lambda)^2\mathbf{u}) }}}\label{34}
\end{align}
where $S_1 = \sigma^2_t\mathbf{s}^H\mathbf{R}^{-1}\mathbf{s}$ is the SNCR in the case of Swerling I target.

\begin{myremark} We observe from \eqref{33} that when given a fixed $P_{\rm fa,pre}$, the asymptotic $P_{\rm d}$ of DL-AMF for Swerling I target depends exclusively on $\mu_0(\mathbf{R},\mathbf{s},\lambda)$ and $\mu_1(\mathbf{R},\mathbf{s},\lambda)$. The former is the asymptotic mean of $|\hat{\alpha}_{\rm DL}(\lambda)|^2$ under $H_0$ whereas the latter is the asymptotic mean of $|\hat{\alpha}_{\rm DL}(\lambda)|^2$ under $H_1$ (in the presence of a Swerling I target). As a result, we arrive at the conclusion that the detection performance of DL-AMF for detecting a Swerling I target is entirely determined by $|\hat{\alpha}_{\rm DL}(\lambda)|^2$ while the random variable $\hat{\beta}_{\rm DL}(\lambda)$ has no influence on the detection performance. 
\end{myremark}

Based on the comments presented in Remarks 1-4,  we now summarize the findings in Sections II-A and II-B in the following propositions.
\begin{myproposition}\label{Proposition 1}
	The asymptotic $P_{\rm fa}$ of DL-AMF relies on the covariance matrix $\mathbf{R}$, the target steering vector $\mathbf{s}$ and the diagonal loading factor ${\lambda}$. Consequently, DL-AMF does not possess the CFAR property with respect to $\mathbf{R}$, $\mathbf{s}$ and $\lambda$.
\end{myproposition}
\begin{myproposition}\label{Proposition 2}
	For both Swerling 0 and Swerling I targets, the detection performance of the DL-AMF under LDR depends solely on $|\hat{\alpha}_{\rm DL}(\lambda)|^2$, with $\hat{\beta}_{\rm DL}(\lambda)$ having no impact on detection performance. In other words,  any DL detector constructed by normalizing $|\hat{\alpha}_{\rm DL}(\lambda)|^2$ with either a deterministic quantity or a random variable that converges almost surely to a deterministic value will share unified ROCs presented in \eqref{27} and \eqref{34}.
\end{myproposition}

\begin{figure*}[t]
	\vspace{-0.5em}
	\centering
	\subfloat[PDF of $\hat{\eta}_{_{\rm DL}}$ at different $\mathbf{R}$ ($\varrho$ variation)]{
		\centering
		\includegraphics[trim=2cm 0cm 1cm 2cm, scale=0.45]{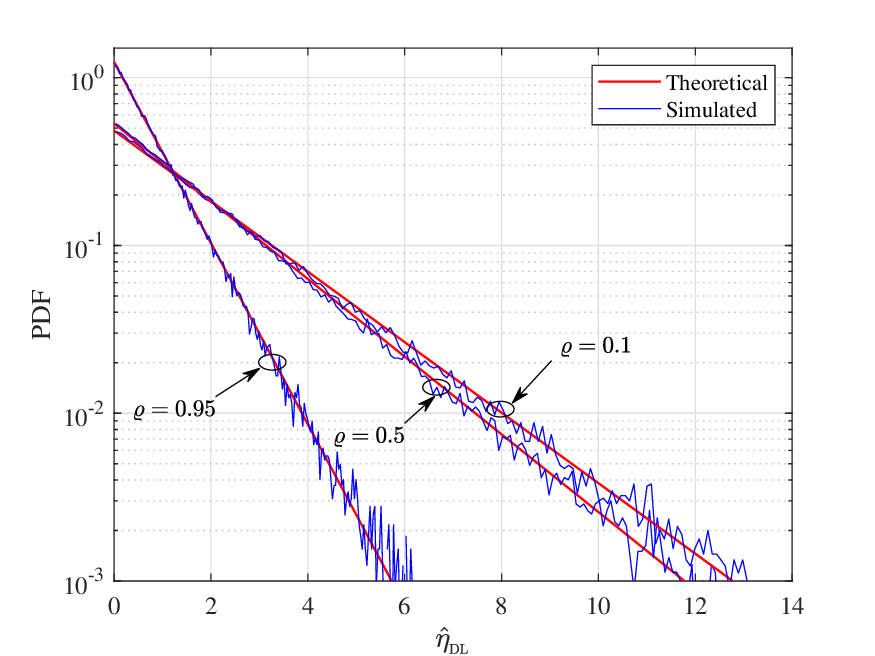}
	}
	\subfloat[PDF of $\hat{\eta}_{_{\rm DL}}$ at different $\mathbf{s}$ ($\theta_t$ variation)]{
		\centering
		\includegraphics[trim=0.5cm 0cm 1cm 2cm, scale=0.45]{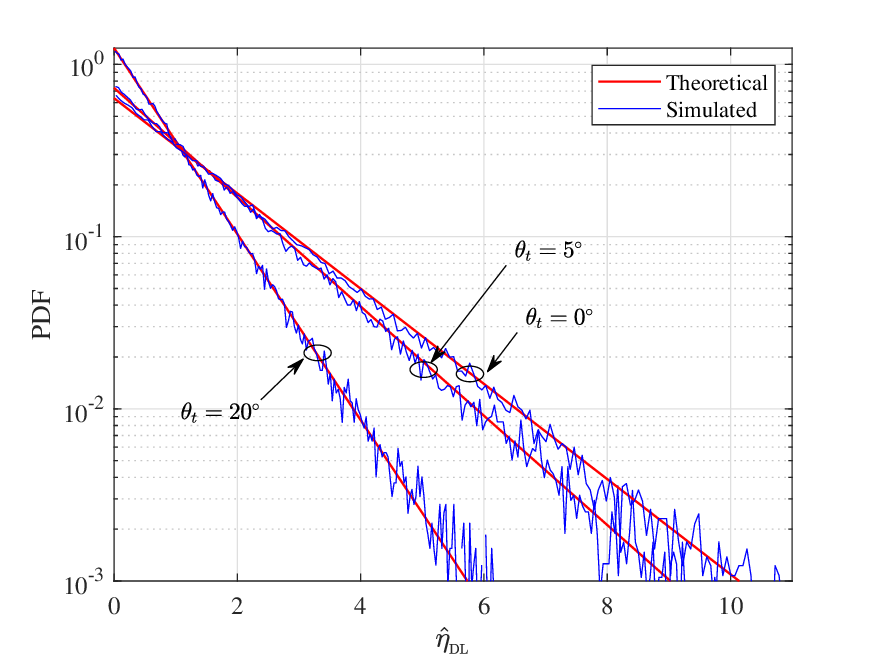}
	}
	\subfloat[PDF of $\hat{\eta}_{_{\rm DL}}$ at different $\lambda$]{
		\centering
		\includegraphics[trim=0.5cm 0cm 1cm 2cm, scale=0.45]{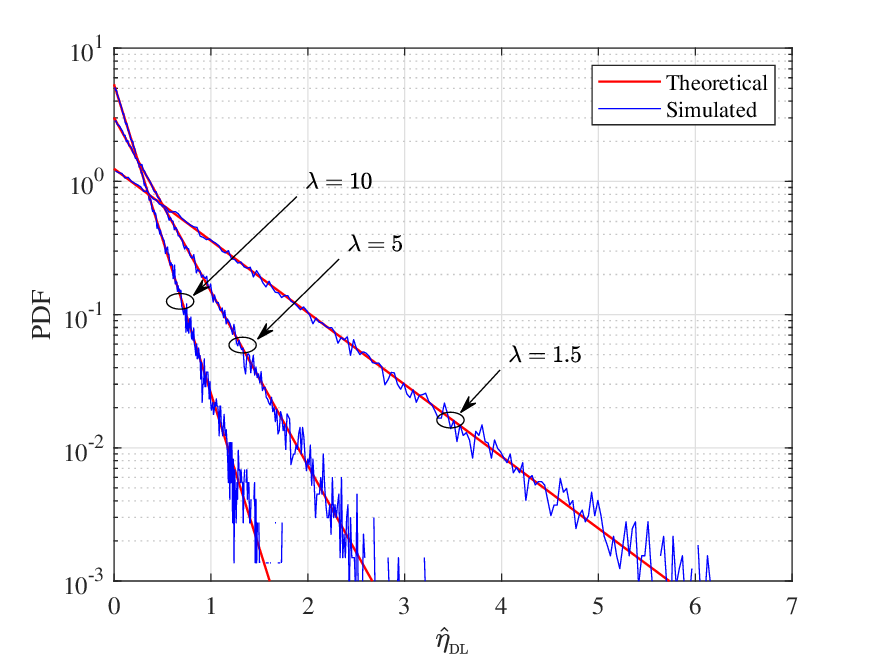}
	}\\
	\caption{Comparison between theoretical asymptotic distribution of $\hat{\eta}_{_{\rm DL}}$ given in \eqref{19-1} and the empirical PDFs obtained from $10^5$ Monte Carlo trials with different $\mathbf{R}$, $\mathbf{s}$ and $\lambda$.}
	\label{Comparisons of theoretical asymptotic distribution of eta DL and the simulated results}
\end{figure*}
\begin{figure*}[t]
	\vspace{-0.5em}
	\centering
	\subfloat[$N=12,K=24$]{
		\centering
		\includegraphics[trim=0.5cm 0.1cm 0 0cm, scale=0.45]{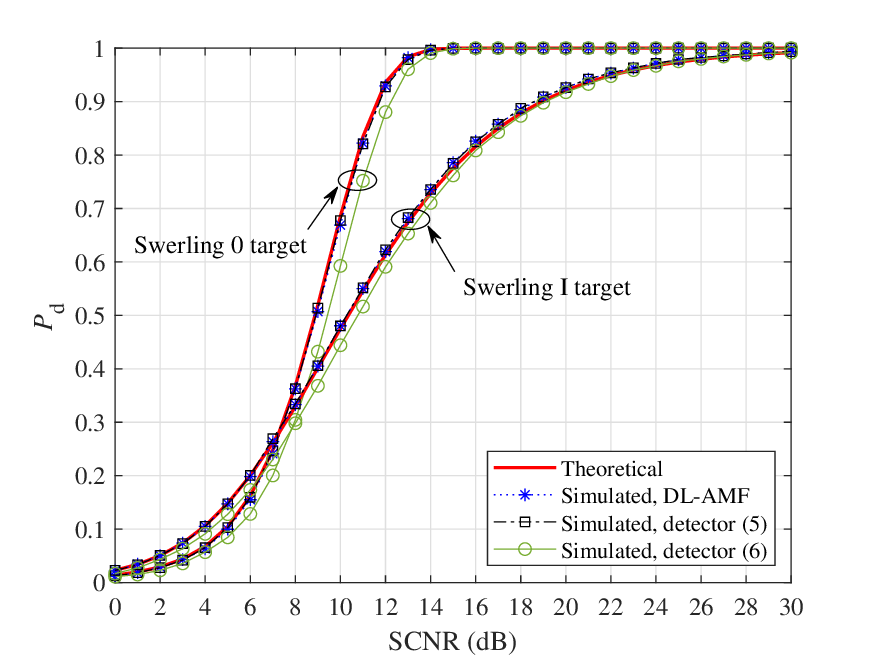}
	}\hspace{2em}
	\subfloat[$N=24,K=48$]{
		\centering
		\includegraphics[trim=0.5cm 0.1cm 0 0cm, scale=0.45 ]{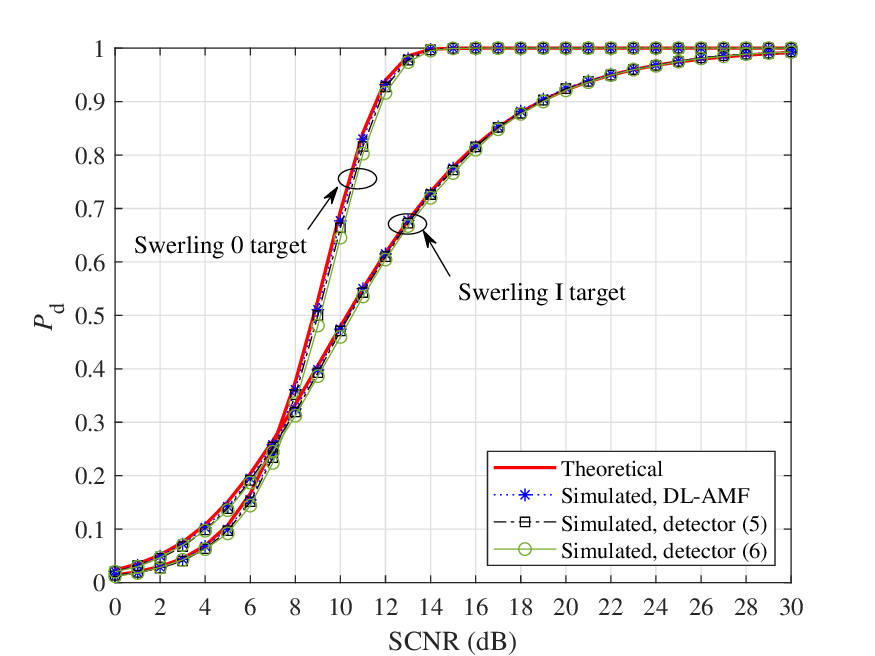}
	}
	\caption{Comparison of the theoretical ROCs presented in \eqref{27} and \eqref{34} and the results obtained through Monte Carlo simulations.}
	\label{Comparisons of the theoretical ROCs presented in 27 and 34 and the simulated results obtained from Monte Carlo trials.}
	\vspace{-1.5em}
\end{figure*}
In the following, we perform some numerical examples to validate the above theoretical results.  Before proceeding, we clarify some parameters in the simulations. We assume that the population covariance matrix is modeled as 
\begin{equation}\label{39}
	\mathbf{R} = \mathbf{R}_c+\sigma^2\mathbf{I}_N
\end{equation} where $\mathbf{R}_c$ represents the covariance matrix of the clutter component, whose $(i,j)$-th element is given by $(\mathbf{R}_c)_{ij} = \sigma^2_c\varrho^{|i-j|}$ with $\sigma^2_c$ denoting the clutter power and $\varrho$ being the one-lag correlation coefficient. The values of $\varrho$ for radar sea clutter typically range from 0.9 to 0.99 \cite{Gini2002Covariance}. The steering vector is modeled as $\mathbf{s} =\frac{1}{\sqrt{N}} [1,\exp(\mathsf{i}\pi\sin(\theta_t)),...,\exp(\mathsf{i}\pi(N-1)\sin(\theta_t))]^T$ where $\theta_t$ is the direction of arrival of the target signal. 

We first validate the asymptotic distribution of $\hat{\eta}_{_{\rm DL}}$ under $H_0$ and the claim in Proposition \ref{Proposition 1}. Fig. \ref{Comparisons of theoretical asymptotic distribution of eta DL and the simulated results} plots the theoretical asymptotic distribution of $\hat{\eta}_{_{\rm DL}}$ given in \eqref{19-1} and the PDF of simulated $\hat{\eta}_{_{\rm DL}}$ from $10^5$ Monte Carlo trials under different configurations of $\mathbf{R}$, $\mathbf{s}$ and $\lambda$. Specifically, in Fig. \ref{Comparisons of theoretical asymptotic distribution of eta DL and the simulated results}(a), we verify the non-CFAR property of DL-AMF with respect to  $\mathbf{R}$ by considering three different correlation coefficients $\varrho = 0.1, 0.5, 0.95$ while fixing $\sigma^2=1,\sigma^2_c=10,\theta_t=20^\circ$ and $\lambda=1.5$; in Fig. \ref{Comparisons of theoretical asymptotic distribution of eta DL and the simulated results}(b), we assess the non-CFAR property with respect to $\mathbf{s}$ by varying the target angle $\theta_t$ ($\theta_t=0^\circ, 5^\circ, 20^\circ$) while keeping $\sigma^2=1,\sigma^2_c=10,\varrho=0.95$ and $\lambda=1.5$ fixed. Finally, in Fig. \ref{Comparisons of theoretical asymptotic distribution of eta DL and the simulated results}(c), we examine the non-CFAR property with respect to the loading factor $\lambda$ using $\lambda=1.5, 5, 10$, with fixed $\sigma^2=1, \sigma^2_c=10, \varrho=0.95$ and $\theta_t=20^\circ$. In all simulations in Fig. \ref{Comparisons of theoretical asymptotic distribution of eta DL and the simulated results}, we set $N=24$ and $K=48$. We can see that the empirical PDFs of the simulated $\hat{\eta}_{_{\rm DL}}$ show excellent agreement with the theoretical distribution given in \eqref{19-1}. Moreover, we observe that the results in Fig. \ref{Comparisons of theoretical asymptotic distribution of eta DL and the simulated results} are fully consistent with the prediction of Proposition \ref{Proposition 1}: the distribution of $\hat{\eta}_{_{\rm DL}}$ varies with $\mathbf{R}$, $\mathbf{s}$ and $\lambda$, demonstrating that the DL-AMF does not achieve the CFAR property with respect to these parameters. 

Next we conduct simulations to verify the theoretical ROCs given in \eqref{27} and \eqref{34} and the statement presented in Proposition \ref{Proposition 2}. Fig. \ref{Comparisons of the theoretical ROCs presented in 27 and 34 and the simulated results obtained from Monte Carlo trials.} compares the theoretical ROCs presented in \eqref{27} and \eqref{34} with the Monte Carlo simulation results. The experimental parameters are configured as follows: $\varrho=0.95$, $\sigma^2=1$, $\sigma^2_c=10$ and $\theta_t=20^\circ$. We set the preassigned $P_{\rm fa}$ as $10^{-3}$. Hence, the thresholds are obtained from $10^5$ Monte Carlo trials and the detection probabilities are obtained from $10^4$ Monte Carlo trials. In Fig. \ref{Comparisons of the theoretical ROCs presented in 27 and 34 and the simulated results obtained from Monte Carlo trials.}(a), we set $N=12$ and $K=24$, while in Fig. \ref{Comparisons of the theoretical ROCs presented in 27 and 34 and the simulated results obtained from Monte Carlo trials.}(b), we proportionally scale both $N$ and $K$ by a factor of two, setting $N = 24$ and $K = 48$ to keep the original ratio $c = 0.5$ unchanged. For DL-AMF and the detectors \eqref{7} and \eqref{6}, the loading factor $\lambda$ is set to 1.5. We observe from Fig. \ref{Comparisons of the theoretical ROCs presented in 27 and 34 and the simulated results obtained from Monte Carlo trials.} that as $N$ and $K$ increase at the same rate, the DL-AMF and the detectors in \eqref{7} and \eqref{6} tend to exhibit nearly identical performance. This phenomenon validates the statement in Proposition \ref{Proposition 2}. Moreover, we observe that the performance of these three detectors approaches the theoretical ROCs presented in \eqref{27} and \eqref{34}, thereby validating the theoretical expressions.
\subsection{Discussions}
Let us now recall the ROCs given in \eqref{27} and \eqref{34}. If we define the function 
\begin{equation}\label{35}
	\kappa(\lambda)=\frac{(1-\gamma(\lambda))(\mathbf{u}^H\mathbf{E}(\lambda)\mathbf{u})^2}{(\mathbf{s}^H\mathbf{R}^{-1}\mathbf{s})( \mathbf{u}^H\mathbf{E}(\lambda)^2\mathbf{u}) },
\end{equation}
then the ROCs in \eqref{27} and \eqref{34} can be  transformed to 
\begin{equation}\label{37}
	\widetilde{P}_{\rm d,DL}^{Swer0}(\lambda) =	Q\left(\sqrt{2S_0\kappa(\lambda)},\sqrt{-2\log {P}_{\rm fa,pre}} \right)
\end{equation}
and 
\begin{equation}\label{38}
	\widetilde{P}_{\rm d,DL}^{Swer1}(\lambda)= \left({P}_{\rm fa,pre}\right)^{\frac{1}{1+S_1\kappa(\lambda)}}.
\end{equation}

Since the Marcum Q-function $Q(a,b)$ is a strictly increasing function in $a$ for all $a\geqslant0$ and $b>0$ \cite{Sun2010On}, $\widetilde{P}_{\rm d,DL}^{Swer0}(\lambda)$ is a strictly increasing function of $\kappa(\lambda)$ for all $\kappa(\lambda)\geqslant0$. Likewise, it is easy to check that $\widetilde{P}_{\rm d,DL}^{Swer1}(\lambda)$ similarly exhibits strict monotonicity in $\kappa(\lambda)$ for all $\kappa(\lambda)\geqslant0$. Hence,  analyzing the $\lambda$-dependence of the asymptotic $P_{\rm d}$ in \eqref{37} and \eqref{38} reduces to characterizing the behavior of $\kappa(\lambda)$. Unfortunately, due to the complexity of $\kappa(\lambda)$, it remains a challenge to give an analytical characterization of the behavior of $\kappa(\lambda)$. We therefore adopt a hybrid approach: first establishing fundamental properties of $\kappa(\lambda)$ through theoretical analysis and then employing graphical analysis to fully characterize its behavior.

Observing that 
\begin{equation}\label{key}
	\lim_{\lambda\to0}\kappa(\lambda) = 1-c,
\end{equation}
we have
\begin{equation}\label{key}
	\lim_{\lambda\to0}\widetilde{P}^{Swer0}_{\rm d,DL}(\lambda)  =	\widetilde{P}_{\rm d, SCM}^{Swer0}
\end{equation}
and 
\begin{equation}\label{key}
		\lim_{\lambda\to0}\widetilde{P}^{Swer1}_{\rm d,DL}(\lambda)  =	\widetilde{P}_{\rm d, SCM}^{Swer1}
\end{equation}
where
\begin{equation}\label{41}
	\widetilde{P}_{\rm d, SCM}^{Swer0}= 
	Q\left(\sqrt{2S_0(1-c)},\sqrt{-2\log {P}_{\rm fa,pre}}\right)
\end{equation}
and
\begin{equation}\label{42}
	\widetilde{P}_{\rm d,SCM}^{Swer1}= \left( {P}_{\rm fa,pre}\right)^{\frac{1}{1+(1-c)S_1}}
\end{equation}
It should be noted that \eqref{41} and \eqref{42} represent the ROCs of the SCM-based AMF under LDR (see \cite[Eqs. (41)-(42)]{Zhou2024On}).  
Examining the test statistic in \eqref{12}, we observe that as $\lambda\to0$, the DL-AMF degenerates into SCM-based AMF. Consequently, the ROCs of DL-AMF naturally reduce to those of the AMF.  This behavior is in agreement with our analytical results.

Furthermore, as shown in \cite{Zhou2024On}, as $c\to0$, the performance of SCM-based AMF tends to that of NP detector which is the theoretically optimal detector. Correspondingly, the ROCs in \eqref{41} and \eqref{42} reduce to those of NP detector, given by
\begin{equation}\label{45-1}
	P_{\rm d,opt}^{Swer0} = Q\left(\sqrt{2S_0},\sqrt{-2\log P_{\rm fa,pre}}\right)
\end{equation}
for Swerling 0 target, and
\begin{equation}\label{46-1}
	P_{\rm d,opt}^{Swer1} = \left(P_{\rm fa,pre}\right)^{\frac{1}{1+S_1}}
\end{equation}
for Swerling I target.

Now we review $\kappa(\lambda)$ presented in \eqref{35}. The Cauchy-Schwarz inequality gives $(\mathbf{u}^H\mathbf{E}(\lambda)\mathbf{u})^2\leqslant(\mathbf{u}^H\mathbf{u})(\mathbf{u}^H\mathbf{E}(\lambda)^2\mathbf{u})$. Then using this fact together with the relation $\gamma(\lambda)>0$, we have
\begin{equation}\label{45}
	\kappa(\lambda)\leqslant (1-\gamma(\lambda))\frac{\mathbf{u}^H\mathbf{u}}{\mathbf{s}^H\mathbf{R}^{-1}\mathbf{s}}=1-\gamma(\lambda)<1.
\end{equation}
From \eqref{45} and the strictly increasing properties of $\widetilde{P}_{\rm d,DL}^{Swer0}(\lambda)$ and $\widetilde{P}_{\rm d,DL}^{Swer1}(\lambda)$ with respect to $\kappa(\lambda)$, we find that
\begin{equation}\label{46}
	\widetilde{P}_{\rm d,DL}^{Swer0}(\lambda)<{P}_{\rm d,opt}^{Swer0}, \quad \text{and} \quad\widetilde{P}_{\rm d,DL}^{Swer1}(\lambda)<{P}_{\rm d,opt}^{Swer1}
\end{equation}
for all $\lambda\geqslant 0$.

\begin{myremark} \label{Remark 5} Eq. \eqref{46} demonstrates that the DL-AMF consistently exhibits inferior performance compared to NP detector across all nonnegative $\lambda$. 
	
Based on the analytical expressions \eqref{37}, \eqref{38}, \eqref{45-1} and \eqref{46-1}, we  can express the detection losses of DL-AMF relative to the NP detector for both Swerling 0 and I targets by a unified explicit formula, given in dB as:
\begin{equation}\label{49}
	\Delta_{\rm NP-DL}(\lambda) = -10\log_{10}\kappa(\lambda) \qquad {\rm (dB)}.
\end{equation} 
Notably, since $\kappa(\lambda)<1$ for all $\lambda\geqslant0$, it holds that $\Delta_{\rm NP-DL}(\lambda)>0$ for all $\lambda\geqslant0$.
\end{myremark}

In addition, we notice that 
\begin{equation}\label{47}
	\left.\frac{d\kappa(\lambda)}{d\lambda}\right|_{\lambda=0} = \frac{2}{K(1-c)}{\rm tr}\mathbf{R}^{-1}>0
\end{equation}
for all $c\in(0,1)$. 

\begin{myremark} Eq. \eqref{47} implies that $\kappa(\lambda)$ is strictly increasing in a sufficiently small neighborhood of $\lambda=0$. Hence,  there exists $\epsilon>0$ such that for all $\lambda\in(0,\epsilon)$, $\kappa(\lambda)>\kappa(0)=1-c$. We notice that the case $\kappa(0)$ actually corresponds the SCM-based AMF. As a result, we conclude that the DL-AMF with a small positive $\lambda$ always outperforms the SCM-based AMF. 
\end{myremark}
Furthermore, we observe that 
\begin{equation}\label{48}
	\lim_{\lambda\to+\infty}\kappa(\lambda) = \frac{1}{(\mathbf{s}^H\mathbf{R}^{-1}\mathbf{s})(\mathbf{s}^H\mathbf{R}\mathbf{s})}\overset{(a)}{\leqslant} 1
\end{equation}
where (a) follows from the Cauchy-Schwarz inequality. A simple analysis shows that the equality in \eqref{48} holds if and only if $\mathbf{s}\in{\rm eig}(\mathbf{R})$.  
%
We should keep in mind that the target steering vector $\mathbf{s}$ and the clutter-plus-noise covariance $\mathbf{R}$ arise from statistically independent mechanisms, making the condition $\mathbf{s}\in{\rm eig}(\mathbf{R})$ a probability-zero event in practice. In addition, even if one artificially constructs $\mathbf{s}\approx\mathbf{v}_i$ with $\mathbf{v}_i\in{\rm eig}(\mathbf{R})$, infinitesimal perturbations, such as the array calibration errors ($\Delta\mathbf{s}$) and clutter non-stationarity ($\Delta\mathbf{R}$), will disrupt the exact alignment, yielding $\mathbf{s}+\Delta\mathbf{s}\notin{\rm eig}(\mathbf{R})$ or $\mathbf{s}\notin{\rm eig}(\mathbf{R}+\Delta\mathbf{R})$ with probability 1. Hence, in all practical radar systems, it holds that $\mathbf{s}\notin{\rm eig}(\mathbf{R})$ almost surely. This fundamental property guarantees that the Eq. \eqref{48} reduces to $\lim_{\lambda\to+\infty}\kappa(\lambda)<1$ in all practical scenarios. 
\begin{figure*}[t]\centering
	\subfloat[$\frac{1}{(\mathbf{s}^H\mathbf{R}^{-1}\mathbf{s})(\mathbf{s}^H\mathbf{R}\mathbf{s})}\geqslant1-c$]{
		\centering
		\includegraphics[width=0.35\textwidth]{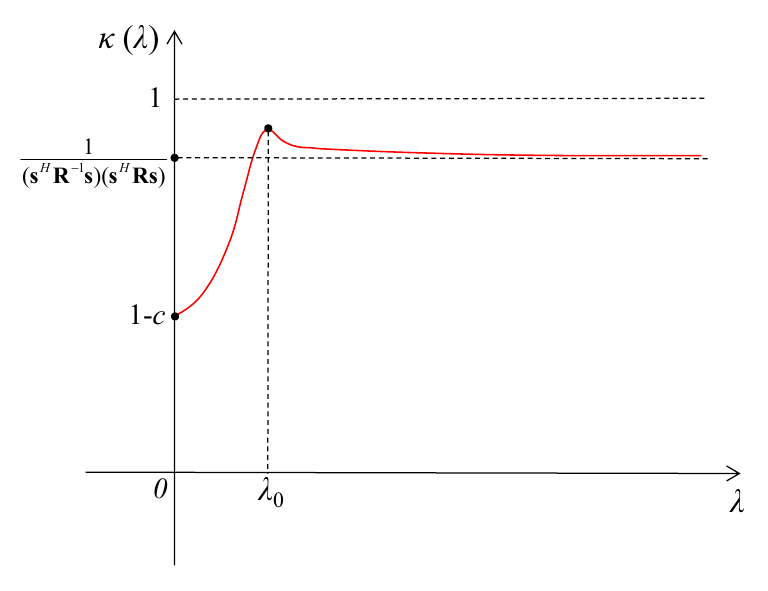}
	}\hspace{3em}
	\subfloat[$\frac{1}{(\mathbf{s}^H\mathbf{R}^{-1}\mathbf{s})(\mathbf{s}^H\mathbf{R}\mathbf{s})}<1-c$]{
		\centering
			\includegraphics[width=0.35\textwidth]{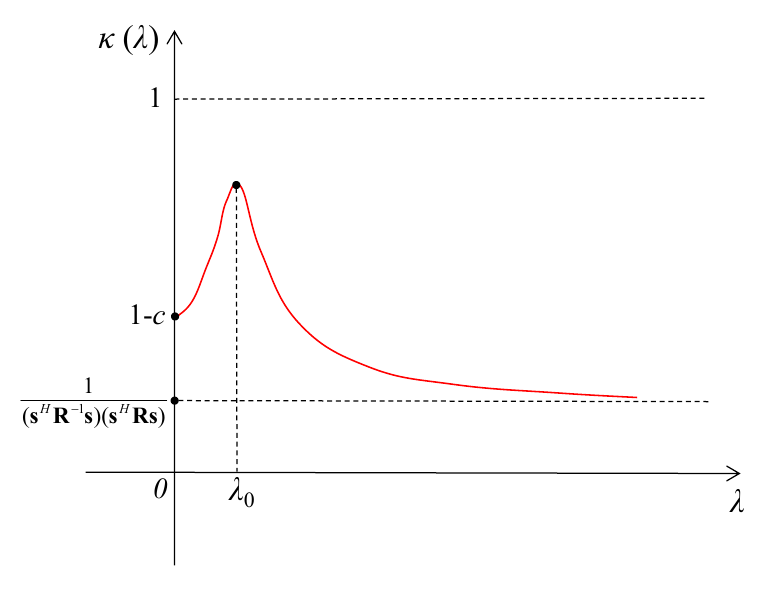}
	}
	\caption{Two typical representations of $\kappa(\lambda)$ when $\mathbf{s}\notin{\rm eig}(\mathbf{R})$.}
	\label{typical representations of kappa}
	\vspace{-1em}
\end{figure*}

Given the above analytical properties of $\kappa(\lambda)$, we are now able to illustrate its general behavior of. In Fig. \ref{typical representations of kappa}, we depict two potential representations of the function $\kappa(\lambda)$ that arise in simulation experiments. We emphasize that while these cases may not constitute an exhaustive set, they encompass all situations observed in our simulation study.  We observe from Fig. \ref{typical representations of kappa} that the behavior of $\kappa(\lambda)$ is fundamentally dependent on the magnitude relationship between $\frac{1}{(\mathbf{s}^H\mathbf{R}^{-1}\mathbf{s})(\mathbf{s}^H\mathbf{R}\mathbf{s})}$ and $1-c$. As stated in \cite{Abramovich2007Modified}, the clairvoyant optimum filter $W_{\rm opt}=\mathbf{R}^{-1}\mathbf{s}$ has an SCNR improvement over the ``white noise optimum'' filter $W_{\rm wn}=\mathbf{s}$ of 
\begin{equation}\label{key}
	\eta = {(\mathbf{s}^H\mathbf{R}^{-1}\mathbf{s})(\mathbf{s}^H\mathbf{R}\mathbf{s})}.
\end{equation} Therefore, the behavior of $\kappa(\lambda)$ is actually related to the relationship between $\eta$ and $1-c$. Specifically:
\begin{itemize}
	\item when $\frac{1}{(\mathbf{s}^H\mathbf{R}^{-1}\mathbf{s})(\mathbf{s}^H\mathbf{R}\mathbf{s})}\geqslant1-c$ or equivalently $\eta\leqslant\frac{1}{1-c}$, it holds that $1-c\leqslant\kappa(\lambda)<1$ for all $\lambda\geqslant0$. Hence, in this case, the DL-AMF with any $\lambda>0$ consistently outperforms the SCM-based AMF;
	\item when $\frac{1}{(\mathbf{s}^H\mathbf{R}^{-1}\mathbf{s})(\mathbf{s}^H\mathbf{R}\mathbf{s})}<1-c$ or equivalently $\eta>                                                                                                                                                                                                                                                                                                                                                                                                                                                                                                                                                                                                                                                                                                                                                                                                                                                                                                                                                                                                                                                                                                                                                                                                                                                                                                                                                                                                                                                                                                                                                                                                                                                                                                                                                                                                                                                                                                                                                                                                                                                                                                                                                                                                                                                                                                                                                                                                                                                                                                                                                                                                                                                                                                                                                                                                                                                                                                                                                                                                                                                                                                                                                                                                                                                                                                                                                                                                                                                                                                                                                                                                                                                                                                                                                                                                                                                                                                                                                                                                                                                                                                                                                                                                                                                                                                                                                                                                                                                                                                                                                                                                                                                                                          \frac{1}{1-c}$, there exists some sufficiently large $\lambda$ such that $\kappa(\lambda)<1-c$. Consequently, in this case, the DL-AMF with an excessively large $\lambda$ performs worse than SCM-based AMF.
\end{itemize}
\begin{figure}
	\centering
	\includegraphics[trim=1cm 0.1cm 0 1cm, scale=0.45]{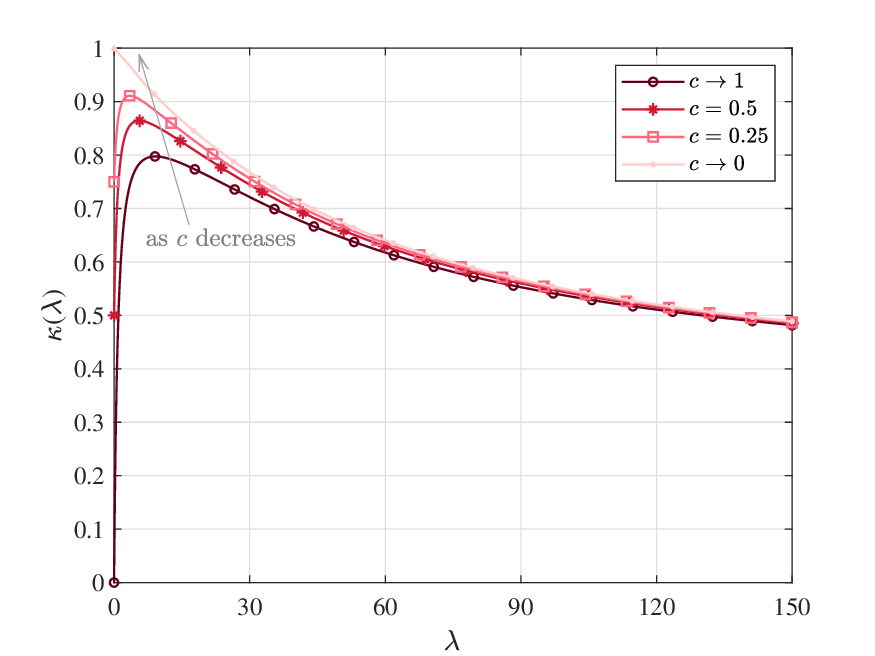}
	\caption{The behavior of $\kappa(\lambda)$ as $c$ decreases from 1 to 0.}
	\label{The behavior of kappa as c decreases from 1 to 0.}
	\vspace{-2em}
\end{figure}
Gathering the above results, we can obtain the following important insights.
\begin{myremark} The performance advantage of DL-AMF over SCM-based AMF is not guaranteed for arbitrary loading factor $\lambda$. This critically depends on the relationship between $\frac{1}{(\mathbf{s}^H\mathbf{R}^{-1}\mathbf{s})(\mathbf{s}^H\mathbf{R}\mathbf{s})}$ and the critical value $1-c$. To ensure superior detection performance of DL-AMF, one should carefully avoid excessively large loading factors.
\end{myremark}

\begin{myremark}The performance advantage of the DL-AMF relative to the SCM-based AMF is also affected by the value of $c$. For smaller $c$ (this corresponds to the large same size case), the probability of $\frac{1}{(\mathbf{s}^H\mathbf{R}^{-1}\mathbf{s})(\mathbf{s}^H\mathbf{R}\mathbf{s})}$ exceeding $1-c$ is low. In such cases, when employing larger loading factors, the DL-AMF is likely to perform worse than the SCM-based detector. The limiting case occurs when $c\to0$. In this case, the function $\kappa(\lambda)$ reaches its maximum at $\lambda=0$. This indicates that diagonal loading becomes unnecessary for detection performance enhancement. This is because, as $c\to0$, the SCM-based AMF can achieve optimal performance without diagonal loading, as we have shown in \eqref{41}-\eqref{46-1}. Conversely, when $c$ is large (i.e., $N$ is close to $K$, corresponding to the insufficient sample scenarios), the probability of $\frac{1}{(\mathbf{s}^H\mathbf{R}^{-1}\mathbf{s})(\mathbf{s}^H\mathbf{R}\mathbf{s})}$ surpassing $1-c$ increases significantly. In such cases, the DL-AMF is highly likely to maintain superior performance over the SCM-based AMF for any choice of $\lambda$. These results demonstrate that the performance advantage of the DL-AMF is particularly prominent in insufficient sample scenarios. 
\end{myremark}
\begin{myremark} We observe that, for all cases depicted in Fig. \ref{typical representations of kappa}, there always exists an optimal loading factor $\lambda_0$ at which $\kappa(\lambda)$ arrives at the maximum, i.e., $\lambda_0 = \arg\max\limits_{\lambda}\kappa(\lambda)$. Hence, the DL-AMF achieves the optimal performance at $\lambda=\lambda_0$ for both Swerling 0 and Swerling I targets.  Substituting $\lambda_0$ into \eqref{37} and \eqref{38}, we can obtain the optimal detection probabilities, given by $\widetilde{P}_{\rm {\rm DL}}^{Swer0}(\lambda_0)$ for Swerling 0 target and $\widetilde{P}_{\rm {\rm DL}}^{Swer1}(\lambda_0)$ for Swerling I target. Notably, for any $\lambda\in(0,\lambda_0]$, we have $\kappa(\lambda)>1-c$. This means that we can take any $\lambda$ within the interval $(0,\lambda_0]$ to ensure that the DL-AMF always outperforms the SCM-based AMF.  
\end{myremark}
\begin{myremark} For fixed $\mathbf{s}$ and $\mathbf{R}$, the value of $\lambda_0$ is related to $c$. As $c$ decreases from 1 to 0, the optimal loading factor $\lambda_0$ decreases, while the corresponding value $\kappa(\lambda_0)$  exhibits an increasing trend, as shown in Fig. \ref{The behavior of kappa as c decreases from 1 to 0.}. This implies that reducing the ratio $c$ (e.g., increasing the sample size $K$ while keeping the dimension $N$ fixed) enhances the detection performance of the optimal DL-AMF (with $\lambda_0$). This is consistent with the expectation that increasing the sample size is in favor of detection performance.
\end{myremark}
\begin{figure*}[t]
	\flushleft
	\subfloat[the behavior of $\kappa(\lambda)$]{
		\centering
		\includegraphics[trim=2cm 0cm 1cm 2cm, scale=0.45]{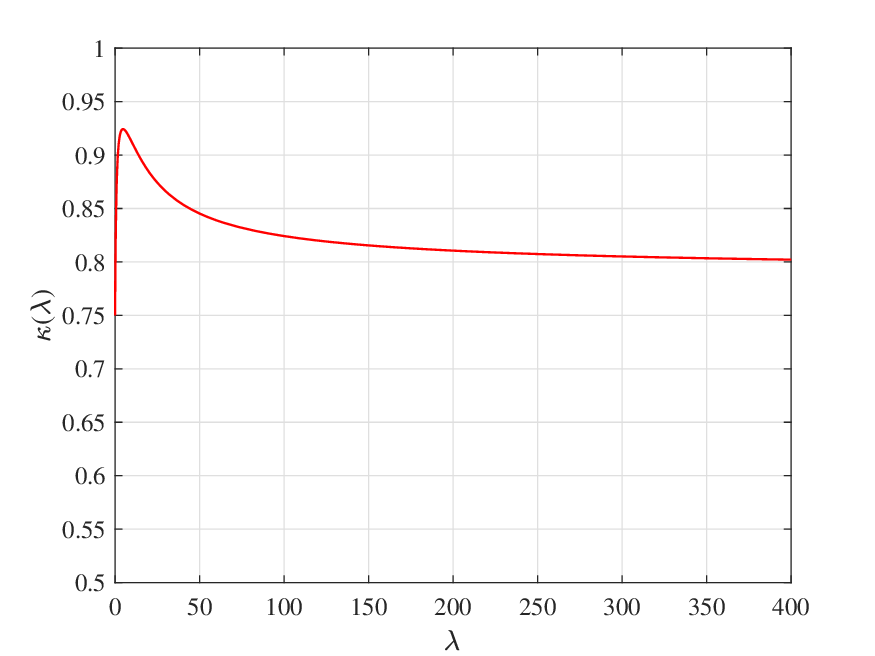}
	}
	\subfloat[Swerling 0 target]{
		\centering
		\includegraphics[trim=0.5cm 0cm 1cm 2cm, scale=0.45]{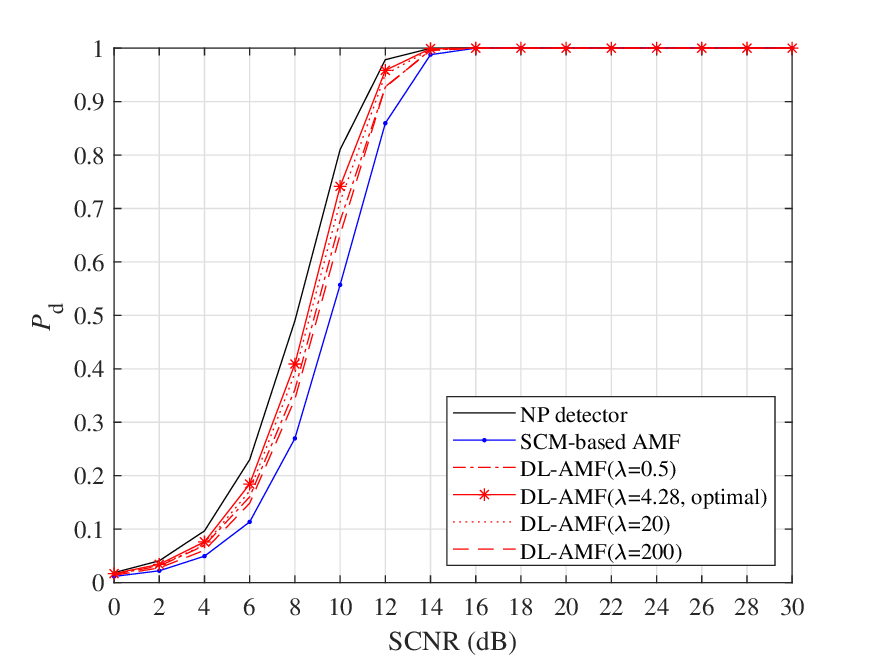}
	}
	\subfloat[Swerling I target]{
		\centering
		\includegraphics[trim=0.5cm 0cm 1cm 2cm, scale=0.45]{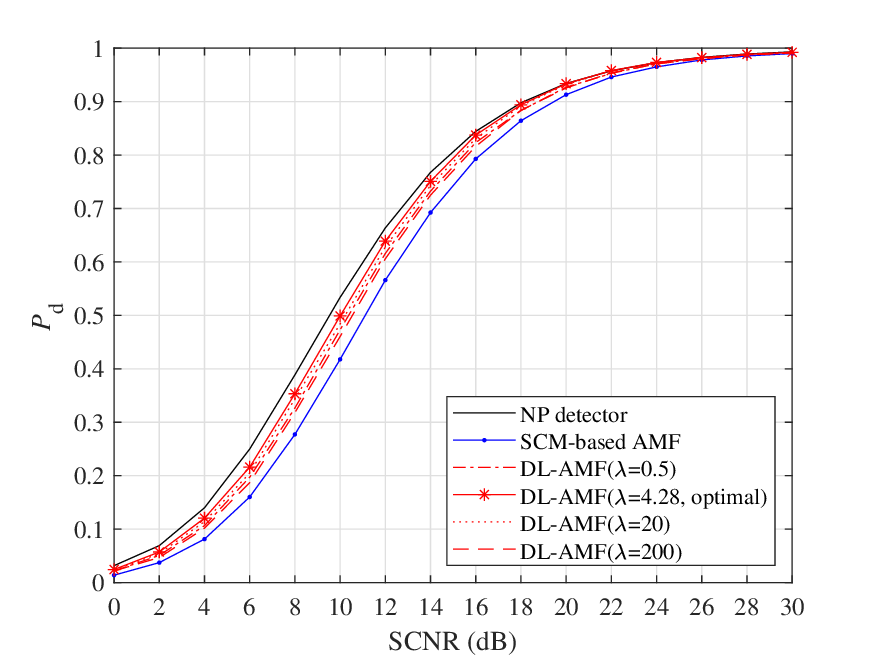}
	}\\
	\caption{The behavior of $\kappa(\lambda)$ and the detection performance of DL-AMF with different loading factor $\lambda$ in the case where $N=12, K=48$ ($c=N/K=0.25$), $\sigma^2=1$, $\sigma_c^2=10$ and $\theta_t=20^\circ$ such that $\frac{1}{(\mathbf{s}^H\mathbf{R}^{-1}\mathbf{s})(\mathbf{s}^H\mathbf{R}\mathbf{s})} \approx 0.79>1-c=0.75$. In this case, the function $\kappa(\lambda)$ achieves the maximum value $0.924$ at $\lambda_0\approx4.28$.}
	\label{The behavior of kappa and the detection performance 1}
	\vspace{-0.5em}
\end{figure*}
\begin{figure*}[t]
	\flushleft
	\subfloat[the behavior of $\kappa(\lambda)$]{
		\centering
		\includegraphics[trim=2cm 0cm 1cm 1cm, scale=0.45]{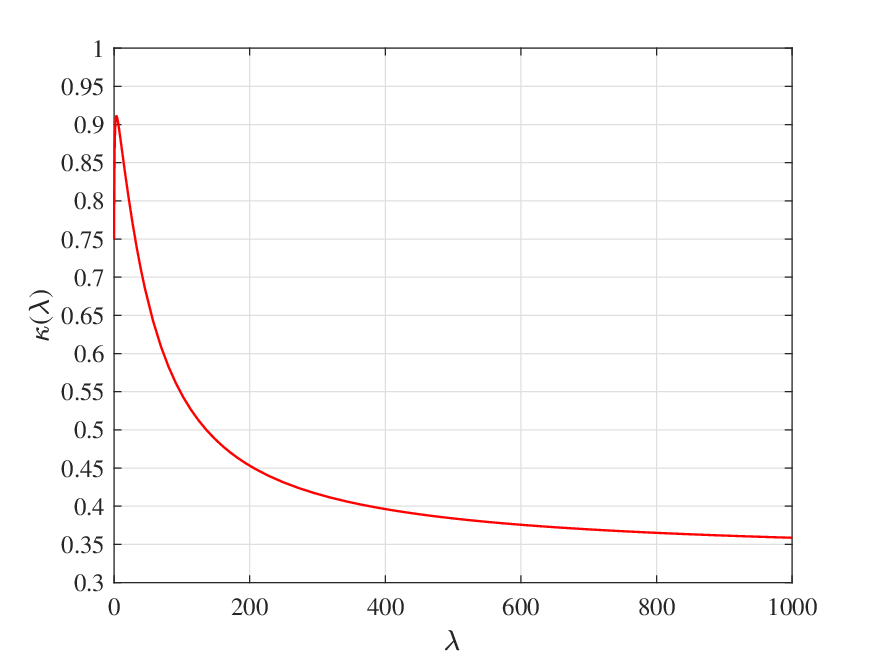}
	}
	\subfloat[Swerling 0 target]{
		\centering
		\includegraphics[trim=0.5cm 0cm 1cm 1cm, scale=0.45]{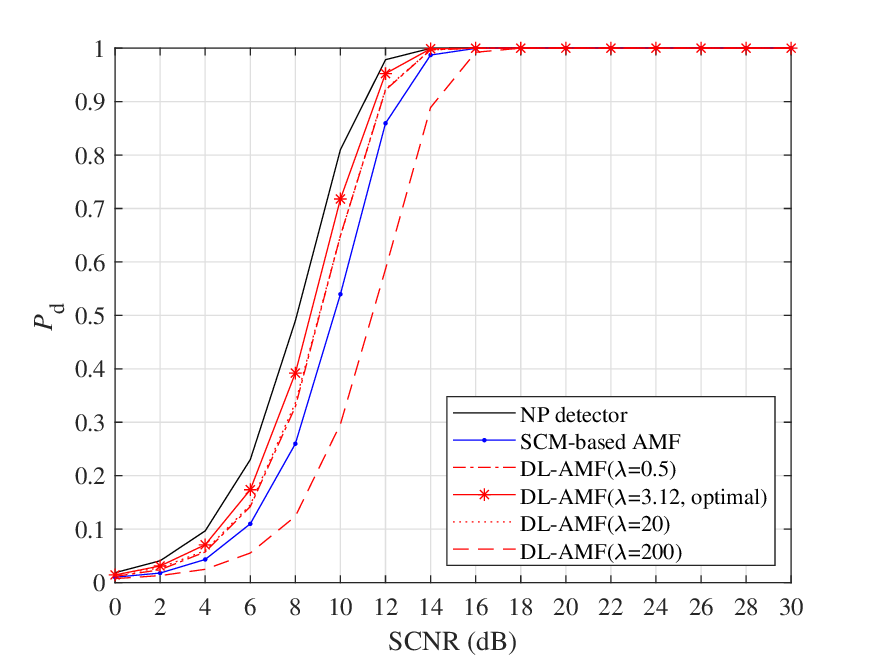}
	}
	\subfloat[Swerling I target]{
		\centering
		\includegraphics[trim=0.5cm 0cm 1cm 1cm, scale=0.45]{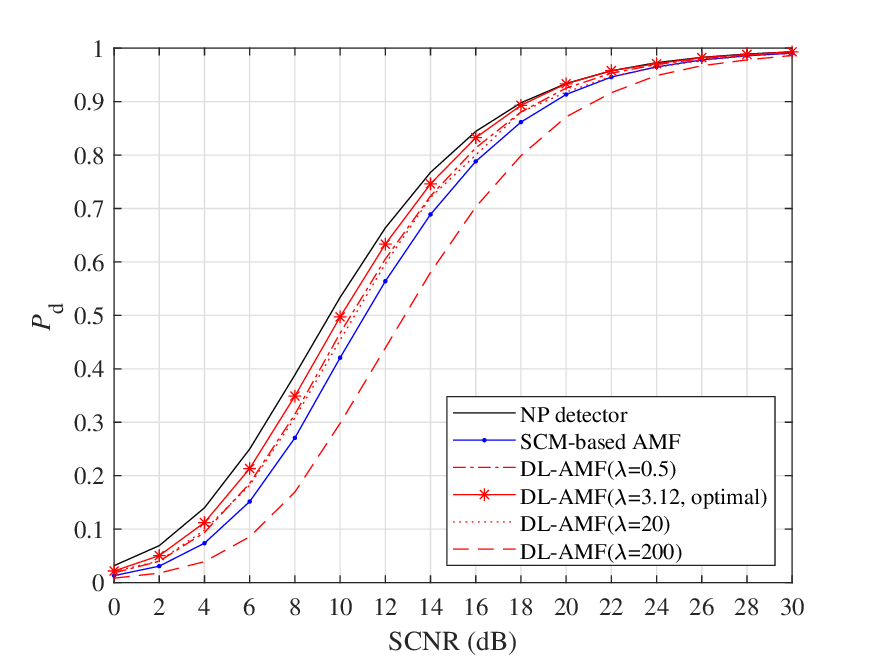}
	}\\
	\caption{The behavior of $\kappa(\lambda)$ and the detection performance of DL-AMF with different loading factor $\lambda$ in the case where $N=12, K=48$ ($c=N/K=0.25$), $\sigma^2=1$, $\sigma_c^2=10$ and $\theta_t=5^\circ$ such that $\frac{1}{(\mathbf{s}^H\mathbf{R}^{-1}\mathbf{s})(\mathbf{s}^H\mathbf{R}\mathbf{s})} \approx 0.33<1-c=0.75$. In this case, the function $\kappa(\lambda)$ achieves the maximum value $0.911$ at $\lambda_0\approx3.12$. }
	\label{The behavior of kappa and the detection performance 2}
	\vspace{-1.5em}
\end{figure*}
\begin{figure*}[t]
	\flushleft
	\subfloat[the behavior of $\kappa(\lambda)$]{
		\centering
		\includegraphics[trim=2cm 0cm 1cm 1cm, scale=0.45]{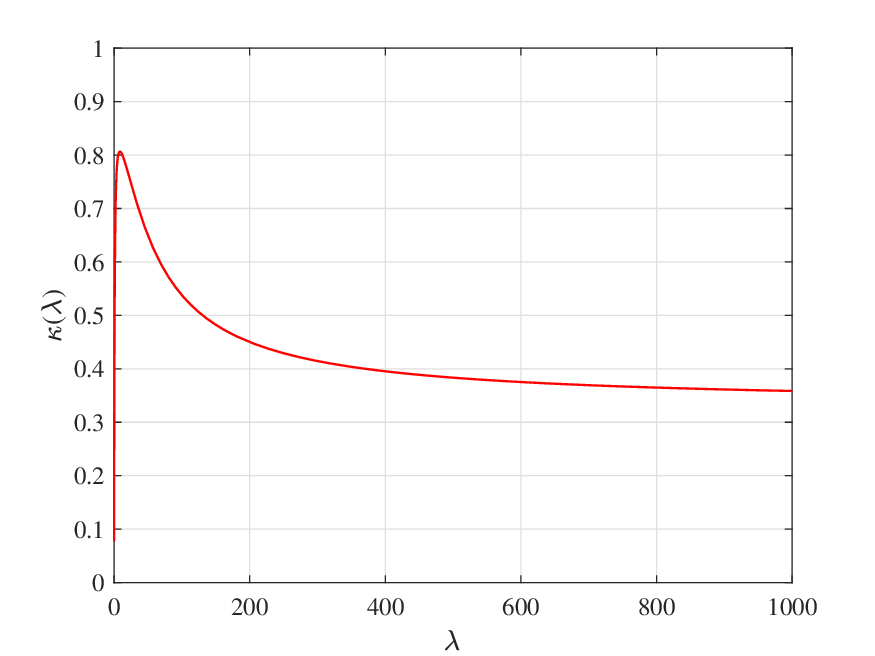}
	}
	\subfloat[Swerling 0 target]{
		\centering
		\includegraphics[trim=0.5cm 0cm 1cm 1cm, scale=0.45]{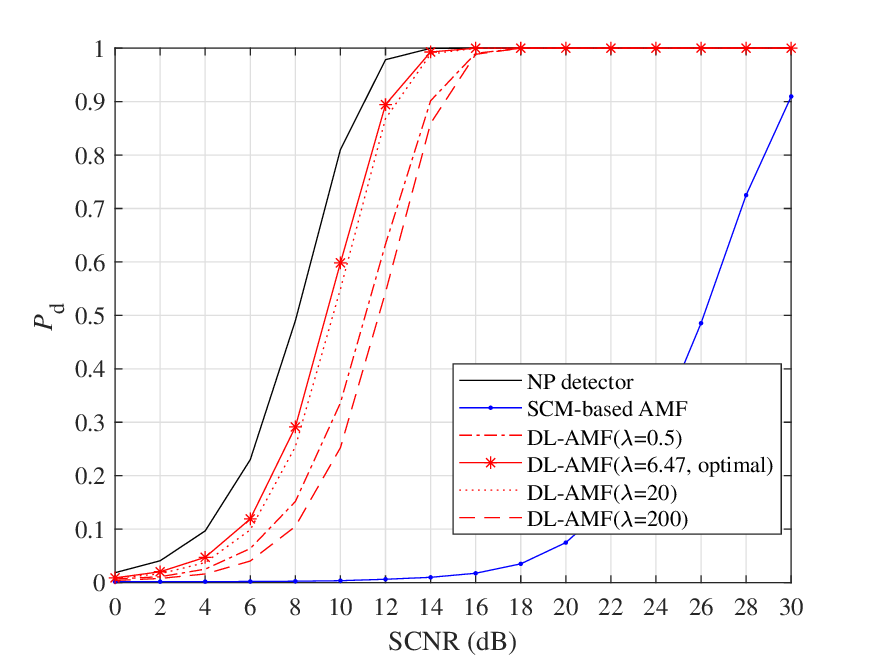}
	}
	\subfloat[Swerling I target]{
		\centering
		\includegraphics[trim=0.5cm 0cm 1cm 1cm, scale=0.45]{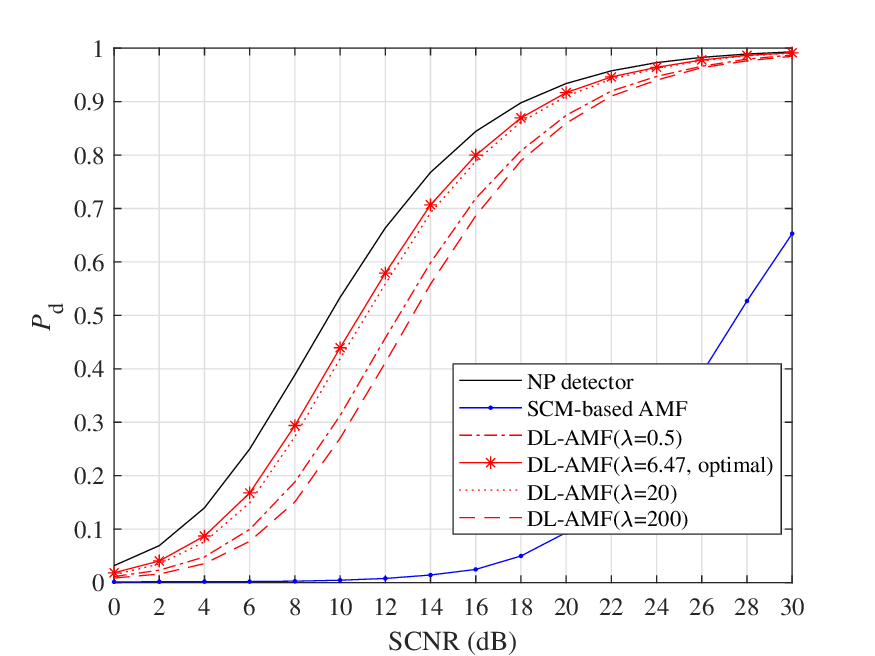}
	}\\
	\caption{The behavior of $\kappa(\lambda)$ and the detection performance of DL-AMF with different loading factor $\lambda$ in the case where $N=12, K=13$ ($c=N/K\approx0.92$), $\sigma^2=1$, $\sigma_c^2=10$ and $\theta_t=5^\circ$ such that $\frac{1}{(\mathbf{s}^H\mathbf{R}^{-1}\mathbf{s})(\mathbf{s}^H\mathbf{R}\mathbf{s})} \approx 0.33>1-c\approx0.08$. In this case, the function $\kappa(\lambda)$ achieves the maximum value $0.81$ at $\lambda_0\approx6.47$.}
	\label{The behavior of kappa and the detection performance 3}
	\vspace{-1.5em}
\end{figure*}
In the remainder of this section, we conduct some numerical simulations to validate above theoretical results and to provide intuitive interpretations of their implications. We adopt the same covariance matrix and steering vector models as those in Fig. \ref{Comparisons of the theoretical ROCs presented in 27 and 34 and the simulated results obtained from Monte Carlo trials.}.

Firstly, we consider the case $\frac{1}{(\mathbf{s}^H\mathbf{R}^{-1}\mathbf{s})(\mathbf{s}^H\mathbf{R}\mathbf{s})}>1-c$.  We evaluate the detection performance of DL-AMF with optimal loading factor $\lambda_0$, as well as three different loading factors, namely $\lambda=0.5,20,100$. We include SCM-based AMF in the the simulations for reference. Additionally,  the experiments also include the NP detector as a benchmark. The results are depicted in Fig. \ref{The behavior of kappa and the detection performance 1}(b) and (c), where we set $\varrho=0.95, \sigma_c^2=10$ and $\sigma^2=1$ such that the clutter-to-noise ratio is 10 dB. The dimension $N$ is 12 and the sample size $K$ is $48$ such that $c=N/K=0.25$. The preassigned $P_{\rm fa}$ is $10^{-3}$. In addition, we set $\theta_t=20^\circ$. The function $\kappa(\lambda)$ is plotted in Fig. \ref{The behavior of kappa and the detection performance 1}(a). In this case, we have $\lim\limits_{\lambda\to+\infty}\kappa(\lambda)=\frac{1}{(\mathbf{s}^H\mathbf{R}^{-1}\mathbf{s})(\mathbf{s}^H\mathbf{R}\mathbf{s})}\approx0.79>1-c=0.75$ and  $\kappa(\lambda)$ achieves the maximum value $0.924$ at $\lambda_0\approx4.28$. Based on our theoretical analysis, the DL-AMF is expected to outperform the SCM-based AMF while remaining inferior to the NP detector, regardless of the $\lambda$ value. This prediction is clearly verified by the results shown in Fig. \ref{The behavior of kappa and the detection performance 1}(b) and (c), where the detection probabilities of DL-AMF with different loading factors consistently lies between these two benchmarks across all tested $\lambda$ values. Notably, the DL-AMF with $\lambda_0\approx4.28$ exhibits optimal performance among all diagonal loading configurations, which further corroborates our theoretical findings regarding the detector's behavior at this critical operating point.

Next we examine the case $\frac{1}{(\mathbf{s}^H\mathbf{R}^{-1}\mathbf{s})(\mathbf{s}^H\mathbf{R}\mathbf{s})}<1-c$. We adjust the target direction $\theta_t$ from $20^\circ$ to $5^\circ$ while maintaining all other experimental parameters unchanged. In this configuration, we have $\lim\limits_{\lambda\to+\infty}\kappa(\lambda) = \frac{1}{(\mathbf{s}^H\mathbf{R}^{-1}\mathbf{s})(\mathbf{s}^H\mathbf{R}\mathbf{s})} \approx 0.33<1-c=0.75$ and $\kappa(\lambda)$ attains its maximum value $0.911$ at $\lambda_0\approx3.12$, as illustrated in Fig. \ref{The behavior of kappa and the detection performance 2}(a). Further computation shows that the equation $\kappa(\lambda)=1-c$ has two real-valued solutions:  one at $\lambda=0$ and another at $\lambda\approx31.2$.  Consequently, based on the theoretical results, we predict that the DL-AMF outperforms the SCM-based AMF for all $0<\lambda<31.2$, but exhibits inferior performance compared to the SCM-based AMF for $\lambda>31.2$. This prediction is confirmed in Fig. \ref{The behavior of kappa and the detection performance 2}(b) and (c): the DL-AMF with $\lambda = 0.5 , 3.12, 20$ exhibits superior detection performance to the SCM-based AMF, whereas the DL-AMF with $\lambda = 200$ incurs significant detection loss compared to the SCM-based AMF (approximately 2.2 dB).  In addition, as expected, the DL-AMF with optimal loading factor $\lambda_0\approx3.12$ shows the best performance among all DL-AMF.

As discussed in Remark 8, the performance advantage of the DL-AMF is most pronounced in insufficient sample scenarios. To validate this claim, we reduce $K$ from 48 to 13 while keeping all other experimental parameters unchanged from those in Fig. \ref{The behavior of kappa and the detection performance 2}. The function $\kappa(\lambda)$ in this case is plotted in Fig. \ref{The behavior of kappa and the detection performance 3}(a). It is clear that $\lim_{\lambda\to+\infty} =  \frac{1}{(\mathbf{s}^H\mathbf{R}^{-1}\mathbf{s})(\mathbf{s}^H\mathbf{R}\mathbf{s})} \approx 0.33>1-c\approx0.08$. Hence, according to our theoretical analysis, the DL-AMF with any positive $\lambda$ should always outperform SCM-based AMF. This has been verified by the results in Fig. \ref{The behavior of kappa and the detection performance 3}(b) and (c). In addition, $\kappa(\lambda)$ reaches the maximum value $0.81$ at $\lambda_0\approx6.47$. Consequently, the DL-AMF with $\lambda_0\approx6.47$ shows the best performance, as illustrated in Fig. \ref{The behavior of kappa and the detection performance 3}(b) and (c). Actually, we can see from Fig. \ref{The behavior of kappa and the detection performance 3}(b) and (c) that in the case $N=12,K=13$, the SCM-based AMF exhibits significant detection loss compared to the NP detector. In contrast, the DL-AMF maintains excellent detection performance, particularly the one using the optimal loading factor $\lambda_0$. Compared to the DL-AMF with $\lambda_0\approx6.47$, the SCM-based AMF suffers more than 10 dB detection loss. Furthermore, even compared to the DL-AMF with an extremely large loading factor, the SCM-based AMF still exhibits $\Delta_{\rm DL-SCM}(+\infty)\approx6.4$ dB detection loss.  It should be emphasized that these values are derived under the LDR framework, which assumes sufficiently large $N$ and $K$. However, for practical scenarios with finite and small $N, K$, as clearly evidenced by Fig. \ref{The behavior of kappa and the detection performance 3}(b) and (c), the actual detection loss of the SCM-based AMF relative to the DL-AMF substantially exceeds the theoretical predictions. For example, in Fig. \ref{The behavior of kappa and the detection performance 3}(b), we observe that the SCM-based AMF shows a detection loss of approximately 16.8 dB relative to the DL-AMF with the optimal loading factor $\lambda_0\approx6.47$. All these results demonstrate the superiority of the DL-AMF over the SCM-based AMF in insufficient sample scenarios, irrespective of whether $N$ and $K$ are large or small.  

Finally, we compare the results in Fig. \ref{The behavior of kappa and the detection performance 2} and Fig. \ref{The behavior of kappa and the detection performance 3}. As $K$ increases from 13 (as shown in Fig. \ref{The behavior of kappa and the detection performance 3}) to 48 (as shown in Fig. \ref{The behavior of kappa and the detection performance 2}), or equivalently as $c$ decreases from 0.92 to 0.25, we observe that the optimal loading factor $\lambda_0$ drops from $6.47$ to $3.12$, while the corresponding $\kappa(\lambda_0)$ increases from 0.81 to 0.911. All these results are consistent with the statements in Remark 10. Based on the previous analysis, a higher $\kappa(\lambda_0)$ value corresponds to a larger detection probability, indicating superior detection performance of DL-AMF. Therefore, we conclude that the optimal DL-AMF at $K=48$ outperforms that at $K=13$. A straightforward computation shows that the optimal DL-AMF in the case $K=48$ achieves a 0.5 dB performance gain (calculated as $10\log_{10}\frac{\kappa(3.12)}{\kappa(6.47)}\approx 0.5$) compared to that in the case $K=13$. This result aligns with the intuitive understanding that increasing the sample size is beneficial to the detection performance.  
\section{Design of CFAR DL Semi-Clairvoyant and Adaptive Matched Filter Detectors}
Although existing DL detectors, particularly ASG's EL-based detectors \cite{Abramovich2007Modified}, exhibit excellent detection performance under insufficient sample scenarios, they suffer from a fundamental limitation: the inherent lack of CFAR property, as we rigorously prove in Section II-A. In this section, our aim is to overcome this limitation.  We will present a novel solution that enables traditional DL detectors to maintain CFAR property while preserving their excellent detection performance.

Based on the theoretical framework established in Sections II-A and II-B, and as formally stated in Propositions \ref{Proposition 1} and \ref{Proposition 2}, we know that the detection performance of DL detectors depends solely on the random variable $|\hat{\alpha}_{\rm DL}(\lambda)|^2$, with $\hat{\beta}_{\rm DL}(\lambda)$ exhibiting no influence. This key insight  inspires us to construct a detector by normalizing $|\hat{\alpha}_{\rm DL}(\lambda)|^2$
with either:
\begin{itemize}
	\item(i) an deterministic quantity, or
	\item(ii) a random variable that converges almost surely to a deterministic quantity,
\end{itemize}
while guaranteeing excellent detection performance under LDR.  By implementing normalization with a properly chosen deterministic quantity, we can ensure that the resulting test statistic's distribution becomes invariant to disturbance parameters, thereby achieving the CFAR property. 

In the following subsections, we will proceed with the normalization of $|\hat{\alpha}(\lambda)|^2$ according to approaches (i) and (ii) to construct CFAR detectors. As will be demonstrated in Section III-A, the normalization with a deterministic quantity yields a test statistic that remains dependent on unknown parameters.  To overcome this limitation, Section III-B explores normalization with a random variable that is the consistent estimate of the deterministic quantity obtained in Section III-A within the LDR framework. 

\subsection{Design of CFAR-DL-SCMF Detector}
In Section II-A, we have proven that $|\hat{\alpha}_{\rm DL}(\lambda)|^2$ converges (in distribution) to an Exponential distribution with scale parameter $\mu_0(\mathbf{R},\mathbf{s},\lambda)$ (see Eq. \eqref{17}) under LDR. Now by normalizing $|\hat{\alpha}_{\rm DL}(\lambda)|^2$ with $\mu_0(\mathbf{R},\mathbf{s},\lambda)$,  we obtain the following detector:
\begin{equation}\label{54}
\hat{\eta}_{_{\rm CFAR\mbox{-}DL\mbox{-}SCMF}} = \frac{|\hat{\alpha}_{\rm DL}(\lambda)|^2}{\mu_0(\mathbf{R},\mathbf{s},\lambda)} \mathop \gtrless \limits_{H_0}^{H_1}\tau
\end{equation}
where $\tau$ is the threshold determined by the preassigned $P_{\rm fa}$. Substituting $\hat{\alpha}_{\rm DL}(\lambda)$ and $\mu_0(\mathbf{R},\mathbf{s},\lambda)$, we get the explicit form of the test statistic $\hat{\eta}_{_{\rm CFAR\mbox{-}DL\mbox{-}SCMF}} $, given by
\begin{equation}\label{55}
	\hat{\eta}_{_{\rm CFAR\mbox{-}DL\mbox{-}SCMF}}  = \frac{1-\gamma(\lambda)}{\mathbf{u}^H\mathbf{E}(\lambda)^2\mathbf{u}}\left|\mathbf{s}^H\left(\hat{\mathbf{R}}+\lambda\mathbf{I}_N\right)^{-1}\mathbf{y}_0\right|^2
\end{equation}
where $\gamma(\lambda)$, $\mathbf{u}$ and $\mathbf{E}(\lambda)$ maintain their original definitions as established in Theorem \ref{Theorem asymptotic distribution of alpha}.

We are now in position to check the performance of the detector in \eqref{54}. The following theorem gives the asymptotic distribution of $\hat{\eta}_{_{\rm CFAR\mbox{-}DL\mbox{-}SCMF}} $ under $H_0$ and provides the asymptotic $P_{\rm fa}$ of the detector in \eqref{54}.
\begin{mytheorem}\label{Theorem Asymptotic distribution of eta CFAR DL under H0}
	Let assumptions (A1)-(A2) hold true. Under $H_0$ and as $N,K\to\infty$, it holds that 
	\begin{equation}\label{58}
			\hat{\eta}_{_{\rm CFAR\mbox{-}DL\mbox{-}SCMF}} \overset{d}{\longrightarrow}{\rm Exp}(1).
	\end{equation}

The asymptotic $P_{\rm fa}$ of the the detector in \eqref{54} is given by
\begin{equation}\label{57}
	\begin{aligned}
		\widetilde{P}_{\rm fa,CFAR\mbox{-}DL}=\exp(-\tau).
	\end{aligned}
\end{equation}
\end{mytheorem}
\noindent\textit{Proof:} Given that $\mu_0(\mathbf{R},\mathbf{s},\lambda)$ is deterministic, it follows from the asymptotic distribution of $|\hat{\alpha}_{\rm DL}(\lambda)|^2$ (see Eq. \eqref{17-1}) that $\hat{\eta}_{_{\rm CFAR\mbox{-}DL\mbox{-}SCMF}}$ converges in distribution to a standard Exponential distribution (scale parameter 1) as $N,K\to\infty, N/K\to c\in(0,1)$, whose PDF is given by
\begin{equation}\label{key}
	f_{\rm CFAR\mbox{-}DL}(x) = \exp(-x)
\end{equation}
from which we obtain the asymptotic $P_{\rm fa}$ of the detector given in \eqref{54}:
\begin{equation}\label{59}
	\begin{aligned}
		\widetilde{P}_{\rm fa,CFAR\mbox{-}DL}= \int_{\tau_4}^\infty f_{\rm CFAR\mbox{-}DL}(x) dx= \exp(-\tau).
	\end{aligned}
\end{equation}
\qed

\begin{myremark}
	We conclude from Theorem \ref{Theorem Asymptotic distribution of eta CFAR DL under H0} that the asymptotic distribution of $\hat{\eta}_{_{\rm CFAR\mbox{-}DL\mbox{-}SCMF}}$  under $H_0$ is invariant to any parameters associated with $\mathbf{R}$, or equivalently the asymptotic $P_{\rm fa}$ is not related to $\mathbf{R}$. This demonstrates that the detector in \eqref{54} possesses the CFAR property with respect to $\mathbf{R}$ under LDR.
\end{myremark} 

\begin{myremark}
It is important to highlight that the asymptotic distribution of $\hat{\eta}_{_{\rm CFAR\mbox{-}DL\mbox{-}SCMF}}$  under $H_0$ is also invariant to both $\mathbf{s}$ and $\lambda$, which implies that the detector in \eqref{54} achieves CFAR property with respect to $\mathbf{s}$ and $\lambda$ under LDR. This CFAR behavior provides a significant practical advantage: the detector’s threshold can be determined without dependence on the choice of target steering vector $\mathbf{s}$ or variations in the loading factor $\lambda$. Notably, the CFAR properties with respect to $\lambda$ is particularly critical for a practical detector. Since $\lambda$ is typically estimated from observations (e.g., via the EL estimation in ASG's EL-AMF), it inherently exhibits random variability. By maintaining CFAR characteristic against $\lambda$'s statistical fluctuations, the detector's threshold remains unchanged even when the estimated $\lambda$ vary significantly.
\end{myremark} 
Unfortunately, as shown in \eqref{55}, the test statistic of this detector relies on the unknown population covariance matrix $\mathbf{R}$, making this detector infeasible for practical implementation. For this reason, we refer to the detector in \eqref{54} as the  \textit{CFAR DL semi-clairvoyant matched filter (CFAR-DL-SCMF) detector} to emphasize its theoretical (but unrealizable) nature.  Although not practically implementable, this detector serves as a crucial theoretical benchmark for constructing the fully adaptive detector in the following subsection.

In the following theorem, we provide the asymptotic distribution of $\hat{\eta}_{_{\rm CFAR\mbox{-}DL\mbox{-}SCMF}}$ under $H_1$ and present the detection performance of  CFAR-DL-SCMF under LDR.
\begin{mytheorem}\label{Theorem Asymptotic distribution of eta CFAR DL under H1}
	Let assumptions (A1)-(A4) hold true. Under $H_1$ and as $N,K\to\infty$, it holds that 
	\begin{equation}\label{62-1}
	\small	\hat{\eta}_{_{\rm CFAR\mbox{-}DL\mbox{-}SCMF}} \overset{d}{\longrightarrow}\begin{cases}
			\chi^2(\nu_{_{\rm CFAR}},\sigma_{_{\rm CFAR}}^2),	& \text{for Swerling 0 target}  \\
			{\rm Exp}(\mu_{_{\rm CFAR}}),	& \text{ for Swerling I target }
		\end{cases},
	\end{equation}
where $\nu_{_{\rm CFAR}} = S_0\kappa(\lambda)$, $\sigma^2_{_{\rm CFAR}}=\frac{1}{2}$ and $\mu_{_{\rm CFAR}} = S_1\kappa(\lambda)+1$ with $\kappa(\lambda)$ defined in \eqref{35}.

Consequently, the asymptotic $P_{\rm d}$ of CFAR-DL-SCMF is given by
	\begin{equation}\label{63}
\small	\begin{cases}
		\widetilde{P}_{\text{\rm d, CFAR\mbox{-}DL} }^{{Swer 0}}(\lambda)=  Q\left(\frac{\sqrt{\nu_{_{\rm CFAR}}}}{\sigma_{_{\rm CFAR}}},\frac{\sqrt{\tau}}{\sigma_{_{\rm CFAR}}}\right),	& \text{ for Swerling 0 target}\\
		\widetilde{P}_{\text{\rm d, CFAR\mbox{-}DL} }^{{Swer 1}}(\lambda) = \exp\left(-\frac{\tau}{\mu_{_{\rm CFAR}}}\right),	& \text{ for Swerling I target}
	\end{cases}.
\end{equation}
\end{mytheorem}
\noindent\textit{Proof:} The proof follows identical procedures to those presented in Section II-B by substituting $\widetilde{\beta}_{\rm DL}$ with $\frac{\mathbf{u}^H\mathbf{E}(\lambda)^2\mathbf{u}}{1-\gamma(\lambda)}$, and hence is omitted for brevity.\qed

Givena preassigned $P_{\rm fa,pre}$, Eq. \eqref{57} yields the threshold  $\tau = -\log P_{\rm fa,pre} $. Substituting $\tau$ into \eqref{63}, we find that the ROCs of the CFAR-DL-SCMF coincide exactly with those given in \eqref{37} and \eqref{38} for Swerling 0 and Swerling I target models, respectively. This result further validates the statement in Proposition \ref{Proposition 2}.

\begin{myremark}
	Theorem \ref{Theorem Asymptotic distribution of eta CFAR DL under H0} together with Theorem \ref{Theorem Asymptotic distribution of eta CFAR DL under H1} demonstrate that the CFAR-DL-SCMF presented in \eqref{54} achieves full CFAR property against $\mathbf{R}$, $\mathbf{s}$ and $\mathbf{\lambda}$ without compromising the original DL-AMF’s detection performance. 
\end{myremark}

In the next subsection, we will construct an implementable CFAR DL detector by developing a consistent estimator of $\mu_0(\mathbf{R},\mathbf{s},\lambda)$.

\subsection{Design of CFAR-DL-AMF Detector}
In this subsection, we are devoted to deriving a fully adaptive version of CFAR-DL-SCMF. The key lies in developing a consistent estimator for $\mu_0(\mathbf{R},\mathbf{s},\lambda)$ under LDR. While the classical approach that replaces $\mathbf{R}$ with the SCM $\hat{\mathbf{R}}$ (yielding $\mu_0(\hat{\mathbf{R}},\mathbf{s},\lambda)$) provides a natural solution, this estimator suffers from a critical theoretical limitation: it only achieves consistency in the traditional large sample size regime ($K\to\infty$ for fixed $N$). This restriction arises from the fact that the elements of $\hat{\mathbf{R}}$ converges almost surely to those of $\mathbf{R}$ only under the large sample size regime. However, the practical utility of DL detectors emerges in insufficient sample situations where $N$ and $K$ have comparable order of magnitude. Under these practically relevant conditions, the classical estimate $\mu_0(\hat{\mathbf{R}},\mathbf{s},\lambda)$ is typically unsatisfactory. Hence, our goal now is to develop an alternative estimator that guarantees almost sure convergence to $\mu_0(\mathbf{R},\mathbf{s},\lambda)$ in the case where $N$ is comparable in magnitude to $K$. 

The following theorem gives the consistent estimator of $\mu_0(\mathbf{R},\mathbf{s},\lambda)$ under LDR.
\begin{mytheorem}\label{Theorem consistent estimate for mu0}
	Let assumptions (A1)-(A3) hold true. Then as $N,K\to \infty$, it holds that
	\begin{equation}\label{key}
		|\widehat{\mu_0(\mathbf{R},\mathbf{s},\lambda)} - \mu_0(\mathbf{R},\mathbf{s},\lambda)|\overset{a.s.}{\longrightarrow}0
	\end{equation}
where 
\begin{equation}\label{65-1}
	\small \widehat{\mu_0(\mathbf{R},\mathbf{s},\lambda)} = \frac{\mathbf{s}^H\left(\hat{\mathbf{R}}+\lambda\mathbf{I}_N\right)^{-1}\hat{\mathbf{R}}\left(\hat{\mathbf{R}}+\lambda\mathbf{I}_N\right)^{-1}\mathbf{s}}{\left[1-\frac{N}{K}+\frac{\lambda}{K}{\rm tr}\left(\hat{\mathbf{R}}+\lambda\mathbf{I}_N\right)^{-1}\right]^2}.
\end{equation}
\end{mytheorem}
\noindent\textit{Proof:} See Appendix D.

Now we normalize $|\hat{\alpha}_{\rm DL}(\lambda)|^2$ with $\widehat{\mu_0(\mathbf{R},\mathbf{s},\lambda)}$ and obtain the following detector
\begin{equation}\label{66}
	\hat{\eta}_{_{\rm CFAR\mbox{-}DL\mbox{-}AMF}} = \frac{|\hat{\alpha}_{\rm DL}(\lambda)|^2}{\widehat{\mu_0(\mathbf{R},\mathbf{s},\lambda)}} \mathop \gtrless \limits_{H_0}^{H_1}\tau
\end{equation}

Substituting \eqref{65-1} into \eqref{66}, the test statistic of the detector in \eqref{66} is specified by
\begin{equation}\label{66-1}
	\small\begin{aligned}
&\hat{\eta}_{_{\rm CFAR\mbox{-}DL\mbox{-}AMF}} =
\\&  \frac{\left[1-\frac{N}{K}+\frac{\lambda}{K}{\rm tr}\left(\hat{\mathbf{R}}+\lambda\mathbf{I}_N\right)^{-1}\right]^2}{\mathbf{s}^H\left(\hat{\mathbf{R}}+\lambda\mathbf{I}_N\right)^{-1}\hat{\mathbf{R}}\left(\hat{\mathbf{R}}+\lambda\mathbf{I}_N\right)^{-1}\mathbf{s}}\left|\mathbf{s}^H\left(\hat{\mathbf{R}}+\lambda\mathbf{I}_N\right)^{-1}\mathbf{y}_0\right|^2.
\end{aligned}
\end{equation} 

The detector in \eqref{66} is the fully adaptive version of CFAR-DL-SCMF presented in \eqref{54} and is thus referred to as \textit{CFAR DL adaptive matched filter (CFAR-DL-AMF) detector}.  

\begin{myremark}\label{Remark 14}
	Since $\widehat{\mu_0(\mathbf{R},\mathbf{s},\lambda)}$ is a consistent estimate for $\mu_0(\mathbf{R},\mathbf{s},\lambda)$,  $\hat{\eta}_{_{\rm CFAR\mbox{-}DL\mbox{-}AMF}}$ shares the same asymptotic distribution as $\hat{\eta}_{_{\rm CFAR\mbox{-}DL\mbox{-}SCMF}}$ (under both $H_0$ and $H_1$, see \eqref{58} and \eqref{62-1} respectively) as $N,K$ go to infinity at the same rate. Consequently, the CFAR-DL-AMF achieves the full CFAR properties with respect to $\mathbf{R}, \mathbf{s}$ and $\lambda$ under LDR and exhibits identical detection performance to the CFAR-DL-SCMF in the LDR. This means that the asymptotic $P_{\rm fa}$ and $P_{\rm d}$ of CFAR-DL-AMF under LDR can also be characterized by \eqref{57} and \eqref{63}, respectively. Moreover, the ROCs of CFAR-DL-AMF under LDR can also be expressed by \eqref{34} (for Swerling 0 target) and \eqref{37} (for Swerling I target).
\end{myremark}
\begin{figure*}[t]
	\vspace{-0.5em}
	\centering
	\subfloat[]{
		\centering
		\includegraphics[trim=2cm 0cm 1cm 2cm, scale=0.45]{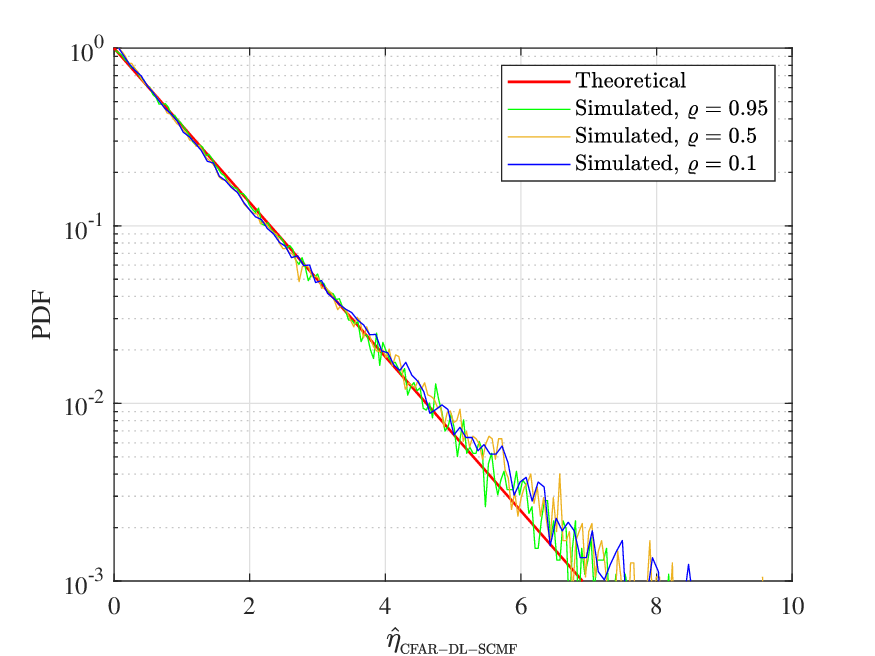}
	}
	\subfloat[]{
		\centering
		\includegraphics[trim=0.5cm 0cm 1cm 2cm, scale=0.45]{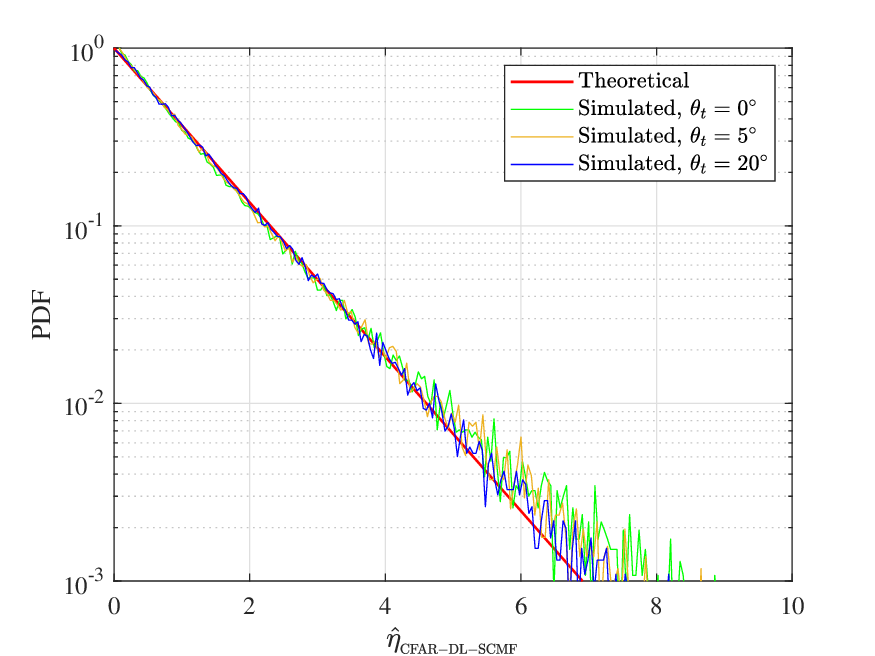}
	}
	\subfloat[]{
		\centering
		\includegraphics[trim=0.5cm 0cm 1cm 2cm, scale=0.45]{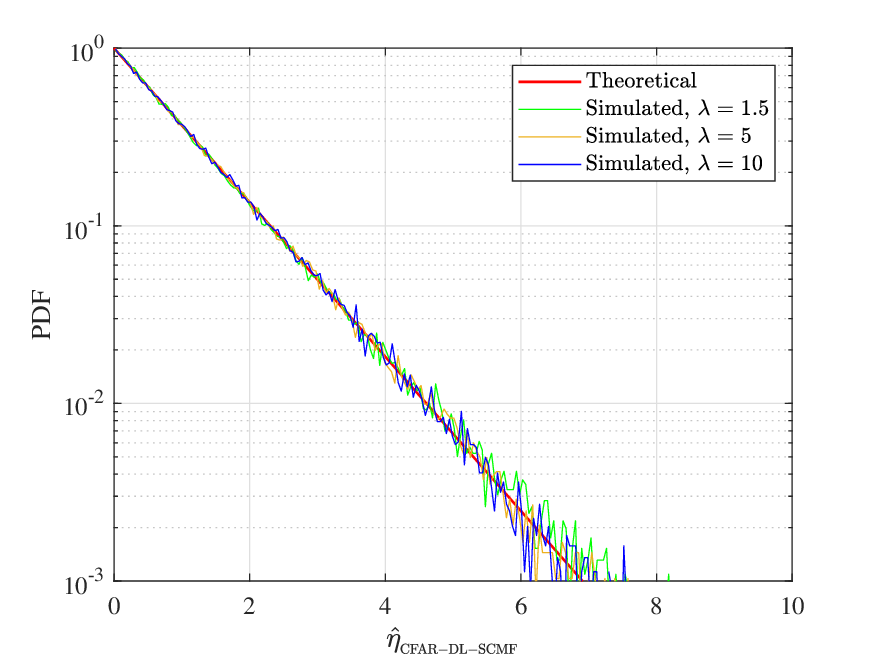}
	}\\
	\caption{Comparison between the theoretical PDFs of $\hat{\eta}_{_{\rm CFAR\mbox{-}DL\mbox{-}SCMF}}$ and its empirical PDFs obtained from $10^5$ Monte Carlo trials under different $\mathbf{R}$, $\mathbf{s}$ and $\lambda$. In (a), $\mathbf{R}$ varies (with varying $\varrho$ and fixed $\sigma^2=1,\sigma^2_c=10$) for fixed $\mathbf{s}(\theta_t=20^\circ)$ and $\lambda=1.5$; in (b), $\mathbf{s}$ varies for fixed $\mathbf{R}(\varrho=0.95,\sigma^2=1,\sigma^2_c=10)$ and $\lambda=1.5$; in (c), $\lambda$ varies for fixed $\mathbf{R}(\varrho=0.95,\sigma^2=1,\sigma^2_c=10)$ and $\mathbf{s}(\theta_t=20^\circ)$. }
	\label{Comparisons between the theoretical PDFs of the test statistics of CFAR-DL-SCMF the empirical PDFs under different R s lambda}
		\vspace{-0.5em}
\end{figure*}
\begin{figure*}[t]
	\centering
	\subfloat[]{
		\centering
		\includegraphics[trim=2cm 0cm 1cm 1cm, scale=0.45]{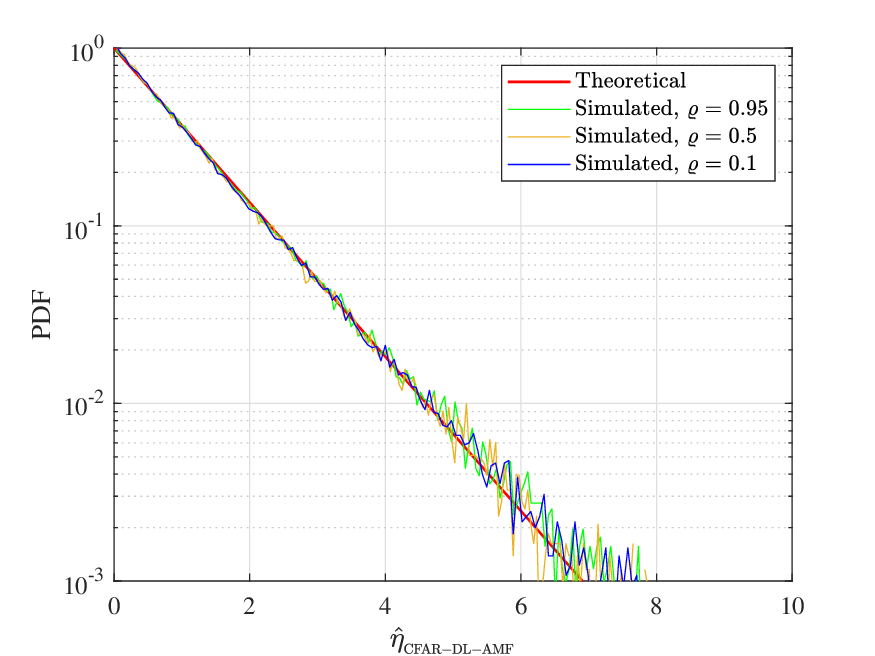}
	}
	\subfloat[]{
		\centering
		\includegraphics[trim=0.5cm 0cm 1cm 1cm, scale=0.45]{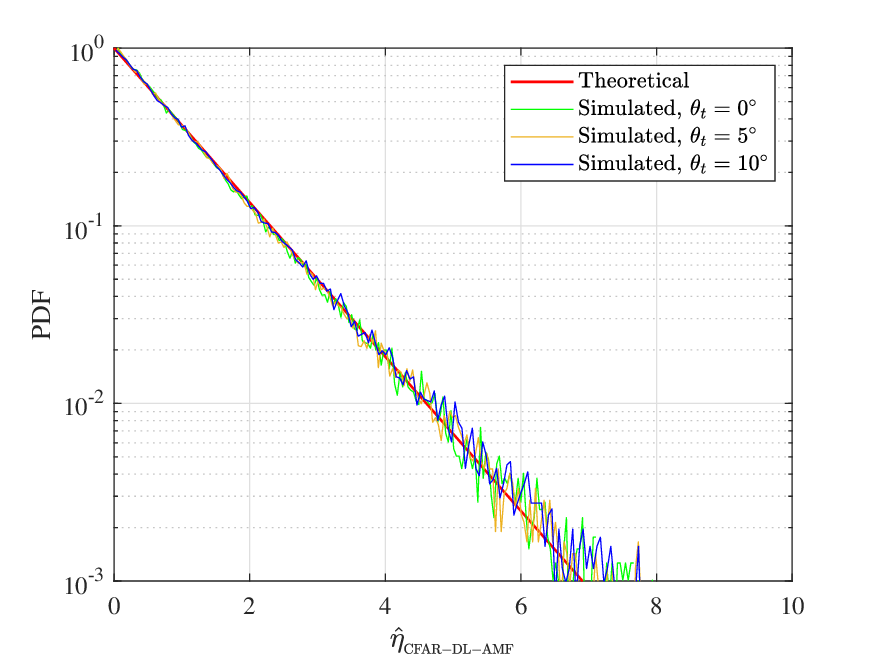}
	}
	\subfloat[]{
		\centering
		\includegraphics[trim=0.5cm 0cm 1cm 1cm, scale=0.45]{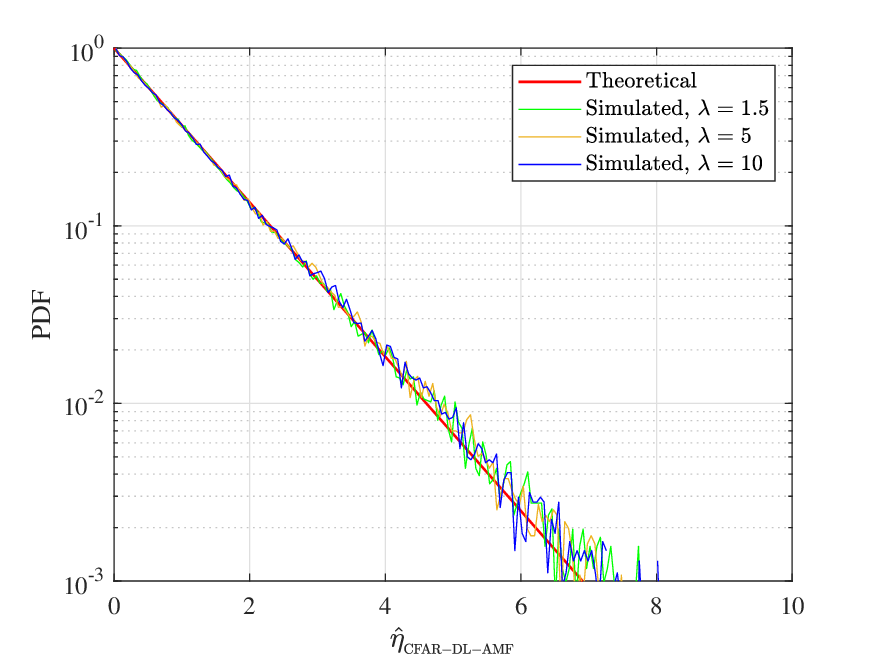}
	}\\
	\caption{Comparison between the theoretical PDFs of $\hat{\eta}_{_{\rm CFAR\mbox{-}DL\mbox{-}AMF}}$ and its empirical PDFs obtained from $10^5$ Monte Carlo trials under different $\mathbf{R}$, $\mathbf{s}$ and $\lambda$. The parameters for $\mathbf{R}$, $\mathbf{s}$ and $\lambda$ in panels (a)-(c) are identical to those in Fig. \ref{Comparisons between the theoretical PDFs of the test statistics of CFAR-DL-SCMF the empirical PDFs under different R s lambda} (a)-(c).}
	\label{Comparisons between the theoretical PDFs of the test statistics of CFAR-DL-AMF the empirical PDFs under different R s lambda}
		\vspace{-0.5em}
\end{figure*}
 \begin{figure}[htbp]
	\centering
	\includegraphics[trim=1cm 0.1cm 0 1cm, scale=0.5]{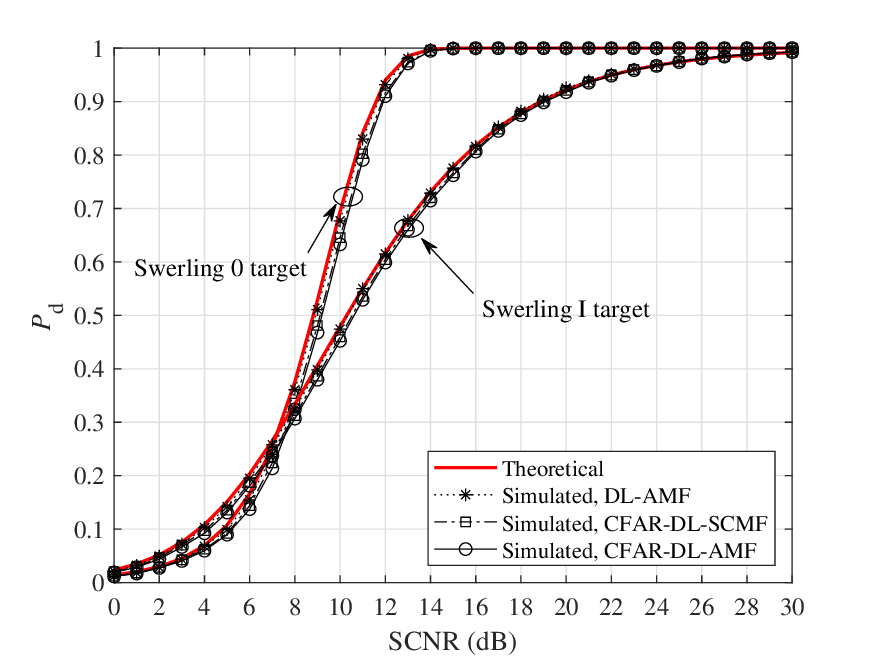}
	\caption{Comparison of the theoretical detection performance of CFAR-DL-SCMF and CFAR-DL-AMF and the simulated results. The simulation parameters are the same as those in Fig. \ref{Comparisons of the theoretical ROCs presented in 27 and 34 and the simulated results obtained from Monte Carlo trials.} (b).}
	\label{Comparison of the theoretical detection performance of CFAR-DL-SCMF and CFAR-DL-AMF and the simulated results obtained through Monte Carlo simulations}
	\vspace{-1.5em}
\end{figure}
In the following, we perform some simulations to validate the theoretical results in Sections III-A and III-B. 

We first validate the CFAR properties of CFAR-DL-SCMF and CFAR-DL-AMF with respect to $\mathbf{R}$, $\mathbf{s}$ and $\lambda$. Figs. \ref{Comparisons between the theoretical PDFs of the test statistics of CFAR-DL-SCMF the empirical PDFs under different R s lambda} and \ref{Comparisons between the theoretical PDFs of the test statistics of CFAR-DL-AMF the empirical PDFs under different R s lambda} present the theoretical and empirical PDFs of the test statistics of CFAR-DL-SCMF and CFAR-DL-AMF under $H_0$  for varying $\mathbf{R}$, $\mathbf{s}$ and $\lambda$. In these experioments, we set $N=24,K=48$. The results in Figs. \ref{Comparisons between the theoretical PDFs of the test statistics of CFAR-DL-SCMF the empirical PDFs under different R s lambda} and \ref{Comparisons between the theoretical PDFs of the test statistics of CFAR-DL-AMF the empirical PDFs under different R s lambda} demonstrate that the test statistics of both CFAR-DL-SCMF and CFAR-DL-AMF follow identical distributions across all configurations, as their empirical PDFs align perfectly with the theoretical standard Exponential distribution (red lines in Figs. \ref{Comparisons between the theoretical PDFs of the test statistics of CFAR-DL-SCMF the empirical PDFs under different R s lambda} and \ref{Comparisons between the theoretical PDFs of the test statistics of CFAR-DL-AMF the empirical PDFs under different R s lambda}). These results are consistent with the theoretical results in Theorem \ref{Theorem Asymptotic distribution of eta CFAR DL under H0} (see \eqref{58}) and Remark \ref{Remark 14}. All these illustrative results indicate that in contrast to DL-AMF (see Fig. \ref{Comparisons of theoretical asymptotic distribution of eta DL and the simulated results}), the test statistics of CFAR-DL-SCMF and CFAR-DL-AMF are invariant to $\mathbf{R}$, $\mathbf{s}$ and $\lambda$. Hence, both detectors achieve full CFAR against $\mathbf{R}$, $\mathbf{s}$ and $\lambda$.

It remains to check whether the CFAR-DL-SCMF and CFAR-DL-AMF maintain identical detection performance to DL-AMF. To do this, we compare their detection performance in Fig. \ref{Comparison of the theoretical detection performance of CFAR-DL-SCMF and CFAR-DL-AMF and the simulated results obtained through Monte Carlo simulations}. The experiments parameters are the same as those in Fig. \ref{Comparisons of the theoretical ROCs presented in 27 and 34 and the simulated results obtained from Monte Carlo trials.}(b). The results in Fig. \ref{Comparison of the theoretical detection performance of CFAR-DL-SCMF and CFAR-DL-AMF and the simulated results obtained through Monte Carlo simulations} reveal that both CFAR-DL-SCMF and CFAR-DL-AMF detectors achieve nearly identical detection performance to the original DL-AMF detector (with negligible performance loss). Moreover, their ROC curves closely approach the theoretical ROCs characterized by \eqref{27} and \eqref{34} (shown as red lines in Fig. \ref{Comparison of the theoretical detection performance of CFAR-DL-SCMF and CFAR-DL-AMF and the simulated results obtained through Monte Carlo simulations}). The minimal performance loss of the CFAR-DL-SCMF (and/or CFAR-DL-AMF) relative to DL-AMF arises inherently from the CFAR constraint, reflecting a fundamental trade-off between the detection performance and CFAR property. In fact, this loss will gradually diminish as the sample size $K$ and dimension $N$ increases, and finally converges to zero when $N,K\to \infty$. In addition, we also observe a very slightly performance loss of CFAR-DL-AMF relative to CFAR-DL-SCMF. This loss comes from the estimation error in the $\widehat{\mu_0(\mathbf{R},\mathbf{s},\lambda)}$ under the  finite $N$ and $K$. When $N,K\to\infty$, this error will converges to 0 and hence the performance loss will also vanish. 

We emphasize that the significance of the CFAR-DL-AMF lies in its capability to enable arbitrary DL detectors to achieve CFAR property with respect to $\mathbf{R}, \mathbf{s}$ and $\lambda$ while maintaining the original detectors' detection performance. In the following subsection, we verify this capability by applying the CFAR-DL-AMF to ASG's EL-AMF \cite{Abramovich2007Modified}, and develop a CFAR version of EL-AMF, referred to as CFAR EL-based AMF (CFAR-EL-AMF).

\subsection{A Direct Application of CFAR-DL-AMF: CFAR Version of ASG's EL-AMF}
ASG's EL-AMF \cite{Abramovich2007Modified} is given by
\begin{equation}\label{key}
 \hat{\eta}_{_{\rm ASG\mbox{-}EL\mbox{-}AMF}}=\frac{|\mathbf{s}^H\left(\hat{\mathbf{R}}+\hat{\lambda}_{\rm EL}\mathbf{I}_N\right)^{-1}\mathbf{y}_0|^2}{\mathbf{s}^H\left(\hat{\mathbf{R}}+\hat{\lambda}_{\rm EL}\mathbf{I}_N\right)^{-1}\mathbf{s}}\mathop\gtrless\limits_{H_0}^{H_1}\tau
\end{equation}
where $\hat{\lambda}_{\rm EL}$ is estimated by the solution of the following equation at $\lambda$:
\begin{equation}\label{key}
	 \frac{{\rm det}\left[\left(\hat{\mathbf{R}}+\lambda\mathbf{I}_N\right)^{-1}\hat{\mathbf{R}}\right]\exp({N})}{\exp\left[{\rm tr}\left(\hat{\mathbf{R}}+\lambda\mathbf{I}_N\right)^{-1}\hat{\mathbf{R}}\right]}=\zeta_{_{0.5}}
\end{equation}
with $\zeta_{_{0.5}}$ being the median of the statistic $\zeta= \frac{{\rm det}(\mathbf{R}^{-1/2}\hat{\mathbf{R}}\mathbf{R}^{-1/2})\exp(N)}{\exp\left[{\rm tr}(\mathbf{R}^{-1/2}\hat{\mathbf{R}}\mathbf{R}^{-1/2})\right]}$. In \cite{Zhou2025On}, we derive the exact expression for $\zeta_{_{0.5}}$ under LDR, given by $\zeta_{_{0.5}} = e^{-N}(1-c)^{-(K-N)}$.

Although ASG's EL-AMF has been demonstrated to perform well in scenarios with limited training samples, it fails to maintain CFAR with respect to the covariance matrix. By applying the CFAR-DL-AMF framework to ASG's EL-AMF, we derive the CFAR-EL-AMF:  
\begin{equation}\label{71-1}
\hat{\eta}_{_{\rm CFAR\mbox{-}EL\mbox{-}AMF}}\mathop\gtrless\limits_{H_0}^{H_1}\tau
\end{equation}
where the test statistic $\hat{\eta}_{_{\rm CFAR\mbox{-}EL\mbox{-}AMF}} $ is given by
\begin{equation}\label{70-1}
	\small \begin{aligned}
	&\hat{\eta}_{_{\rm CFAR\mbox{-}EL\mbox{-}AMF}} =
	\\& \frac{\left[1-\frac{N}{K}+\frac{\hat{\lambda}_{\rm EL}}{K}{\rm tr}\left(\hat{\mathbf{R}}+\hat{\lambda}_{\rm EL}\mathbf{I}_N\right)^{-1}\right]^2\left|\mathbf{s}^H\left(\hat{\mathbf{R}}+\hat{\lambda}_{\rm EL}\mathbf{I}_N\right)^{-1}\mathbf{y}_0\right|^2}{\mathbf{s}^H\left(\hat{\mathbf{R}}+\hat{\lambda}_{\rm EL}\mathbf{I}_N\right)^{-1}\hat{\mathbf{R}}\left(\hat{\mathbf{R}}+\hat{\lambda}_{\rm EL}\mathbf{I}_N\right)^{-1}\mathbf{s}}.
\end{aligned}
\end{equation}
\begin{myremark}
	According to the analysis in Sections III-A and III-B, it is not difficult to find that the test statistic of the CFAR-EL-AMF, i.e., $\hat{\eta}_{_{\rm CFAR\mbox{-}EL\mbox{-}AMF}}$, asymptotically follows  a standard Exponential distribution (see \eqref{58}) under $H_0$.  Hence, the CFAR-EL-AMF achieves CFAR behavior with respect to $\mathbf{R}$ and $\mathbf{s}$. In addition, as we stated in Proposition \ref{Proposition 2}, normalizing $|\hat{\alpha}_{\rm DL}(\hat{\lambda}_{\rm EL})|$ with a random variable that converges almost surely to a deterministic quantity does not affect the detection performance under LDR. Hence, the CFAR-EL-AMF  preserves the original detection performance of ASG's EL-AMF under LDR.
\end{myremark}
\begin{figure*}[t]
	\vspace*{-0.5em}
	\centering
	\subfloat[ASG's EL-AMF]{
		\centering
		\includegraphics[trim=0.5cm 0.1cm 0 1cm, scale=0.5]{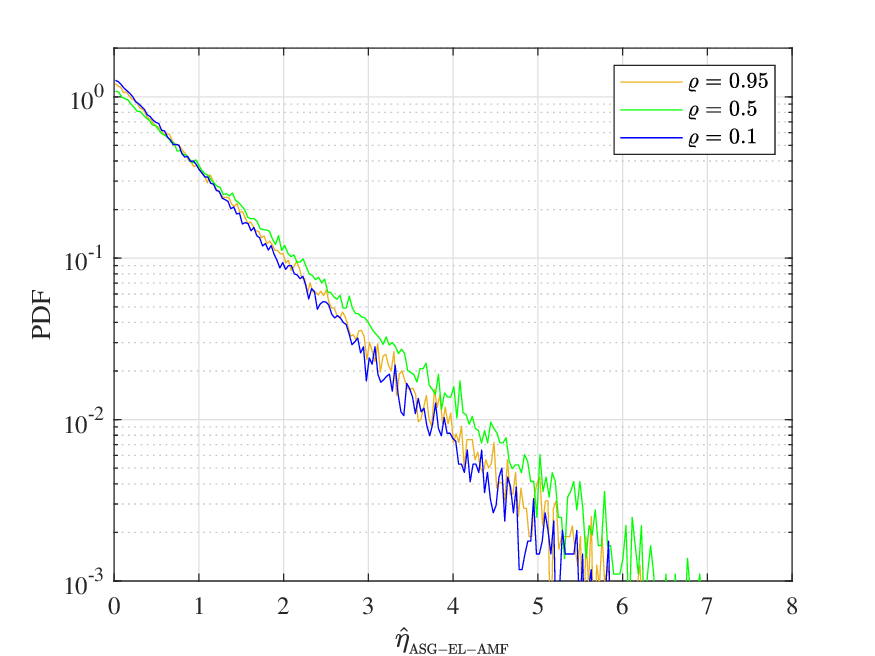}
	}\hspace{2em}
	\subfloat[CFAR-EL-AMF]{
		\centering
		\includegraphics[trim=0.5cm 0.1cm 0 1cm, scale=0.5]{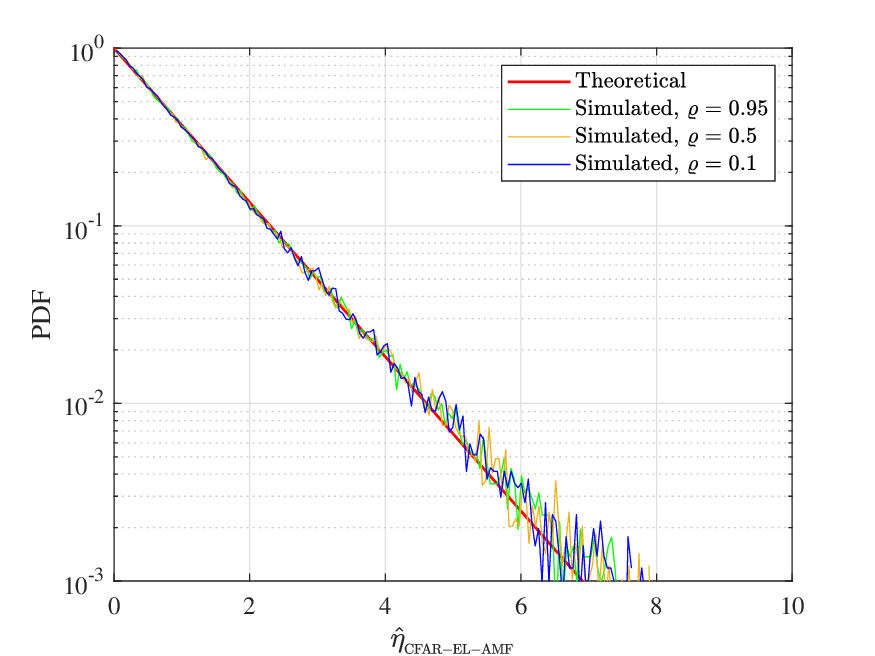}
	}\\
	\subfloat[ASG's EL-AMF]{
		\centering
		\includegraphics[trim=0.5cm 0.1cm 0 0.8cm, scale=0.5]{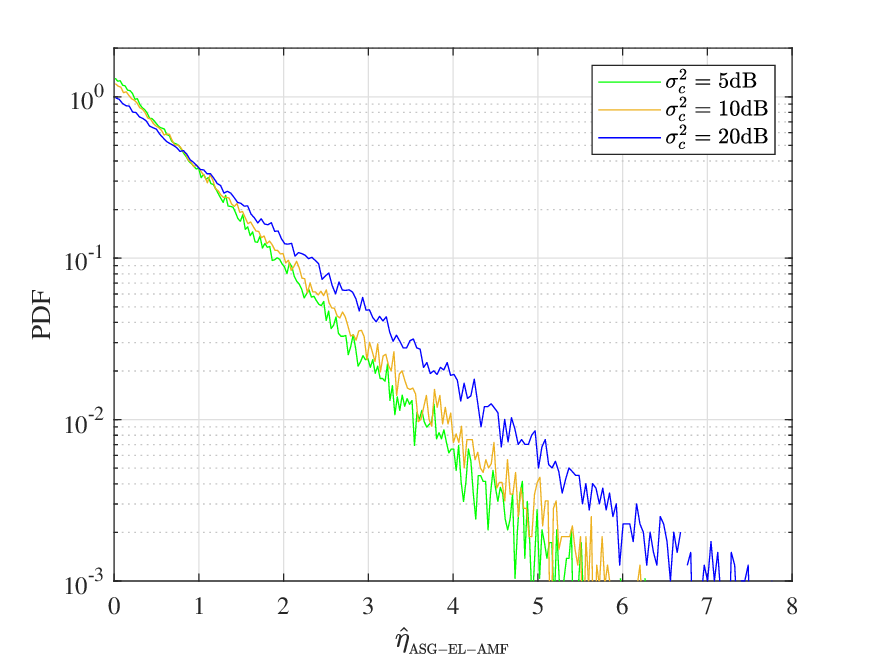}
	}\hspace{2em}
	\subfloat[CFAR-EL-AMF]{
		\centering
		\includegraphics[trim=0.5cm 0.1cm 0 0.8cm, scale=0.5]{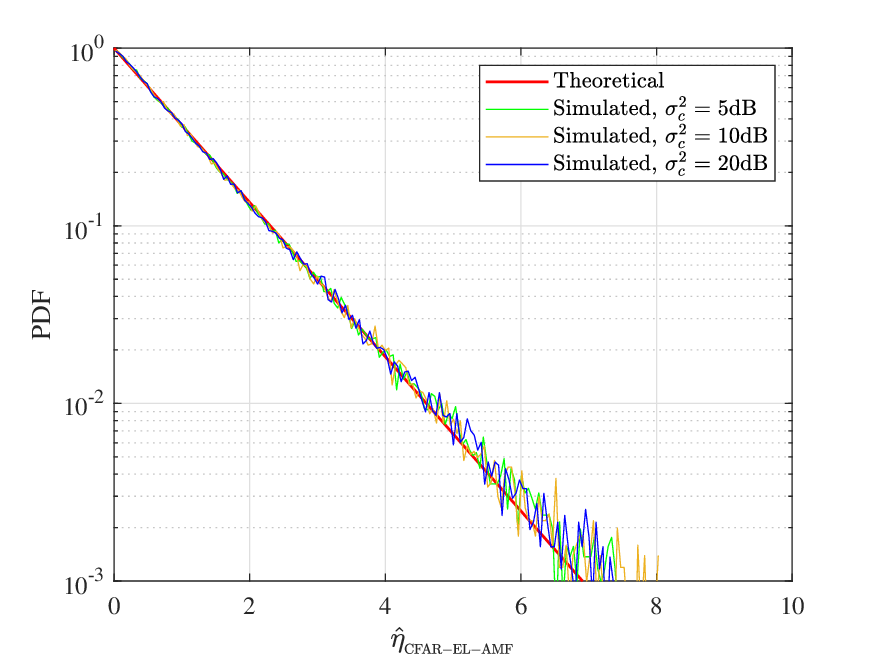}
	}
	\caption{The PDFs of the test statistics of ASG's EL-AMF and CFAR-EL-AMF under $H_0$ at different configurations of $\mathbf{R}$. In (a) and (b), $\varrho$ varies for fixed $\sigma^2=1,\sigma^2_c=10$; In (c) and (d), $\sigma^2_c$ varies for fixed $\varrho=0.95, \sigma^2=1$. In all these simulations, $\theta_t=20^\circ$, $N=24,K=48$. The red lines in (b) and (d) denote the standard Exponential distribution.}
	\label{The PDFs of the test statistics of ASG's EL-AMF and CFAR-EL-AMF under H_0}
\end{figure*}
\begin{figure}[t]
	\centering
	\includegraphics[trim=1cm 0.1cm 0 1cm, scale=0.5]{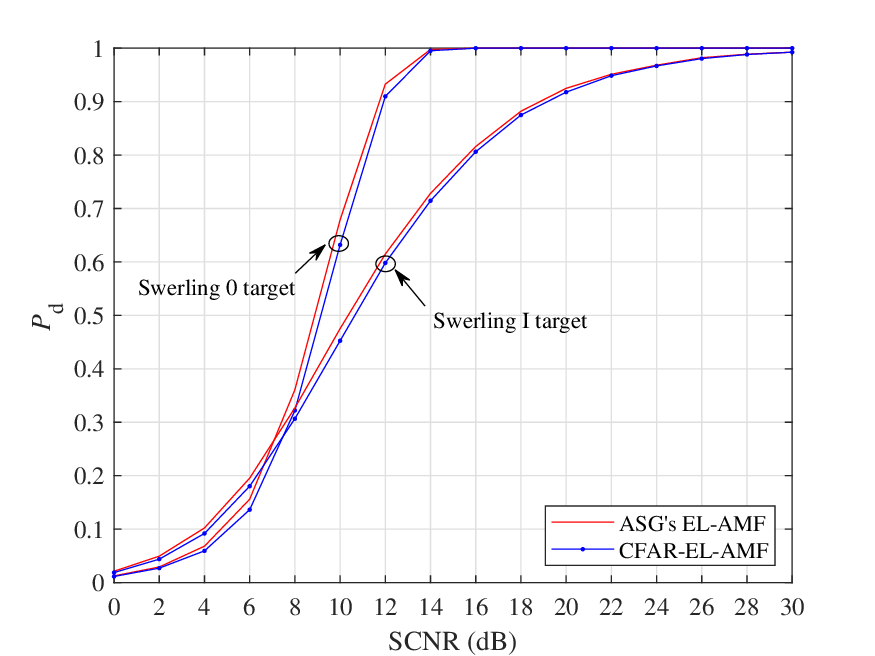}
	\caption{Detection performance comparison between ASG's EL-AMF and CFAR-EL-AMF. The parameter configuration is identical to that in Fig.\ref{Comparisons of the theoretical ROCs presented in 27 and 34 and the simulated results obtained from Monte Carlo trials.}(b).}
	\label{Performance comparison between ASG's EL-AMF and CFAR-EL-AMF.}
	\vspace{-1.5em}
\end{figure}

We now validate the CFAR property of CFAR-EL-AMF through simulations. Fig. \ref{The PDFs of the test statistics of ASG's EL-AMF and CFAR-EL-AMF under H_0} plots the PDFs of the test statistics of ASG's EL-AMF and CFAR-EL-AMF under $H_0$ with varying $\mathbf{R}$. The simulation setup follows these configurations: in Fig. \ref{The PDFs of the test statistics of ASG's EL-AMF and CFAR-EL-AMF under H_0}(a) and (b), we vary $\varrho$ while fixing $\sigma^2_c=10,\sigma^2=1$; in Fig.\ref{The PDFs of the test statistics of ASG's EL-AMF and CFAR-EL-AMF under H_0}(c) and (d), we vary $\sigma^2_c$ while maintaining $\varrho=0.95,\sigma^2=1$ unchanged.  The red curves in Fig. \ref{The PDFs of the test statistics of ASG's EL-AMF and CFAR-EL-AMF under H_0}(b) and (d) represent the standard Exponential distribution (see Eq. \eqref{58}), while all other empirical PDFs are obtained from $10^5$ Monte Carlo trials. As shown in Fig. \ref{The PDFs of the test statistics of ASG's EL-AMF and CFAR-EL-AMF under H_0}, the PDFs of the test statistics of ASG's EL-AMF vary with $\varrho$ and $\sigma^2_c$, whereas those of CFAR-EL-AMF remain invariant under different $\varrho$ and $\sigma^2_c$. This confirms that CFAR-EL-AMF achieves CFAR with respect to $\mathbf{R}$. Additionally, as we expect, the standard Exponential distribution (red lines) provides an excellent prediction to the empirical PDFs of the test statistic of CFAR-EL-AMF for different $\mathbf{R}$. In fact, the test statistic of CFAR-EL-AMF for different steering vectors $\mathbf{s}$ also follows the standard Exponential distribution (we do not present here for brevity). These results fully support our theoretical analysis.

Next, we present numerical examples to compare the detection performance of ASG's EL-AMF and CFAR-EL-AMF. The results are shown in Fig. \ref{Performance comparison between ASG's EL-AMF and CFAR-EL-AMF.}, in which the simulation parameters are the same as those in Fig. \ref{Comparisons of the theoretical ROCs presented in 27 and 34 and the simulated results obtained from Monte Carlo trials.}(b).  We observe that both detectors exhibit nearly identical detection performance, with the CFAR version exhibiting only a marginal detection loss of  approximately 0.3 dB (a practically negligible loss) relative to the original ASG's EL-AMF (for both Swerling 0 and I targets). This slight detection loss is the inevitable cost of achieving CFAR characteristic. In fact, as both $N$ and $K$ increases, the loss incurred by CFAR property will decrease. Consequently, for large $N$ and $K$, the CFAR-EL-AMF will exhibit comparable detection performance to ASG's EL-AMF. 
\section{Design of Optimal CFAR DL Semi-Clairvoyant and Adaptive Matched Filter Detectors}
The CFAR properties of CFAR-DL-SCMF and CFAR-EL-AMF with respect to $\lambda$ enable us to establish an optimal selection criterion for the loading factor $\lambda$ that achieves two critical objectives simultaneously: (1) maximizing detection probabilities while (2) rigorously maintaining CFAR characteristics with respect to both $\mathbf{R}$ and $\mathbf{s}$. This crucial decoupling allows $\lambda$ to be optimized purely for detection enhancement without compromising the false alarm regulation. 

In Section IV-A, we will develop the theoretical framework for determining optimal $\lambda$ in CFAR-DL-SCMF, yielding the opt-CFAR-DL-SCMF that maximizes the detection probability. However, the opt-CFAR-DL-SCMF depends on the unknown covariance matrix $\mathbf{R}$ and is therefore unimplementable in practice. To solve this problem, Section IV-B develops a practical approach for estimating the optimal $\lambda$ directly from observations. This approach yields the opt-CFAR-DL-AMF detector that achieves asymptotically optimal detection performance under LDR while preserving original CFAR property.
\subsection{Design of opt-CFAR-DL-SCMF Detector}
Based on Theorems \ref{Theorem Asymptotic distribution of eta CFAR DL under H0} and \ref{Theorem Asymptotic distribution of eta CFAR DL under H1}, we can express the ROCs of CFAR-DL-SCMF for Swerling 0 and I targets by \eqref{37} and \eqref{38}, respectively. Hence, we can determine the optimal $\lambda$, denoted by $\lambda_{\rm opt}$, by maximizing the asymptotic detection probabilities:
\begin{equation}\label{73}
	\lambda_{\rm opt} = \arg\max_{\lambda}	\widetilde{P}_{\rm d,DL}^{Swer0}(\lambda) 
\end{equation} for Swerling 0 target,
and
\begin{equation}\label{74}
	\lambda_{\rm opt} = \arg\max_{\lambda}	\widetilde{P}_{\rm d,DL}^{Swer1}(\lambda) 
\end{equation} for Swerling I target.   Recall that both $\widetilde{P}_{\rm d,DL}^{Swer0}(\lambda) $ and $\widetilde{P}_{\rm d,DL}^{Swer1}(\lambda)$ (see \eqref{37} and \eqref{38}, respectively) are strictly increasing functions of $\kappa(\lambda)$. Hence, $\lambda_{\rm opt}$ for both Swerling 0 and I targets can be determined by maximizing $\kappa(\lambda)$. Reviewing $\kappa(\lambda)$ given in \eqref{35}, we can see that the term $\mathbf{s}^H\mathbf{R}^{-1}\mathbf{s}$ in denominator is not related to $\lambda$. Hence, determining $\lambda_{\rm opt}$ reduces to the following optimization:
\begin{equation}\label{75}
	\lambda_{\rm opt} = \arg\max_{\lambda} \underline{\kappa} (\lambda)
\end{equation}
where
\begin{equation}\label{key}
\underline{\kappa} (\lambda)= \frac{(1-\gamma(\lambda))(\mathbf{u}^H\mathbf{E}(\lambda)\mathbf{u})^2}{\mathbf{u}^H\mathbf{E}(\lambda)^2\mathbf{u}}.
\end{equation}

Substituting $\lambda_{\rm opt}$ in \eqref{54} yields the opt-CFAR-DL-SCMF detector:
\begin{equation}\label{77}
	\hat{\eta}_{_{\rm opt\mbox{-} CFAR\mbox{-}DL\mbox{-}SCMF}}\mathop\gtrless\limits_{H_0}^{H_1}\tau
\end{equation}
where the test statistic $\hat{\eta}_{_{\rm opt\mbox{-} CFAR\mbox{-}DL\mbox{-}SCMF}}$ is given by
\begin{equation}\label{76-1}
	 \begin{aligned}
		\hat{\eta}_{_{\rm opt\mbox{-} CFAR\mbox{-}DL\mbox{-}SCMF}}=
	 \frac{1-\gamma(\lambda_{\rm opt})}{\mathbf{u}^H\mathbf{E}(\lambda_{\rm opt})^2\mathbf{u}}\left|\mathbf{s}^H\left(\hat{\mathbf{R}}+\lambda_{\rm opt}\mathbf{I}_N\right)^{-1}\mathbf{y}_0\right|^2.
	\end{aligned}
\end{equation}

We notice that the optimization problems in \eqref{73} and \eqref{74} strictly adhere to the NP criterion, i.e., maximizing $P_{\rm d}$ while maintaining a prescribed $P_{\rm fa}$. Consequently, the resulting detector in \eqref{77} achieves optimality in the sense of NP criterion. Of course, this optimality is established within the LDR framework.

Furthermore, we can see that the optimal loading factor $\lambda_{\rm  opt}$ coincides exactly with $\lambda_0$ introduced in Section II-C. This equivalence means that the opt-CFAR-DL-SCMF exhibits identical performance to DL-AMF with $\lambda_0$ that was analyzed previously. Hence we omit redundant performance analysis here. 

We now formally present the asymptotic performance of opt-CFAR-DL-SCMF under LDR in the following theorem (proof omitted for brevity).
\begin{mytheorem}\label{Theorem performance of optimal CFAR-DL-SCMF}
Let assumptions (A1)-(A4) hold true. As $N,K\to\infty$, the asymptotic $P_{\rm fa}$ of opt-CFAR-DL-SCMF is characterized by $\eqref{57}$. For a given preassigned $P_{\rm fa}$, the ROCs of opt-CFAR-DL-SCMF under LDR are given by 
\begin{equation}\label{77-1}
		\small \widetilde{P}_{\rm d,DL}^{Swer0}(\lambda_{\rm opt}) =	Q\left(\sqrt{2S_0\kappa(\lambda_{\rm opt})},\sqrt{-2\log {P}_{\rm fa,pre}} \right),
	\end{equation}
	for Swerling 0 target, and 
	\begin{equation}\label{78}
		\widetilde{P}_{\rm d,DL}^{Swer1}(\lambda_{\rm opt})= \left({P}_{\rm fa,pre}\right)^{\frac{1}{1+S_1\kappa(\lambda_{\rm opt})}}
	\end{equation}
	for Swerling I target.
\end{mytheorem}
\subsection{Design of opt-CFAR-DL-AMF Detector}
We now estimate $\lambda_{\rm opt}$ from finite observations. The key challenge lies in deriving a consistent estimator for $\underline{\kappa} (\lambda)$ under LDR. The following theorem establishes such an estimator.
\begin{mytheorem}\label{Theorem Consistent estimator of kappa}
	Let assumptions (A1)-(A3) hold true. Then as $N,K\to\infty$, it holds that
	\begin{equation}\label{key}
	\widehat{\underline{\kappa} (\lambda)}\overset{a.s.}{\longrightarrow}\underline{\kappa} (\lambda)
	\end{equation}
where
\begin{equation}\label{key}
\small\widehat{\underline{\kappa} (\lambda)}= \frac{\left(1-\frac{N}{K}+\frac{\lambda}{K}{\rm tr}(\hat{\mathbf{R}}+\lambda\mathbf{I}_N)^{-1}\right)^2\left(\mathbf{s}^H\left(\hat{\mathbf{R}}+\lambda\mathbf{I}_N\right)^{-1}\mathbf{s}\right)^2}{\mathbf{s}^H\left(\hat{\mathbf{R}}+\lambda\mathbf{I}_N\right)^{-1}\hat{\mathbf{R}}\left(\hat{\mathbf{R}}+\lambda\mathbf{I}_N\right)^{-1}\mathbf{s}}.
\end{equation}
\end{mytheorem}
\noindent\textit{Proof:} See Appendix E.\qed

By using $\widehat{\kappa}(\lambda)$, we can estimate $\lambda_{\rm opt}$ by 
\begin{equation}\label{key}
	\hat{\lambda}_{\rm opt} =\arg\max_{\lambda}\widehat{\underline{\kappa}(\lambda)}.
\end{equation}

Replacing $\lambda$ with $\hat{\lambda}_{\rm opt}$ in \eqref{66}, we obtain the opt-CFAR-DL-AMF, given by
\begin{equation}\label{79-1}
	\hat{\eta}_{_{\rm opt\mbox{-} CFAR\mbox{-}DL\mbox{-}AMF}}\mathop\gtrless\limits_{H_0}^{H_1}\tau
\end{equation} 
where
\begin{equation}\label{81}
\small	\begin{aligned}
	&\hat{\eta}_{_{\rm opt\mbox{-} CFAR\mbox{-}DL\mbox{-}AMF}} =
	\\& \frac{\left[1-\frac{N}{K}+\frac{\hat{\lambda}_{\rm opt}}{K}{\rm tr}\left(\hat{\mathbf{R}}+\hat{\lambda}_{\rm opt}\mathbf{I}_N\right)^{-1}\right]^2\left|\mathbf{s}^H\left(\hat{\mathbf{R}}+\hat{\lambda}_{\rm opt}\mathbf{I}_N\right)^{-1}\mathbf{y}_0\right|^2}{\mathbf{s}^H\left(\hat{\mathbf{R}}+\hat{\lambda}_{\rm opt}\mathbf{I}_N\right)^{-1}\hat{\mathbf{R}}\left(\hat{\mathbf{R}}+\hat{\lambda}_{\rm opt}\mathbf{I}_N\right)^{-1}\mathbf{s}}.
\end{aligned}
\end{equation}

Thanks to the  CFAR property of original CFAR-DL-AMF with respect to $\lambda$, the test statistic of opt-CFAR-DL-AMF remains invariant to $\hat{\lambda}_{\rm opt}$. Hence, $\hat{\eta}_{_{\rm opt\mbox{-} CFAR\mbox{-}DL\mbox{-}AMF}}$ (under $H_0$) still asymptotically follows a standard Exponential distribution under LDR. This indicates that the opt-CFAR-DL-AMF maintains CFAR property with respect to $\mathbf{R}$ and $\mathbf{s}$. Furthermore, due to the consistency of $\widehat{\underline{\kappa}(\lambda)}$, the opt-CFAR-DL-AMF will exhibit equivalent detection performance to opt-CFAR-DL-SCMF under LDR. 

The following theorem formally characterizes the asymptotic performance of the opt-CFAR-DL-SCMF under LDR (proof omitted for brevity).
\begin{mytheorem}
	Let assumptions (A1)-(A4) hold true. As $N,K\to\infty$, the performance (including false alarm and detection performance) of opt-CFAR-DL-AMF converges to those of opt-CFAR-DL-SCMF. This means that the asymptotic $P_{\rm fa}$ of opt-CFAR-DL-AMF can also be characterized by $\eqref{57}$. For a given preassigned $P_{\rm fa}$, the ROCs of opt-CFAR-DL-AMF under LDR are given by $\widetilde{P}_{\rm d,DL}^{Swer0}(\lambda_{\rm opt})$ for Swerling 0 target (see \eqref{77-1}) and $\widetilde{P}_{\rm d,DL}^{Swer1}(\lambda_{\rm opt}) $ for Swerling I target (see \eqref{78}), respectively.
\end{mytheorem}
\begin{figure*}[t]
	\centering
	\subfloat[opt-CFAR-DL-SCMF]{
		\centering
		\includegraphics[trim=0.5cm 0.1cm 0 1cm, scale=0.5]{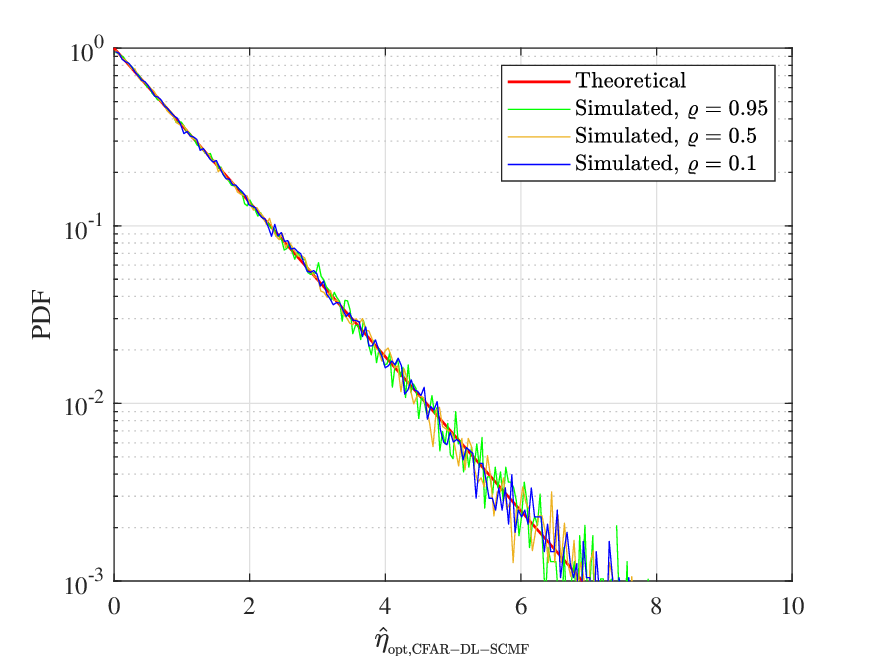}
	}\hspace{2em}
	\subfloat[opt-CFAR-DL-AMF]{
		\centering
		\includegraphics[trim=0.5cm 0.1cm 0 1cm, scale=0.5]{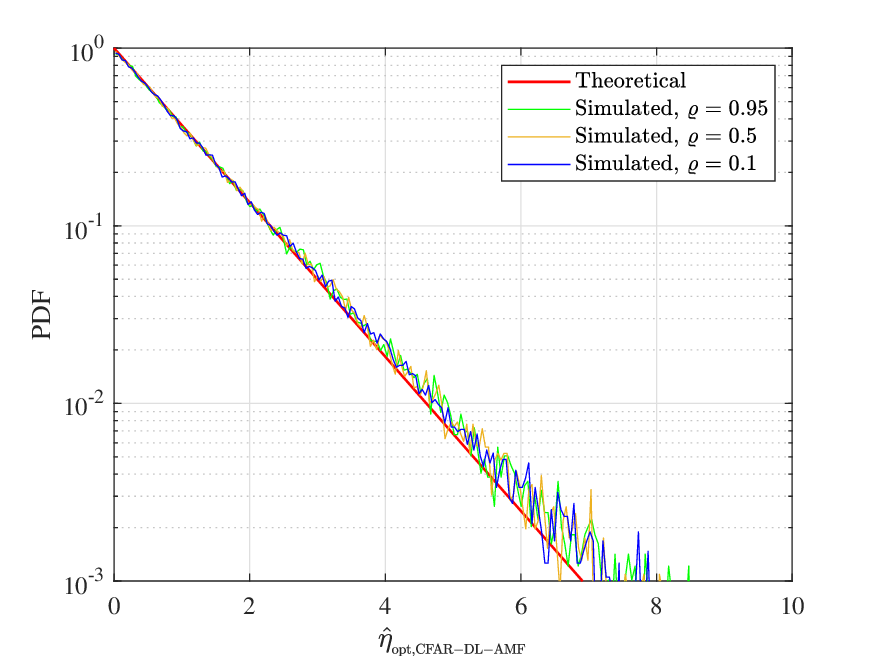}
	}
	\caption{Comparison between the theoretical PDFs of the test statistics of opt-CFAR-DL-SCMF and opt-CFAR-DL-AMF and their empirical PDFs obtained from $10^5$ Monte Carlo trials under different $\mathbf{R}$ (varying $\varrho$ while fixing $\sigma^2=1,\sigma^2_c=10$) and fixed $\mathbf{s}(\theta_t=20^\circ)$.}
	\label{Comparison between the theoretical PDFs of the test statistics of optimal CFAR-DL-SCMF and optimal CFAR-DL-AMF and their empirical PDFs obtained from Monte Carlo trials under different R.}
	\vspace{-0.5em}
\end{figure*}
\begin{figure}[t]
	\centering
	\includegraphics[trim=1cm 0.1cm 0 1cm, scale=0.5]{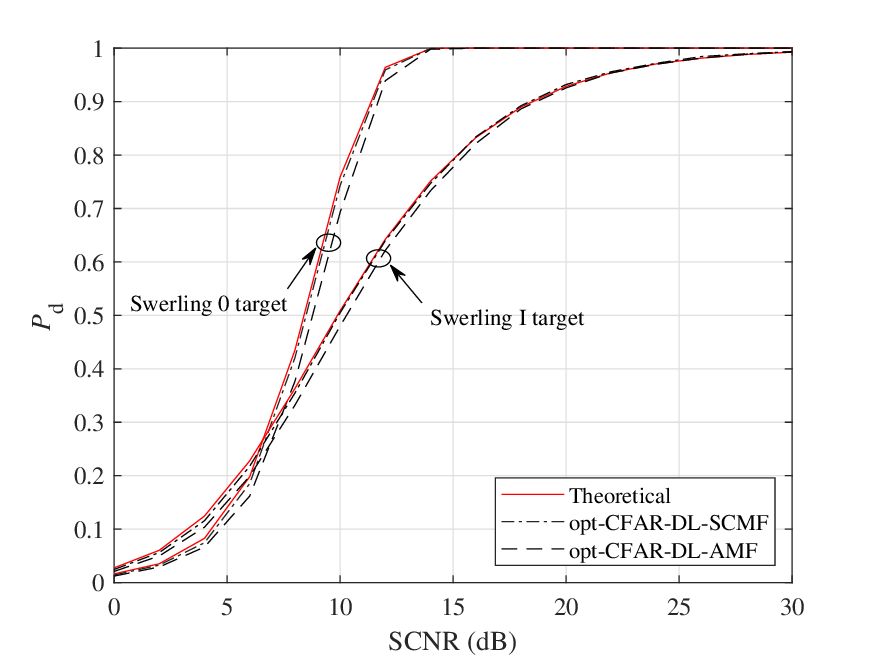}
	\caption{Detection performance comparison between the opt-CFAR-DL-SCMF and opt-CFAR-DL-AMF. The simulation parameters are the same as those in Fig. \ref{Comparisons of the theoretical ROCs presented in 27 and 34 and the simulated results obtained from Monte Carlo trials.}(b).}
	\label{Detection performance comparison between the optimal CFAR-DL-SCMF and optimal CFAR-DL-AMF.}
\end{figure}

Next, we conduct Monte Carlo simulations to verify the aforementioned theoretical results. 

We first validate the CFAR properties of opt-CFAR-DL-SCMF and opt-CFAR-DL-AMF with respect to $\mathbf{R}$. Fig. \ref{Comparison between the theoretical PDFs of the test statistics of optimal CFAR-DL-SCMF and optimal CFAR-DL-AMF and their empirical PDFs obtained from Monte Carlo trials under different R.}(a) and (b) respectively show the PDFs of the test statistics of opt-CFAR-DL-SCMF and opt-CFAR-DL-AMF under different $\mathbf{R}$. We set $N=24$ and $K=48$ in these simulations. It can be seen that the distributions of both test statistics are invariant to $\mathbf{R}$, indicating that both detectors achieve CFAR with respect to $\mathbf{R}$. We also observe that the distributions of both test statistics are predicted well by the standard Exponential distribution (red lines in Fig. \ref{Comparison between the theoretical PDFs of the test statistics of optimal CFAR-DL-SCMF and optimal CFAR-DL-AMF and their empirical PDFs obtained from Monte Carlo trials under different R.}). This further validates our theoretical analysis. If we change $\mathbf{s}$ while fixing $\mathbf{R}$, we will also observe that the distributions of both test statistics are invariant and align perfectly with the standard Exponential distribution, implying that both detectors achieve CFAR with respect to $\mathbf{s}$. For the sake of brevity, we do not present the experimental results of CFAR characteristics with respect to $\mathbf{s}$ here.

We further compare the detection performance between the opt-CFAR-DL-SCMF and the opt-CFAR-DL-AMF. The results are shown in Fig. \ref{Detection performance comparison between the optimal CFAR-DL-SCMF and optimal CFAR-DL-AMF.}. For reference, we also plot the theoretical $\widetilde{P}_{\rm d,DL}^{Swer0}(\lambda_{\rm opt}) $ and $\widetilde{P}_{\rm d,DL}^{Swer1}(\lambda_{\rm opt})$ (red lines). We see that the theoretical curves show excellent agreement with the opt-CFAR-DL-SCMF. We also observe that the opt-CFAR-DL-AMF suffers a slight detection loss compared to the opt-CFAR-DL-SCMF. This minor loss comes from the estimation error in $\hat{\lambda}_{\rm opt}$ due to finite $N$ and $K$. As both $N$ and $K$ increase, this loss is expected to diminish. Ultimately, for sufficiently large $N$ and $K$, the opt-CFAR-DL-AMF will achieve performance equivalent to that of the opt-CFAR-DL-SCMF.

\section{Performance Comparisons}
In this section, we conduct a comparative study to validate the superiority of the proposed opt-CFAR-DL-SCMF and opt-CFAR-DL-AMF detectors. The following competitors are included for comparison:
\begin{itemize}
	\item The NP detector as an optimal benchmark,
	\item 
	The CFAR version of ASG's EL-AMF, i.e., CFAR-EL-AMF (see \eqref{71-1}),
	\item The persymmetric AMF \cite{Pailloux2011Persymmetric}, recognized for its strong performance in sample-deficient scenarios, and
	\item RFKN’s AMF \cite{Robey1992A} (referred to as SCM-based AMF in our experiments) for additional baseline comparison.
\end{itemize}
Notably, all these detectors except NP detector maintain CFAR property with respect to $\mathbf{R}$. We evaluate these detectors under two distinct scenarios: full-rank clutter plus noise and low-rank clutter plus noise.
\subsection{Comparison in Full-Rank Clutter Plus Noise}
The full-rank clutter typically originates from extended sources comprising multiple distributed scatters with diverse scattering properties. 
A canonical example is sea clutter, whose covariance matrix $\mathbf{R}_c$ is usually modeled as a Toeplitz structure with the $(i,j)$th element given by $(\mathbf{R}_c)_{ij} = \sigma^2_c\varrho^{|i-j|}$. Here, $\sigma^2_c$ is the clutter power and $\varrho$ is the correlation coefficient (typically $\varrho\in[0.9,0.99]$ for sea clutter) \cite{Gini2002Covariance}. Hence, in this subsection, we model the clutter-plus-noise covariance matrix by 
\begin{equation}\label{key}
	\mathbf{R}=\mathbf{R}_c+\sigma^2\mathbf{I}_N
\end{equation}
where $\sigma^2$ represents the noise power.
\begin{figure*}[t]
	\centering
	\subfloat[Swerling 0 target]{
		\centering
		\includegraphics[trim=0.5cm 0.1cm 0 1cm, scale=0.5]{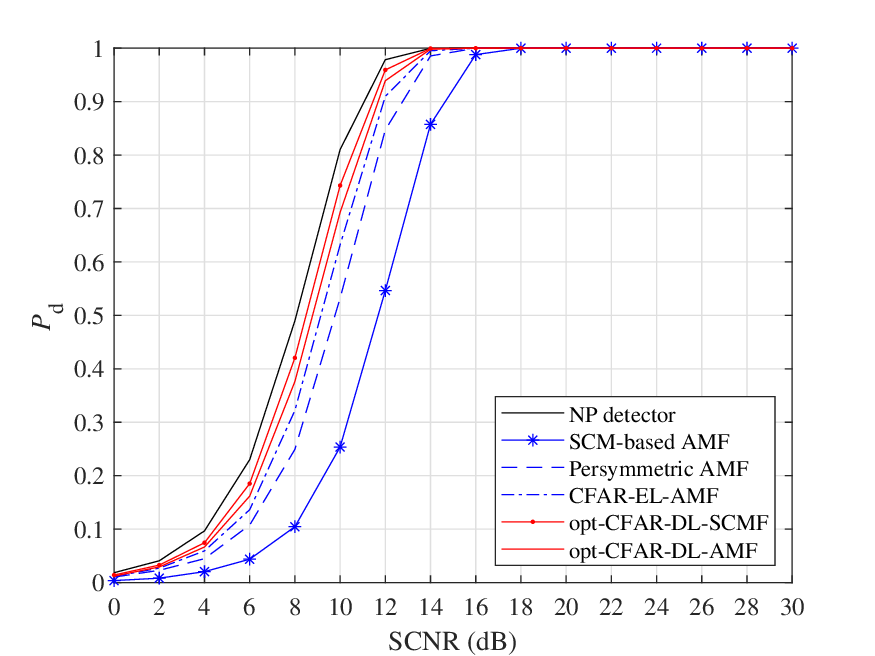}
	}\hspace{2em}
	\subfloat[Swerling I target]{
		\centering
		\includegraphics[trim=0.5cm 0.1cm 0 1cm, scale=0.5]{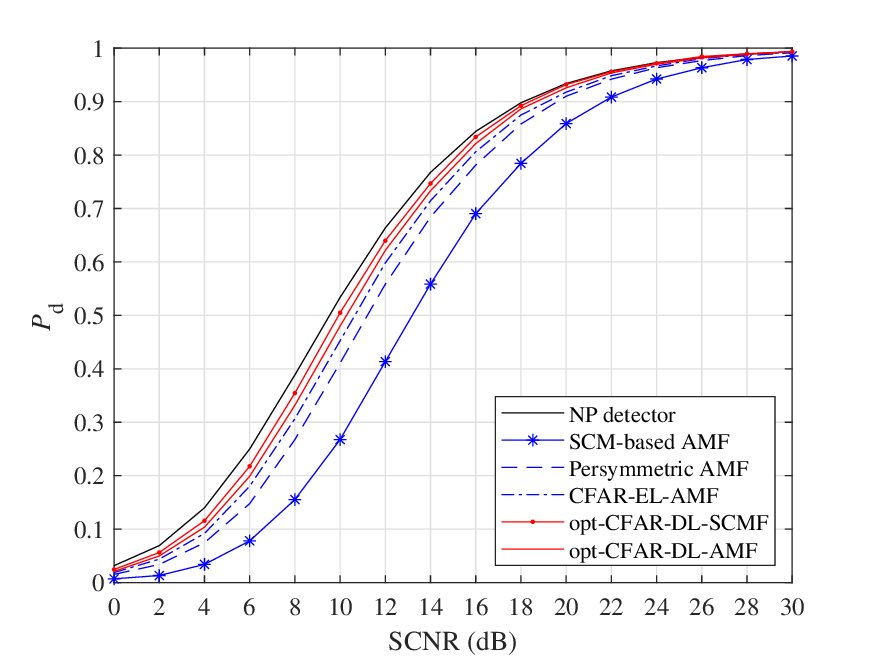}
	}
	\caption{Detection performance comparisons between proposed detectors and other detectors in full-rank clutter plus noise when $N=24,K=48$.}
	\label{Detection performance comparisons between proposed detectors and other detectors in full-rank clutter-plus-noise when N=24,K=48.}
\end{figure*}
\begin{figure*}[t]
	\centering
	\subfloat[Swerling 0 target]{
		\centering
		\includegraphics[trim=0.5cm 0.1cm 0 1cm, scale=0.5]{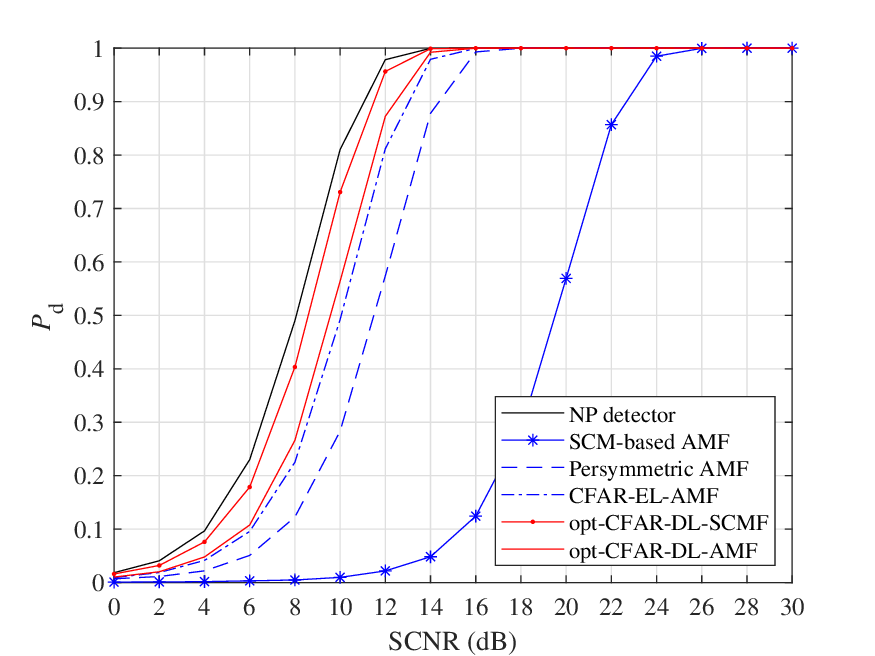}
	}\hspace{2em}
	\subfloat[Swerling I target]{
		\centering
		\includegraphics[trim=0.5cm 0.1cm 0 1cm, scale=0.5]{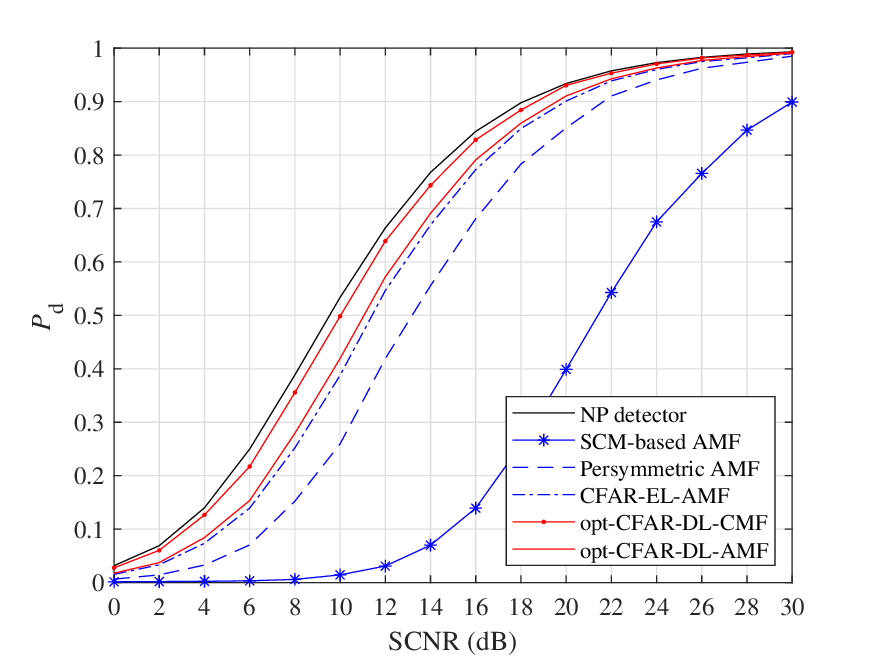}
	}
	\caption{Detection performance comparisons between proposed detectors and other detectors in full-rank clutter plus noise when $N=24,K=28$.}
	\label{Detection performance comparisons between proposed detectors and other detectors in full-rank clutter-plus-noise when N=24, K=28.}
	\vspace{-1em}
\end{figure*}

Fig. \ref{Detection performance comparisons between proposed detectors and other detectors in full-rank clutter-plus-noise when N=24,K=48.} evaluates the detection performance of the proposed detectors and other detectors in full-rank clutter plus noise background. The experimental parameters are set as follows: $N=24,K=48$ (yielding $c=N/K=0.5$), $\varrho=0.95, \sigma^2=1, \sigma^2_c=10$ and $\theta_t=20^\circ$. Under these parameters, Eq. \eqref{75} yields an optimal loading factor $\lambda_{\rm opt} = 11.92$ for opt-CFAR-DL-SCMF. The preassigned $P_{\rm fa}$ is set to $10^{-3}$ and hence the thresholds are determined using $10^5$ Monte Carlo trials, while the detection probabilities are obtained from $10^4$ Monte Carlo trials. Furthermore, to quantitatively assess detectors' performance, we list the detection losses (in dB) of all CFAR detectors relative to the NP detector at $P_{\rm d}=0.5$ in Table \ref{Table Detection losses (dB)  of CFAR detectors relative to NP detector in full-rank clutter plus noise background.} .

 Table \ref{Table Detection losses (dB)  of CFAR detectors relative to NP detector in full-rank clutter plus noise background.} reveals an interesting phenomenon: each detector shows identical detection losses (relative to NP detector) for Swerling 0 and Swerling I targets. This observation aligns perfectly with our theoretical predictions. We further observe from Fig. \ref{Detection performance comparisons between proposed detectors and other detectors in full-rank clutter-plus-noise when N=24,K=48.} that as expected, the SCM-based AMF performs the worst, exhibiting a substantial detection loss of about 3.6 dB compared to the NP detector. In contrast, the other CFAR detectors show significant performance improvement compared to SCM-based AMF. Particularly, the opt-CFAR-DL-SCMF achieves the best performance among all CFAR detectors, with only a slight detection loss of about 0.4 dB relative to the NP detector. It should be noted that this loss is inherent to DL detectors, because the performance of DL detectors is always inferior to that of the NP detector, as we have analyzed in Remark \ref{Remark 5}. Based on our theoretical results in Theorem \ref{Theorem performance of optimal CFAR-DL-SCMF}, for large $N$ and $K$, the ROCs of the opt-CFAR-DL-SCMF for Swerling 0 and I targets are characterized by $\widetilde{P}_{\rm d,DL}^{Swer0}(\lambda_{\rm opt}) $ and $\widetilde{P}_{\rm d,DL}^{Swer1}(\lambda_{\rm opt})$, respectively. Hence, from \eqref{49}, the theoretical detection loss of the opt-CFAR-DL-SCMF relative to the NP detector is $\Delta_{\rm NP-DL}(\lambda_{\rm opt}=11.92)=0.3543\cdots {\rm dB} \approx 0.4$ dB. Although this loss is derived under the asymptotic assumption where $N$ and $K$ approach infinity, we observe that for finite values of $N=24$ and $K=48$ in our simulations, the theoretical loss still provides a reasonably accurate prediction of the actual loss.  We should keep in mind that the opt-CFAR-DL-SCMF is actually non-realizable in practice due to its reliance on unknown $\mathbf{R}$. Instead, its adaptive version, i.e., the opt-CFAR-DL-AMF, is more practically relevant. As shown in Fig. \ref{Detection performance comparisons between proposed detectors and other detectors in full-rank clutter-plus-noise when N=24,K=48.}, the opt-CFAR-DL-AMF outperforms all other CFAR adaptive detectors (note that the opt-CFAR-DL-SCMF is not adaptive). This fact validates the superiority of the proposed opt-CFAR-DL-AMF and provides an evidence that both EL-based AMF (i.e., CFAR-EL-AMF) and persymmetric AMF are not optimal. From Table \ref{Table Detection losses (dB)  of CFAR detectors relative to NP detector in full-rank clutter plus noise background.}, it is easy to get that the CFAR-EL-AMF incurs an approximately 0.4 dB detection loss compared to the opt-CFAR-DL-AMF. The persymmetric AMF shows inferior performance, with a detection loss of about 0.9 dB relative to the proposed opt-CFAR-DL-AMF. 
 
We are more interested in the detection performance of the proposed detectors in insufficient sample scenarios. We reduce the sample size $K$ from 48 to 28, while maintaining all other parameters unchanged from those in Fig. \ref{Detection performance comparisons between proposed detectors and other detectors in full-rank clutter-plus-noise when N=24,K=48.}. In this case, it follows from \eqref{75} that the optimal loading factor for CFAR-DL-SCMF is given by $\lambda_{\rm opt} = 18.14$. Furthermore, according to \eqref{49}, we can predict that the theoretical loss of opt-CFAR-DL-SCMF is $\Delta_{\rm NP-DL}(\lambda_{\rm opt}=18.14)=0.4782\cdots{\rm dB} \approx 0.5 $ dB. The comparison results are shown in Fig. \ref{Detection performance comparisons between proposed detectors and other detectors in full-rank clutter-plus-noise when N=24, K=28.}. As we expect, the performance of SCM-based AMF deteriorate seriously, with a significant detection loss of 11.7 dB relative to NP detector (see Table \ref{Table Detection losses (dB)  of CFAR detectors relative to NP detector in full-rank clutter plus noise background.}). The performance of other detectors shows no significant degradation. Specifically, we see from Table \ref{Table Detection losses (dB)  of CFAR detectors relative to NP detector in full-rank clutter plus noise background.} that the opt-CFAR-DL-SCMF exhibits 0.5 dB detection loss relative to the NP detector, which matches the theoretical prediction.  Furthermore, the opt-CFAR-DL-SCMF remains the best one among all CFAR detectors and keeps nearly identical performance to the case $K=48$. This fact reflects the critical role of prior knowledge about $\mathbf{R}$ in achieving optimal detection performance. Instead, other adaptive detectors that do not utilize any prior knowledge, including persymmetric AMF, CFAR-EL-AMF and opt-CFAR-DL-AMF, show relatively larger detection loss compared to the case $K=48$. The proposed opt-CFAR-DL-AMF still outperforms all other adaptive detectors. Additionally, Table \ref{Table Detection losses (dB)  of CFAR detectors relative to NP detector in full-rank clutter plus noise background.} revals an interesting observation: when $K$ decreases from 48 to 28, the detection loss of CFAR-EL-AMF relative to the opt-CFAR-DL-AMF remains unchanged (approximately 0.4 dB). This phenomenon suggests that the loading factor $\hat{\lambda}_{\rm EL}$ derived by maximizing the expected likelihood criterion preserves an intrinsic relationship with the theoretically optimal loading factor $\hat{\lambda}_{\rm opt}$, and remarkably, this relationship appears to be invariant to the sample size $K$. Consequently, the detection loss of CFAR-EL-AMF relative to the opt-CFAR-DL-AMF remains constant across different sample size  $K$. As for persymmetric AMF, its detection loss relative to the opt-CFAR-DL-AMF increases from 0.9 dB to 2 dB as $K$ decreases from 48 to 28. This pronounced sensitivity to sample size $K$ makes the persymmetric AMF unsuitable for practical implementation under insufficient sample scenarios.
\begin{table}[t]
	\vspace{-1em}
	\flushleft
	\caption{Detection losses (in dB)  of CFAR detectors relative to NP detector at $P_{\rm d}=0.5$ in full-rank clutter plus noise background.}
	\label{Table Detection losses (dB)  of CFAR detectors relative to NP detector in full-rank clutter plus noise background.}
	\resizebox{0.48\textwidth}{!}{%
		\renewcommand{\arraystretch}{1.25} 
		\begin{tabular}{ccccc}
			\toprule[1.5pt]
			\multicolumn{1}{c}{\multirow{2}{*}{Detectors}} &
			\multicolumn{2}{c}{$N=24, K=48$} &
			\multicolumn{2}{c}{$N=24, K=28$} \\ \cline{2-5} 
			\multicolumn{1}{c}{} &
			\multicolumn{1}{c}{Swerling 0} &
			\multicolumn{1}{c}{Swerling I} &
			\multicolumn{1}{c}{Swerling 0} &
			\multicolumn{1}{c}{Swerling I} \\ \hline
			SCM-based AMF       & 3.6 & 3.6 & 11.7 & 11.7 \\
			Persymmetric AMF     & 1.7 & 1.7 & 3.5  & 3.5  \\
			CFAR-EL-AMF      & 1.2 & 1.2 & 1.9  & 1.9  \\
			opt-CFAR-DL-SCMF & 0.4 & 0.4 & 0.5  & 0.5  \\ 
			opt-CFAR-DL-AMF & 0.8 & 0.8 & 1.5  & 1.5  \\ 
			 \bottomrule[1.5pt]
		\end{tabular}%
	}
	\begin{tablenotes}
		\item[*] {\scriptsize Note that all numerical results are rounded to one decimal place.}
	\end{tablenotes}
\vspace{-1em}
\end{table}

\subsection{ Comparison in Low-Rank Clutter Plus Noise}
\begin{table}[]
	\flushleft
	\vspace{-0.5em}
	\caption{Detection losses (in dB)  of CFAR detectors relative to NP detector at $P_{\rm d}=0.5$ in low-rank clutter plus noise background.}
	\label{Table Detection losses (dB)  of CFAR detectors relative to NP detector in low-rank clutter plus noise background.}
	\resizebox{0.48\textwidth}{!}{%
		\renewcommand{\arraystretch}{1.25} 
		\begin{tabular}{ccccc}
			\toprule[1.5pt]
			\multicolumn{1}{c}{\multirow{2}{*}{Detectors}} &
			\multicolumn{2}{c}{$N=24, K=48$} &
			\multicolumn{2}{c}{$N=24, K=28$} \\ \cline{2-5} 
			\multicolumn{1}{c}{} &
			\multicolumn{1}{c}{Swerling 0} &
			\multicolumn{1}{c}{Swerling I} &
			\multicolumn{1}{c}{Swerling 0} &
			\multicolumn{1}{c}{Swerling I} \\ \hline
			SCM-based AMF       & 3.7 & 3.7 & 11.6 & 11.6 \\
			Persymmetric AMF     & 1.6 & 1.6 & 3.5  & 3.5  \\
			CFAR-EL-AMF      & 1.2 & 1.2 & 1.8  & 1.8  \\
			opt-CFAR-DL-SCMF & 0.3 & 0.3 & 0.3  & 0.3  \\ 
			opt-CFAR-DL-AMF & 0.7 & 0.7 & 1.3  & 1.3  \\ 
			\bottomrule[1.5pt]
		\end{tabular}%
	}
	\begin{tablenotes}
		\item[*] {\scriptsize Note that all numerical results are rounded to one decimal place.}
	\end{tablenotes}
\vspace{-1em}
\end{table}

Low-rank clutter is widely encountered in practical radar systems due to the inherent correlation in environmental backscattering (e.g., ground/sea clutter exhibiting spatial-temporal coherence) and system-induced effects (e.g., array coupling or pulse repetition). Its covariance matrix often admits a low-rank structure, particularly in scenarios like airborne space time adaptive processing (where the clutter rank follows Brennan’s rule). The covariance matrix of low-rank clutter plus noise can be modeled by
\begin{equation}\label{key}
	\mathbf{R} =  \sum_{i=1}^r\sigma^2_{c_i}\mathbf{s}(\theta_{c_i})\mathbf{s}(\theta_{c_i})^H +\sigma^2\mathbf{I}_N
\end{equation}
where $\sigma^2$ is the noise power, $r$ represents the clutter rank, $\sigma^2_{c_i}$ denotes the power of $i$th clutter patch and $
	\mathbf{s}(\theta_{c_i}) = \frac{1}{\sqrt{N}}\left[1,\exp(\mathsf{i}\pi\sin(\theta_{c_i})), ..., \exp(\mathsf{i}\pi\sin(\theta_{c_i})(N-1))\right]^T$
for $i=1,2,...,r$.
\begin{figure*}[t]
	\centering
	\subfloat[Swerling 0 target]{
		\centering
		\includegraphics[trim=0.5cm 0.1cm 0 1cm, scale=0.5]{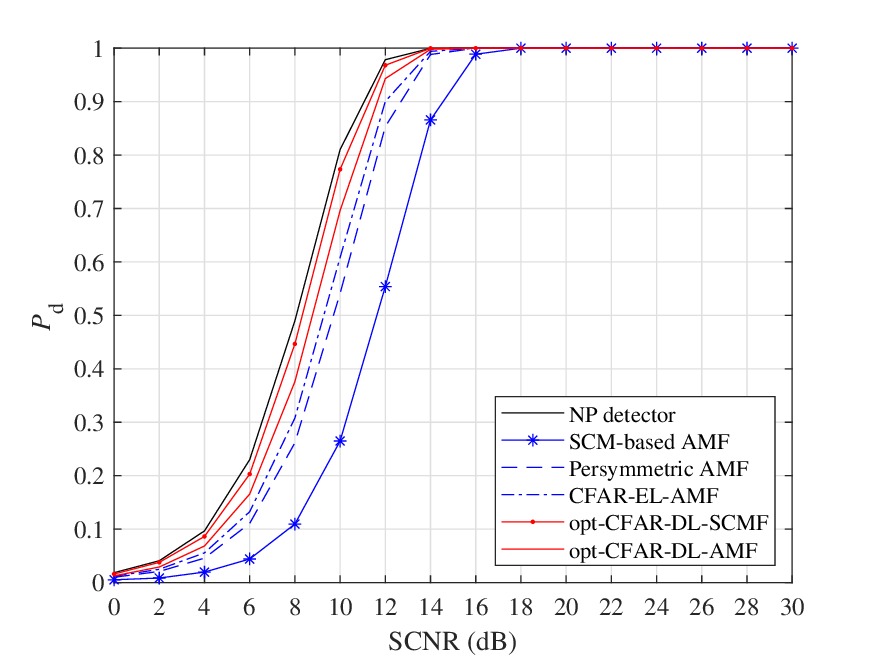}
	}\hspace{2em}
	\subfloat[Swerling I target]{
		\centering
		\includegraphics[trim=0.5cm 0.1cm 0 1cm, scale=0.5]{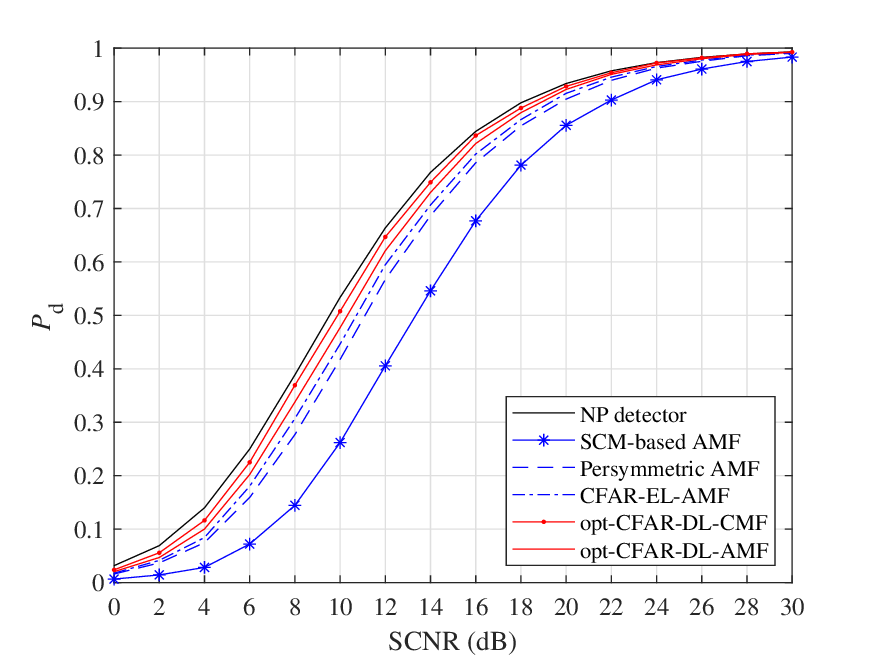}
	}
	\caption{Detection performance comparisons between proposed detectors and other detectors in low-rank clutter plus noise when $N=24,K=48$.}
	\label{Detection performance comparisons between proposed detectors and other detectors in low-rank clutter-plus-noise when N=24, K=48.}
\end{figure*}
\begin{figure*}[t]
	\centering
	\subfloat[Swerling 0 target]{
		\centering
		\includegraphics[trim=0.5cm 0.1cm 0 1cm, scale=0.5]{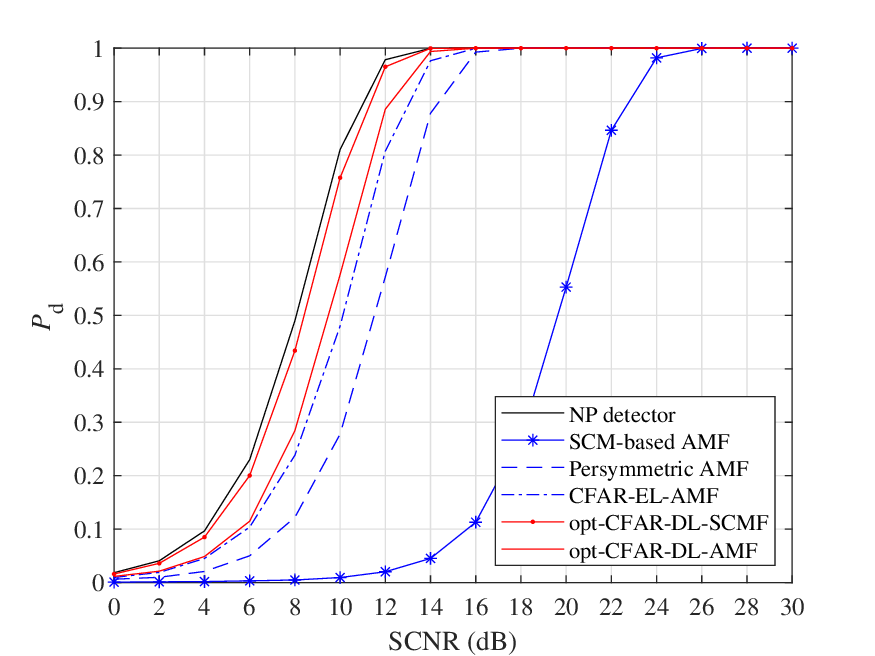}
	}\hspace{2em}
	\subfloat[Swerling I target]{
		\centering
		\includegraphics[trim=0.5cm 0.1cm 0 1cm, scale=0.5]{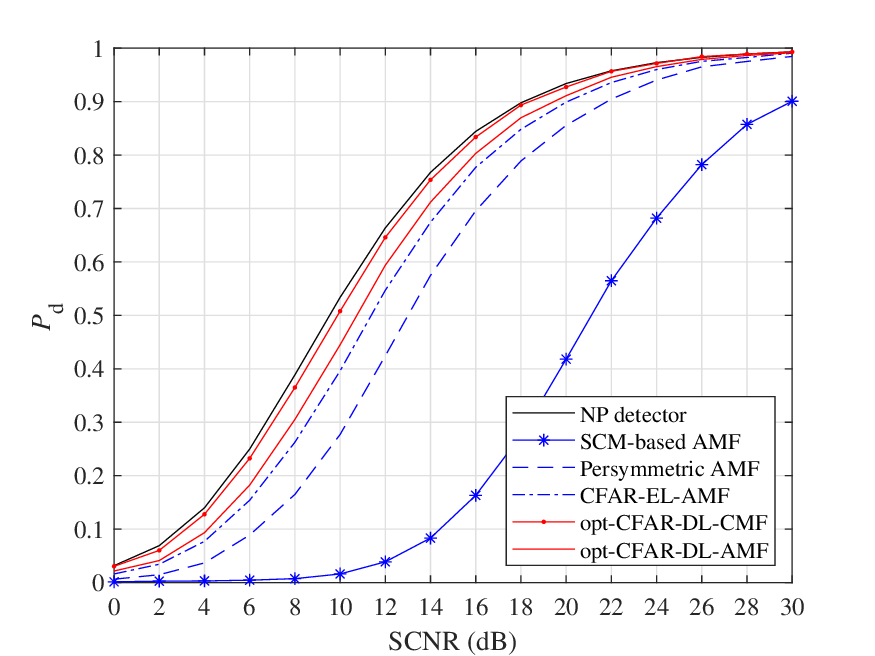}
	}
	\caption{Detection performance comparisons between proposed detectors and other detectors in low-rank clutter plus noise when $N=24,K=28$.}
	\label{Detection performance comparisons between proposed detectors and other detectors in low-rank clutter-plus-noise when N=24, K=28.}
	\vspace{-1em}
\end{figure*}

In the experiments conducted in this subsection, we assume the low-rank clutter originates from nine distinct sources ($r=9$), with their arrival angles specified as $[\theta_{c_1},\theta_{c_2},...,\theta_{c_9}] = [0^\circ, 5^\circ, 5^\circ, 10^\circ, 25^\circ, 25^\circ, 30^\circ, 30^\circ, 60^\circ]$. Each clutter patch has equal power  $\sigma^2_{c_i} = 10$ for $i=1,2,...,9$ and the noise power is $\sigma^2=1$.  Further, the direction arrival of target is $\theta_t=20^\circ$.

We first consider the case with sufficient samples. We set $N=24, K=48$. In this case, it follows from \eqref{75} that the optimal loading factor for CFAR-DL-SCMF is $\lambda_{\rm opt} \approx 18.75$ and the theoretical detection loss of opt-CFAR-DL-SCMF relative to the NP detector is calculated by $\Delta_{\rm NP-DL}(\lambda_{\rm opt}=18.75)=0.2521\cdots {\rm dB}\approx 0.3 $ dB (based on \eqref{49}). The comparison results are illustrated in Fig. \ref{Detection performance comparisons between proposed detectors and other detectors in low-rank clutter-plus-noise when N=24, K=48.}. Further, Table \ref{Table Detection losses (dB)  of CFAR detectors relative to NP detector in low-rank clutter plus noise background.} lists the detection losses of all CFAR detectors relative to NP detector at $P_{\rm d}=0.5$ under the low-rank clutter-plus-noise background. We can see that the results are very similar to those in the case of full-rank clutter. Each detector show identical detection losses for Swerling 0 and I targets. The SCM-based AMF has the worst performance, suffering a 3.7 dB detection loss compared to the NP detector. In contrast, the opt-CFAR-DL-SCMF performs best among all CFAR detectors, with only a 0.3 dB detection loss relative to the NP detector. This loss is in good agreement with the theoretical value. As expected, although the opt-CFAR-DL-AMF exhibits minor degradation compared to opt-CFAR-DL-SCMF due to the estimation error, it outperforms all other adaptive CFAR detectors, offering 0.5 dB and 0.9 dB detection gains over the CFAR-EL-AMF and persymmetric AMF, respectively.

We next examine detectors' performance under sample-deficient conditions by reducing the sample size from $K=48$ to $K=2$8, while keeping all other parameters consistent with those in Fig. \ref{Detection performance comparisons between proposed detectors and other detectors in low-rank clutter-plus-noise when N=24, K=48.}. The illustrative curves of $P_{\rm d}$ versus SCNR are plotted in Fig. \ref{Detection performance comparisons between proposed detectors and other detectors in low-rank clutter-plus-noise when N=24, K=28.} and the quantitative detection losses are summarized in Table \ref{Table Detection losses (dB)  of CFAR detectors relative to NP detector in low-rank clutter plus noise background.}. As expected, a severe performance degradation is immediately observed in the SCM-based AMF. The opt-CFAR-DL-SCMF still maintains the best performance, without obvious degradation compared to the case $K=48$. This result is analytically justified as the follows. In the case $K=28$ and under previous given parameters regarding $\mathbf{R}$ and $\mathbf{s}$, by using \eqref{75}, we readily obtain the optimal loading factor for CFAR-DL-SCMF, given by $\lambda_{\rm opt}\approx 31.51$. Based on \eqref{49}, the corresponding detection loss relative to the NP detector is $\Delta_{\rm NP-DL}(\lambda_{\rm opt}=31.51)= 0.3231\cdots {\rm dB}\approx 0.3 $ dB. We can see that when $K$ decreases from 48 to $28$, the opt-CFAR-DL-SCMF only exhibits approximately $0.071$ dB detection loss (calculated by $\Delta_{\rm NP-DL}(\lambda_{\rm opt}=31.51) -\Delta_{\rm NP-DL}(\lambda_{\rm opt}=18.75)$), which is practically negligible. Compared to the opt-CFAR-DL-SCMF, its adaptive version, namely opt-CFAR-DL-AMF,  exhibits certain performance decline but still outperforms other CFAR adaptive detectors. Specifically, quantitative analysis reveals that the CFAR-EL-AMF shows a 0.5 dB detection loss compared to the opt-CFAR-DL-AMF, which is the same as that in the case $K=48$.  This remarkable consistency provides an empirical evidence that the performance gap between CFAR-EL-AMF and the opt-CFAR-DL-AMF seems to be insensitive to the variations in the sample size $K$. Notice that this phenomenon is consistent with that previously observed in full-rank clutter. Finally, similar to the full-rank clutter scenario, the persymmetric AMF exhibits a significant increase in detection loss as $K$ decreases from 48 to 28 (rising from 0.9 dB to 2.2 dB relative to the opt-CFAR-DL-AMF). Thus, the persymmetric AMF is not recommended in insufficient sample scenarios.
\section{Conclusion and Future Work}
In this paper, we systematically investigate the asymptotic performance of DL-AMF detector under a large dimensional regime (namely LDR) where both the dimension $N$ and sample size $K$ tend to infinity whereas their quotient $N/K$ converges to a constant $c\in(0,1)$ . We derive an explicit expression for the asymptotic $P_{\rm fa}$ of DL-AMF under LDR. It is found that DL-AMF fails to achieve CFAR properties with respect to the covariance matrix $\mathbf{R}$, target steering vector $\mathbf{s}$ and the loading factor $\lambda$. Furthermore, we derive the asymptotic $P_{\rm d}$ for DL-AMF under both Swerling 0 and Swerling I target models. By analyzing the asymptotic $P_{\rm d}$, we obtain the following crucial insights:
\begin{itemize}
	\item For all nonnegative loading factor, the DL-AMF is always inferior to the NP detector;
	\item Not all loading factor will produce a DL-AMF that performs better than the SCM-based AMF. To ensure that the DL-AMF outperforms SCM-based AMF, it is recommended to avoid excessively large loading factors;
	\item  The detection performance of DL-AMF under LDR is determined solely by $|{\hat{\alpha}}_{\rm DL}(\lambda)|^2$ while $\hat{\beta}_{\rm DL}(\lambda)$ has no effect. This implies that any DL detectors constructed by normalizing $|\hat{\alpha}_{\rm DL}(\lambda)|^2$ with a deterministic quantity (or a random variable that converges almost surely to a deterministic value) will perform equivalently under LDR. 
\end{itemize}

The third insight is critically important and motivates the construction of a CFAR detector by normalizing $|{\hat{\alpha}}_{\rm DL}(\lambda)|^2$ with an appropriate deterministic quantity or a random variable while maintaining the detection performance unchanged. Building on this idea, we have successfully established two CFAR DL detectors: CFAR-DL-SCMF and CFAR-DL-AMF. The former relies on the unknown covariance matrix $\mathbf{R}$ and is thus impractical, while the latter is its adaptive counterpart and can be implemented in practice. We prove that the asymptotic distribution of both detectors' test statistics under $H_0$  follows a standard Exponential distribution within the LDR framework. Consequently, both achieve CFAR properties with respect to $\mathbf{R}$, $\mathbf{s}$ and $\mathbf{\lambda}$.  The key importance of CFAR-DL-AMF lies in its ability to enforce CFAR properties on arbitrary DL detectors while keeping the original detectors' detection performance. As an example, we apply it to ASG's EL-AMF \cite{Abramovich2007Modified}, yielding CFAR-EL-AMF.  It is shown that the CFAR-EL-AMF achieves CFAR property with respect to arbitrary covariance matrix $\mathbf{R}$ at the cost of practically negligible detection loss compared to the original ASG's EL-AMF under finite $N$ and $K$. 

We further address the problem of determining optimal loading factor $\lambda_{\rm opt}$. Using our derived explicit expressions for the asymptotic $P_{\rm d}$, we are able to determine $\lambda_{\rm opt}$ via the NP criterion, i.e., by maximizing the asymptotic probability detection probabilities for a given false alarm probability. By using  $\lambda_{\rm opt}$ in CFAR-DL-SCMF, we obtain the opt-CFAR-DL-SCMF, which achieves theoretical optimality in the sense of NP criterion under LDR. Furthermore, we provide a consistent estimate for  $\lambda_{\rm opt}$ under LDR, denoted by $\hat{\lambda}_{\rm opt}$. Using $\hat{\lambda}_{\rm opt}$ in CFAR-DL-AMF, we obtain the opt-CFAR-DL-AMF. Both the theoretical analysis and experiments results show that the opt-CFAR-DL-AMF tend to exhibit identical detection performance to opt-CFAR-DL-SCMF as $N,K$ tend to infinity at the same rate.  Finally, the numerical simulations demonstrate that the proposed opt-CFAR-DL-SCMF and opt-CFAR-DL-AMF consistently outperform EL-based CFAR AMF detector (i.e., CFAR-EL-AMF) and persymmetric AMF whenever in both full-rank and low-rank clutter-plus-noise environments.

Future research will focus on two key directions to advance this work. First, a rigorous theoretical analysis of the function $\kappa(\lambda)$ should be conducted to thoroughly characterize its behavior. Furthermore, the formal proofs of the existence and uniqueness of the optimal loading factor ($\lambda_0$ or $\lambda_{\rm opt}$) is needed. Second, future work should address more realistic non-Gaussian environments, particularly the compound Gaussian clutter. This extension includes the following three fundamental aspects: deriving the asymptotic $P_{\rm fa}$ and $P_{\rm d}$ for DL detectors under non-Gaussian conditions, developing CFAR version of the DL detectors, and establishing optimization criteria for the optimal loading factor.
\section*{Appendix A. Proof of Theorem \ref{Theorem asymptotic distribution of alpha}}
We emphasize that the proof relies on the integration by parts formula for Gaussian vectors. Let $\mathbf{x}$ be an $N$-dimensional vector following a complex Gaussian distribution with mean $\mathbf{0}$ and covariance matrix $\mathbf{I}_N$ and let $f$ be a continuously differentiable function with polynomially bounded partial derivatives, then the integration by parts formula for Gaussian vectors writes
\begin{equation}\label{62}
\mathbb{E}[\mathbf{x}_{n1}f(\mathbf{x},\mathbf{x}^H)] = \mathbb{E}\left[\frac{\partial f(\mathbf{x},\mathbf{x}^H)}{\partial \overline{\mathbf{x}_{n1}}}\right].
\end{equation}
Notice that this formula has been widely used in RMT literature \cite{Hachem2008A,Rubio2012A}

Now let us start the proof. Recall $\hat{\alpha}_{\rm DL}(\lambda)$ defined in \eqref{4}. If we take the transformations $\mathbf{u} = \mathbf{R}^{-1/2}\mathbf{s}$ and $\mathbf{x}_0=\mathbf{R}^{-1/2}\mathbf{y}_0$, then it is easy to get $\hat{\alpha}_{\rm DL}(\lambda) = \mathbf{u}^H\mathbf{Q}(\lambda)\mathbf{x}_0$ with $\mathbf{Q}(\lambda) = \left(\mathbf{X}\mathbf{X}^H+\lambda\mathbf{R}^{-1}\right)^{-1}$. Here, $\mathbf{X}=\frac{1}{\sqrt{K}}[\mathbf{x}_1,\mathbf{x}_2,...,\mathbf{x}_K]$ with $\mathbf{x}_k = \mathbf{R}^{-1/2}\mathbf{y}_k$ ($k=1,2,..,K$) following $N$-dimensional complex Gaussian distribution with mean $\mathbf{0}$ and covariance matrix $\mathbf{I}_N$.

We define the characteristic function 
\begin{equation}\label{key}
 \phi(\omega) = \mathbb{E}\left[\Psi(\omega)\right]
\end{equation}
where $\Psi(\omega) = \exp\left[\mathsf{i}\omega{\rm Re}(\hat{\alpha}_{\rm DL}(\lambda))\right]$. Then taking the derivative of $\phi(\omega)$ with respect to $\omega$, we have
	\begin{equation}\label{64}
		\frac{\partial\phi(\omega)}{\partial \omega} = \mathsf{i}\mathbb{E}\left[{\rm Re}(\hat{\alpha}_{\rm DL}(\lambda))\Psi(\omega)\right]
	\end{equation}
	
Notice that $\mathbf{Q}(\lambda)=\mathbf{Q}(\lambda)^H$. Hence, we have
\begin{equation}\label{65}
	{\rm Re}(\hat{\alpha}_{\rm DL}(\lambda)) = \frac{1}{2}\left[\mathbf{u}^H\mathbf{Q}(\lambda)\mathbf{x}_0+\mathbf{x}_0^H\mathbf{Q}(\lambda)\mathbf{u}\right].
\end{equation}

Substituting \eqref{65} into \eqref{64} yields
	\begin{equation}\label{key}
	\frac{\partial{\phi(\omega)}}{\partial{\omega}}= \frac{\mathsf{i}}{2}\mathbb{E}\left[\mathbf{u}^H\mathbf{Q}(\lambda)\mathbf{x}_0\Psi(\omega)+\mathbf{x}_0^H\mathbf{Q}(\lambda)\mathbf{u}\Psi(\omega)\right]
	\end{equation}
	
The remaining task is to develop the terms $\mathbb{E}\left[\mathbf{u}^H\mathbf{Q}(\lambda)\mathbf{x}_0\Psi(\omega)\right]$ and $\mathbb{E}\left[\mathbf{x}_0^H\mathbf{Q}(\lambda)\mathbf{u}\Psi(\omega)\right]$. 

Notice that
\begin{equation}\label{key}
	\mathbb{E}\left[\mathbf{u}^H\mathbf{Q}(\lambda)\mathbf{x}_0\Psi(\omega)\right] = \sum_{m,n}\mathbb{E}\left[\overline{\mathbf{u}_{m1}}(\mathbf{Q}(\lambda))_{mn}(\mathbf{x}_0)_{n1}\Psi(\omega)\right].
\end{equation}

Using the integration by parts formula \eqref{62}, we have
\begin{equation}\label{68}
	\begin{aligned}
	&\mathbb{E}\left[\mathbf{u}^H\mathbf{Q}(\lambda)\mathbf{x}_0\Psi(\omega)\right] 
	\\&= \sum_{m,n} \mathbb{E}\left[\overline{\mathbf{u}_{m1}}(\mathbf{Q}(\lambda))_{mn}\frac{\partial\Psi(\omega)}{\partial \overline{(\mathbf{x}_0)_{n1}}}\right]
	\\& = \mathsf{i}\omega\sum_{m,n} \mathbb{E}\left[\overline{\mathbf{u}_{m1}}(\mathbf{Q}(\lambda))_{mn}\frac{\partial{\rm Re}(\hat{\alpha}_{\rm DL}(\lambda))}{\partial \overline{(\mathbf{x}_0)_{n1}}}\Psi(\omega)\right]
	\\& = \frac{\mathsf{i}\omega}{2}\mathbb{E}\left[\mathbf{u}^H\mathbf{Q}^2(\lambda)\mathbf{u}\Psi(\omega)\right]
	\\& = \frac{\mathsf{i}\omega}{2}\mathbb{E}\left[\mathbf{u}^H\mathbf{Q}^2(\lambda)\mathbf{u}\right]\mathbb{E}\left[\Psi(\omega)\right] +\frac{\mathsf{i}\omega}{2}K\mathbb{E}\left[\overset{\circ}{\eta}_1\overset{\circ}{\eta}_2\right]
\end{aligned}
\end{equation}
where $\overset{\circ}{\eta}_1 = \mathbf{u}^H\mathbf{Q}^2(\lambda)\mathbf{u} - \mathbb{E}\left[\mathbf{u}^H\mathbf{Q}^2(\lambda)\mathbf{u}\right]$ and $\overset{\circ}{\eta}_2 = \frac{1}{K}\Psi(\omega) -\frac{1}{K} \mathbb{E}\left[\Psi(\omega)\right]$.

We need to use the following lemma.
\begin{mylemma}\label{Lemma variance cocntrol} Let assumptions (A1)-(A3) hold true. As $N,K\to\infty$, it holds that
	\begin{equation}\label{69}
		{\rm Var}\left(\mathbf{u}^H\mathbf{Q}^2(\lambda)\mathbf{u}\right) =O\left(\frac{1}{K}\right)
	\end{equation}
and 
\begin{equation}\label{70}
		{\rm Var}\left(\frac{1}{K}\Psi(\omega)\right)= O\left(\frac{\omega^2}{K^2}\right).
\end{equation}
\end{mylemma}
\noindent\textit{Proof:}
Equation \eqref{69} has been established in \cite[Lemma 3]{Rubio2012A}. The proof of \eqref{70} is deferred to Appendix B.
\qed

Using Lemma \ref{Lemma variance cocntrol} and the Cauchy-Schwarz inequality, we have
\begin{equation}\label{71}
	\small \mathbb{E}\left[\overset{\circ}{\eta}_1\overset{\circ}{\eta}_2\right]\leqslant \sqrt{	{\rm Var}\left(\mathbf{u}^H\mathbf{Q}^2(\lambda)\mathbf{u}\right) }\sqrt{{\rm Var}\left(\frac{1}{K}\Psi(\omega)\right)} = O\left(\frac{\omega}{K^{3/2}}\right).
\end{equation}

Substituting \eqref{71} into \eqref{68}, we have
\begin{equation}\label{72}
\small	\begin{aligned}
		\mathbb{E}\left[\mathbf{u}^H\mathbf{Q}(\lambda)\mathbf{x}_0\Psi(\omega)\right] 
		= \frac{\mathsf{i}\omega}{2}\mathbb{E}\left[\mathbf{u}^H\mathbf{Q}^2(\lambda)\mathbf{u}\right]\mathbb{E}\left[\Psi(\omega)\right] +O\left(\frac{\omega^2}{\sqrt{K}}\right).
	\end{aligned}
\end{equation}

Using the same procedures, it is easy to get $\mathbb{E}\left[\mathbf{u}^H\mathbf{Q}(\lambda)\mathbf{x}_0\Psi(\omega)\right] = \mathbb{E}\left[\mathbf{x}_0^H\mathbf{Q}(\lambda)\mathbf{u}\Psi(\omega)\right]$. Then using this equality and \eqref{72} in \eqref{64}, we have
\begin{equation}\label{key}
	\frac{\partial \phi(\omega)}{\partial \omega} = -\frac{\omega}{2}\mathbb{E}\left[\mathbf{u}^H\mathbf{Q}^2(\lambda)\mathbf{u}\right]\phi(\omega) + O\left(\frac{\omega^2}{\sqrt{K}}\right).
\end{equation}

We find that this equation is a classical nonhomogeneous linear differential equation of the first order. Hence, we obtain
\begin{equation}\label{key}
	\phi(\omega) = \exp\left(-\frac{\bar{\sigma}^2}{2}\omega^2\right) +o(1)
\end{equation}
where $\bar{\sigma}^2 = \frac{1}{2}\mathbb{E}\left[\mathbf{u}^H\mathbf{Q}^2(\lambda)\mathbf{u}\right]$.

According to Levy's continuity theorem, we get $\frac{{\rm Re}(\hat{\alpha}_{\rm DL}(\lambda))}{\sqrt{\bar{\sigma}^2}}\overset{d}{\longrightarrow}\mathcal{N}(0,1)$. In addition, we know form \cite[Lemma 1]{Rubio2012A} that $\mathbb{E}\left[\mathbf{u}^H\mathbf{Q}^2(\lambda)\mathbf{u}\right]\to\frac{\mathbf{u}^H\mathbf{E}(\lambda)^2\mathbf{u}}{1-\gamma(\lambda)}$ as $N,K\to \infty$ with $\mathbf{u}$, $\mathbf{E}(\lambda)$ and $\gamma(\lambda)$ defined in Theorem \ref{Theorem asymptotic distribution of alpha}. Hence, we finally get $\bar{\sigma}^2\to\sigma^2_{\hat{\alpha},H_0}$.

Now we have established that ${\rm Re}(\hat{\alpha}_{\rm DL}(\lambda))$ converges in distribution to a real Gaussian distribution. By employing analogous procedures, we can further prove that ${\rm Im}(\hat{\alpha}_{\rm DL}(\lambda))$ also converges in distribution to the identical Gaussian distribution as that of ${\rm Re}(\hat{\alpha}_{\rm DL}(\lambda))$.

In what follows, we will show that ${\rm Re}(\hat{\alpha}_{\rm DL}(\lambda))$ and ${\rm Im}(\hat{\alpha}_{\rm DL}(\lambda))$ are uncorrelated, which is equivalent to proving that the the covariance of ${\rm Re}(\hat{\alpha}_{\rm DL}(\lambda))$ and ${\rm Im}(\hat{\alpha}_{\rm DL}(\lambda))$ is 0. We first notice that due to the zero mean of $\mathbf{x}_0$, we have $\mathbb{E}\left[{\rm Re}(\hat{\alpha}_{\rm DL}(\lambda))\right] = \mathbb{E}\left[{\rm Im}(\hat{\alpha}_{\rm DL}(\lambda))\right]=0$. Hence, the covariance between ${\rm Re}(\hat{\alpha}_{\rm DL}(\lambda))$ and ${\rm Im}(\hat{\alpha}_{\rm DL}(\lambda))$ is given by
\begin{equation}\label{key}
	\begin{aligned}
		&	{\rm Cov}\left[{\rm Re}(\hat{\alpha}_{\rm DL}(\lambda)),{\rm Im}(\hat{\alpha}_{\rm DL}(\lambda))\right] 
		\\&= \mathbb{E}\left[{\rm Re}(\hat{\alpha}_{\rm DL}(\lambda)){\rm Im}(\hat{\alpha}_{\rm DL}(\lambda))\right]
		\\& = \frac{1}{4\mathsf{i}}\mathbb{E}\left[(\mathbf{u}^H\mathbf{Q}(\lambda)\mathbf{x}_0)^2-(\mathbf{x}_0^H\mathbf{Q}(\lambda)\mathbf{u})^2\right].
	\end{aligned}
\end{equation}

Notice that $\mathbf{u}^H\mathbf{Q}(\lambda)\mathbf{x}_0$ is a scalar, hence we have $\mathbf{u}^H\mathbf{Q}(\lambda)\mathbf{x}_0 = \mathbf{x}_0^T\mathbf{Q}^T(\lambda)\mathbf{u}^*$. Then, we have 
\begin{equation}\label{key}
	\begin{aligned}
		\mathbb{E}\left[(\mathbf{u}^H\mathbf{Q}(\lambda)\mathbf{x}_0)^2\right]& = \mathbb{E}\left[\mathbf{u}^H\mathbf{Q}(\lambda)\mathbf{x}_0\mathbf{x}_0^T\mathbf{Q}^T(\lambda)\mathbf{u}^*\right]
		\\& \overset{(a)}{=} \mathbf{u}^H\mathbb{E}\left[\mathbf{Q}(\lambda)\mathbb{E}(\mathbf{x}_0\mathbf{x}_0^T)\mathbf{Q}^T(\lambda)\right]\mathbf{u}^*
		\\& \overset{(b)}{=} \mathbf{u}^H\mathbb{E}\left[\mathbf{Q}(\lambda)\mathbf{0}\mathbf{Q}^T(\lambda)\right]\mathbf{u}^*
		\\& = 0
	\end{aligned}
\end{equation}
where $(a)$ is due to the independence between $\mathbf{Q}(\lambda)$ and $\mathbf{x}_0$, and $(b)$ is due to the circular symmetric property of $\mathbf{x}_0$ (i.e., $\mathbf{E}(\lambda)\left[\mathbf{x}_0\mathbf{x}_0^T\right]=0$). Proceeding similarly, one can show that $\mathbb{E}\left[(\mathbf{x}_0^H\mathbf{Q}(\lambda)\mathbf{u})^2\right]=0$. Hence, ${\rm Cov}\left[{\rm Re}(\hat{\alpha}_{\rm DL}(\lambda)),{\rm Im}(\hat{\alpha}_{\rm DL}(\lambda))\right] =0$, which implies that ${\rm Re}(\hat{\alpha}_{\rm DL}(\lambda))$ and ${\rm Im}(\hat{\alpha}_{\rm DL}(\lambda))$ are uncorrelated. Now we complete the proof of Theorem \ref{Theorem asymptotic distribution of alpha}.

\section*{Appendix B. Proof of Theorem \ref{Theorem Almost Convergence of Beta}}
The proof relies on the following well-known result regarding the almost sure convergence of bilinear forms in RMT.
\begin{mytheorem}\cite[Theorem 1]{Mestre2008On}\label{Theorem almost convergence of M(z)}
	Define $\hat{\mathcal{M}}(z) =\mathbf{s}^H(\hat{\mathbf{R}}-z\mathbf{I}_N)^{-1}\mathbf{s} $. Then under the assumptions (A1)-(A3) and as $N,K\to\infty$, it holds that $|\hat{\mathcal{M}}(z)-\mathcal{M}(z)|\overset{a.s.}{\longrightarrow}0$ for all $z\in\mathbb{C}$ except for a segment of the positive real axis, where $\mathcal{M}(z) = \mathbf{s}^H\left[\left(1-c-czm(z)\right)\mathbf{R}-z\mathbf{I}_N\right]^{-1}\mathbf{s}^H$ and $m(z)=m$ is the unique solution to the equation $m = \frac{1}{N}{\rm tr}\left[\left(1-c-czm\right)\mathbf{R}-z\mathbf{I}_N\right]^{-1}$.
\end{mytheorem}

We observe that $\hat{\beta}_{\rm DL}(\lambda) = \hat{\mathcal{M}}(z)|_{z=-\lambda}$. Hence, applying Theorem \ref{Theorem almost convergence of M(z)} directly yields  
\begin{equation}\label{83}
	\small \hat{\beta}_{\rm DL}(\lambda)\overset{a.s.}{\longrightarrow}\mathcal{M}(z)|_{z=-\lambda} = \mathbf{s}^H\left[(1-c+c\lambda m(-\lambda))\mathbf{R}+\lambda\mathbf{I}_N\right]^{-1}\mathbf{s}
\end{equation}
 with $m(-\lambda)=m$ satisfying the equation 
\begin{equation}\label{84}
	m = \frac{1}{N}{\rm tr}\left[\left(1-c+c\lambda m\right)\mathbf{R}+\lambda\mathbf{I}_N\right]^{-1}.
\end{equation}

We further define $\delta = 1-c+c\lambda m(-\lambda)$. Then substituting $\delta$ into \eqref{83}, it is easy to find that 
\begin{equation}\label{key}
	\begin{aligned}
	\mathcal{M}(z)|_{z=-\lambda} = \mathbf{s}^H\left[\delta\mathbf{R}+\lambda\mathbf{I}_N\right]^{-1}\mathbf{s} 
	= \mathbf{u}^H\mathbf{E}(\lambda)\mathbf{u}
\end{aligned}
\end{equation}
with $\mathbf{E}(\lambda) =\left[\delta\mathbf{I}_N+\lambda\mathbf{R}^{-1}\right]^{-1} $ and $\mathbf{u}=\mathbf{R}^{-1/2}\mathbf{s}$. In addition, substituting $\delta = 1-c+c\lambda m(-\lambda)$ into \eqref{84}, we obtain 
\begin{equation}\label{key}
	\begin{aligned}
	&\qquad\quad\frac{\delta+c-1}{c\lambda} = \frac{1}{N}{\rm tr}\left[\delta \mathbf{R}+\lambda\mathbf{I}_N\right]^{-1}
	\\&\overset{(a)}{\iff} \delta+\frac{\lambda}{N}{\rm tr}\left[(\lambda\mathbf{I}_N)^{-1}-(\delta\mathbf{R}+\lambda\mathbf{I}_N)^{-1}\right] = 1
	\\&\overset{(b)}{\iff} \text{Eq. \eqref{14}}.
\end{aligned}
\end{equation}
where $(a)$ is due to $1 = \frac{\lambda}{N}{\rm tr}(\lambda\mathbf{I}_N)^{-1}$ and $(b)$ uses the identity $\mathbf{A}^{-1}-\mathbf{B}^{-1} = \mathbf{A}^{-1}(\mathbf{B}-\mathbf{A})\mathbf{B}^{-1}$ with $\mathbf{A} = \lambda \mathbf{I}_N$ and $\mathbf{B} = \delta\mathbf{R}+\lambda\mathbf{I}_N$.

\section*{Appendix C. Proof of Lemma \ref{Lemma variance cocntrol} }
For simplicity, we will denote $\mathbf{Q}(\lambda)$ as $\mathbf{Q}$.
The proof relies on the Poincare-Nash inequality. By using the Poincare-Nash inequality and noticing that $\partial \Psi(\omega)/\partial \mathbf{X}_{ij} =\partial \Psi(\omega)/\partial \overline{\mathbf{X}_{ij}} $, we have
\begin{equation}\label{79}
	\begin{aligned}
		{\rm Var}\left(\frac{1}{K}\Psi(\omega)\right)&\leqslant\frac{2}{K^3}\sum_{i,j}\mathbb{E}\left|\frac{\partial \Psi(\omega)}{\partial \mathbf{X}_{ij}}\right|^2
		\\&\leqslant \frac{2\omega^2}{K^3}\mathbb{E}\left|\frac{\partial {\rm Re}(\hat{\alpha}_{\rm DL}(\lambda))}{\partial \mathbf{X}_{ij}}\Psi(\omega)\right|^2
	\end{aligned}
\end{equation}

Taking $ {\rm Re}(\hat{\alpha}_{\rm DL}(\lambda)$ derivative with respect to $\mathbf{X}_{ij}$ yields
\begin{equation}\label{80}
	\frac{\partial  {\rm Re}(\hat{\alpha}_{\rm DL}(\lambda)}{\partial \mathbf{X}_{ij}} = -\frac{1}{2}\left((\mathbf{X}^H\mathbf{Q}\mathbf{x}_0\mathbf{u}^H\mathbf{Q})_{ji}+(\mathbf{X}^H\mathbf{Q}\mathbf{u}\mathbf{x}_0^H\mathbf{Q})_{ji}\right).
\end{equation}

Substituting \eqref{80} into \eqref{79} and noticing that $|\Psi(\omega)|^2=1$ and $\mathbf{Q}=\mathbf{Q}^H$, we have
\begin{equation}\label{key}
	\begin{aligned}
		{\rm Var}\left(\frac{1}{K}\Psi(\omega)\right)&\leqslant\frac{\omega^2}{2K^3}\mathbb{E}\left[{\rm tr}(\mathbf{X}^H\mathbf{Q}\mathbf{x}_0\mathbf{u}^H\mathbf{Q}^2\mathbf{u}\mathbf{x}_0^H\mathbf{Q}\mathbf{X})\right]
		\\&+\frac{\omega^2}{2K^3}\mathbb{E}\left[{\rm tr}(\mathbf{X}^H\mathbf{Q}\mathbf{u}\mathbf{x}_0^H\mathbf{Q}^2\mathbf{x}_0\mathbf{u}^H\mathbf{Q}\mathbf{X})\right]
	\end{aligned}
\end{equation}
In addition, by using the fact
\begin{equation}\label{76}
	\mathbf{X}\mathbf{X}^H\mathbf{Q} = \mathbf{I}_N-\lambda\mathbf{R}^{-1}\mathbf{Q}
\end{equation}
we have
\begin{equation}\label{key}
	\small \begin{aligned}
		{\rm Var}\left(\frac{1}{K}\Psi(\omega)\right)&\leqslant\frac{\omega^2}{2K^3}\mathbb{E}\left[{\rm tr}(( \mathbf{I}_N-\lambda\mathbf{R}^{-1}\mathbf{Q})\mathbf{x}_0\mathbf{u}^H\mathbf{Q}^2\mathbf{u}\mathbf{x}_0^H\mathbf{Q})\right]
		\\&+\frac{\omega^2}{2K^3}\mathbb{E}\left[{\rm tr}(( \mathbf{I}_N-\lambda\mathbf{R}^{-1}\mathbf{Q})\mathbf{u}\mathbf{x}_0^H\mathbf{Q}^2\mathbf{x}_0\mathbf{u}^H\mathbf{Q})\right]
		\\&\leqslant \frac{\omega^2}{2K^3}\mathbb{E}\left[{\rm tr}(\mathbf{x}_0\mathbf{u}^H\mathbf{Q}^2\mathbf{u}\mathbf{x}_0^H\mathbf{Q})\right]
		\\&+\frac{\omega^2\lambda}{2K^3}\mathbb{E}\left[{\rm tr}(\mathbf{R}^{-1}\mathbf{Q}\mathbf{x}_0\mathbf{u}^H\mathbf{Q}^2\mathbf{u}\mathbf{x}_0^H\mathbf{Q})\right]
		\\&+\frac{\omega^2}{2K^3}\mathbb{E}\left[{\rm tr}(\mathbf{u}\mathbf{x}_0^H\mathbf{Q}^2\mathbf{x}_0\mathbf{u}^H\mathbf{Q})\right]
		\\&+\frac{\omega^2\lambda}{2K^3}\mathbb{E}\left[{\rm tr}(\mathbf{R}^{-1}\mathbf{Q}\mathbf{u}\mathbf{x}_0^H\mathbf{Q}^2\mathbf{x}_0\mathbf{u}^H\mathbf{Q})\right].
	\end{aligned}
\end{equation}

Then by using the inequality $|{\rm tr}(\mathbf{A}\mathbf{B})|\leqslant \Vert\mathbf{B}\Vert{\rm tr}\mathbf{A}$ for an Hermitian nonnegative matrix $\mathbf{A}$ and a square matrix $\mathbf{B}$ and the facts that $\Vert\mathbf{Q}\Vert\leqslant \frac{C_{\rm max}}{\lambda}$ and $\Vert\mathbf{R}^{-1}\Vert\leqslant \frac{1}{C_{\rm min}}$, we have
\begin{equation}\label{key}
	\begin{aligned}
		&	{\rm Var}\left(\frac{1}{K}\Psi(\omega)\right)\\&\leqslant \frac{\omega^2}{K^2}\left(\frac{C_{\rm max}}{\lambda}\right)^3\left(C_{\rm max}+\frac{1}{C_{\rm min}}\right)\frac{1}{K}\mathbb{E}\left[{\rm tr}(\mathbf{u}\mathbf{x}_0^H\mathbf{x}_0\mathbf{u}^H)\right].
	\end{aligned}
\end{equation}

Notice that $\frac{1}{K}\mathbb{E}{\rm tr}(\mathbf{u}\mathbf{x}_0^H\mathbf{x}_0\mathbf{u}^H) = c{\rm tr}(\mathbf{u}\mathbf{u}^H)<\infty$. Hence, we obtain $	{\rm Var}\left(\frac{1}{K}\Psi(\omega)\right)=O\left(\frac{1}{K^2}\right)$.
\section*{Appendix D. Proof of Theorem \ref{Theorem consistent estimate for mu0}}

We emphasize that the proof heavily relies on Girko's pioneer work on the generalized statistic analysis ($G$-analysis) \cite{Girko1987G2, Girko1995Statistical,GirkoTen,Girko1990G25}. The $G$-analysis provides a strong mathematical tool for deriving the consistent estimators for the traces of the resolvents of large dimensional random matrices. Precisely, here we frequently use the $G_2$ and $G_{25}$ estimators for the real Stieltjes transform. Now let us briefly review  Girko's $G_2$ and $G_{25}$ estimators. 

If we define the real Stieltjes transform of the spectra of $\mathbf{R}$ by $b(x) = \frac{1}{N}{\rm tr}(\mathbf{I}_N+x\mathbf{R})^{-1}$, then the $G_2$ estimator for $b(x)$ is given by \cite{Girko1987G2, Girko1995Statistical, GirkoTen}
\begin{equation}\label{111}
	\hat{b}(x) = \frac{1}{N}{\rm tr}\left(\mathbf{I}_N+\theta(x)\hat{\mathbf{R}}\right)^{-1}
\end{equation}
where $\theta(x)=\theta$ is the unique solution to the equation
\begin{equation}\label{112}
	\theta\left[1-c+\frac{c}{N}{\rm tr}\left(\mathbf{I}_N+\theta\hat{\mathbf{R}}\right)^{-1}\right] = x.
\end{equation}

Define the quadratic form $B(x) = \mathbf{s}^H\left(\mathbf{I}_N+x\mathbf{R}\right)^{-1}\mathbf{s}$. The $G_{25}$ estimator for $B(x)$ is given by \cite{Girko1990G25,Girko1995Statistical,GirkoTen}
\begin{equation}\label{key}
	\hat{B}(x) = \mathbf{s}^H\left(\mathbf{I}_N+\theta(x)\hat{\mathbf{R}}\right)^{-1}\mathbf{s}
\end{equation}
where $\theta(x)$ is the unique solution to the equation \eqref{112}.

It has been shown by Girko that both $\hat{b}(x)$ and $\hat{B}(x)$ are consistent under LDR. Furthermore, we notice that Mestre and Lagunas \cite{Mestre2006Finite} have employed the $G_2$ and $G_{25}$ estimators to estimate the optimal loading factor through maximization of the output signal-to-noise-plus-interference ratio.

In what follows, we use $G_2$ and $G_{25}$ to derive a consistent estimator for $\mu_0(\mathbf{R},\mathbf{s},\lambda)$. Recalling from \eqref{17} that $\mu_0(\mathbf{R},\mathbf{s},\lambda)$ depends on $\gamma(\lambda)$ and $\mathbf{u}^H\mathbf{E}(\lambda)^2\mathbf{u}$, we focus on estimating these two quantities. 

\noindent$\bullet$  \textbf{Consistent estimator for $\gamma(\lambda)$.}

 Reviewing the definition of $\gamma(\lambda)$ given in Theorem \ref{Theorem asymptotic distribution of alpha}, we have
		\begin{equation}\label{130}
			\begin{aligned}
			\gamma(\lambda) = \frac{\delta^2}{K}{\rm tr}\mathbf{E}(\lambda)^2& = \frac{1}{K}{\rm tr}\left[\mathbf{R}^2\left(\mathbf{R}+\frac{\lambda}{\delta}\mathbf{I}_N\right)^{-2}\right]
			\\& = \frac{N}{K}\left[1-{b}(x)\left|_{x=\frac{\delta}{\lambda}}\right.+\frac{\delta}{\lambda}\frac{\partial b(x)}{\partial x}\left|_{x=\frac{\delta}{\lambda}}\right.\right].
		\end{aligned}
		\end{equation}
	
	Then using the consistent $G_2$ estimator for $b(x)$, we can estimate $\gamma(\lambda)$ by
	\begin{equation}\label{115}
		\widehat{\gamma(\lambda)} = \frac{N}{K}\left[1-\hat{b}(x)\left|_{x=\frac{\delta}{\lambda}}\right.+\frac{\delta}{\lambda}\frac{\partial \hat{b}(x)}{\partial x}\left|_{x=\frac{\delta}{\lambda}}\right.\right].
	\end{equation}

It remains to develop the term $\frac{\partial \hat{b}(x)}{\partial x}$. From \eqref{111}, we have
\begin{equation}\label{116}
\frac{\partial \hat{b}(x)}{\partial x} = -\frac{\theta'(x)}{N}{\rm tr}\left[\hat{\mathbf{R}}\left(\mathbf{I}_N+\theta(x)\hat{\mathbf{R}}\right)^{-2}\right]
\end{equation}
where $\theta'(x)$ is the derivative of $\theta(x)$ and satisfies the following equation 
\begin{equation}\label{117}
\small 	\theta'\left[1-c+\frac{c}{N}{\rm tr}\left(\mathbf{I}_N+\theta\hat{\mathbf{R}}\right)^{-1}\right] - \theta\theta'\frac{c}{N}{\rm tr}\left[\hat{\mathbf{R}}\left(\mathbf{I}_N+\theta\hat{\mathbf{R}}\right)^{-2}\right] = 1.    
\end{equation}

From \eqref{117}, we can express $\theta'(x)$ by
\begin{equation}\label{118}
	\small \begin{aligned}
	\theta'(x) &= \frac{1}{1-c+\frac{c}{N}{\rm tr}\left(\mathbf{I}_N+\theta\hat{\mathbf{R}}\right)^{-1}-\theta\frac{c}{N}{\rm tr}\left[\hat{\mathbf{R}}\left(\mathbf{I}_N+\theta\hat{\mathbf{R}}\right)^{-2}\right]}
	\\& = \frac{1}{1-c+\frac{c}{N}{\rm tr}\left(\mathbf{I}_N+\theta\hat{\mathbf{R}}\right)^{-2}}.
\end{aligned}
\end{equation}
Substituting \eqref{111}, \eqref{116}, and \eqref{118} into \eqref{115}, we get
\begin{equation}\label{119}
	\small \begin{aligned}
	&\widehat{\gamma(\lambda)} = \\&\frac{N}{K}\left[1-\frac{1}{N}{\rm tr}\left(\mathbf{I}_N+\theta(\delta/\lambda)\hat{\mathbf{R}}\right)^{-1} - \frac{\frac{\delta}{\lambda}\frac{1}{N}{\rm tr}\left[\hat{\mathbf{R}}\left(\mathbf{I}_N+\theta(\delta/\lambda)\hat{\mathbf{R}}\right)^{-2}\right]}{1-\frac{N}{K}+\frac{1}{K}{\rm tr}\left(\mathbf{I}_N+\theta(\delta/\lambda)\hat{\mathbf{R}}\right)^{-2}}\right].
\end{aligned}
\end{equation}

From \eqref{112}, we know that $\theta(\delta/\lambda)$ satisfies the following equation
\begin{equation}\label{key}
		\theta(\delta/\lambda)\left[1-c+c\hat{b}(\delta/\lambda)\right] = \frac{\delta}{\lambda}.
\end{equation}
Since $\hat{b}(\delta/\lambda)\overset{a.s.}{\longrightarrow}b(\delta/\lambda)$, we have
\begin{equation}\label{121}
	\theta(\delta/\lambda)\left[1-c+c{b}(\delta/\lambda)\right] = \frac{\delta}{\lambda}.
\end{equation}

In addition, we notice from \eqref{14} that $\delta$ satisfies the following equation
\begin{equation}\label{122}
	\delta = 1-c+cb(\delta/\lambda)
\end{equation}

Substituting \eqref{122} into \eqref{121} yields
\begin{equation}\label{123}
	\theta(\delta/\lambda) = \frac{1}{\lambda}.
\end{equation}

To obtain the consistent estimate of $\gamma(\lambda)$, it remains to derive the consistent estimate for $\delta$. From \eqref{122}, we can estimate $\delta$ by 
\begin{equation}\label{key}
	\hat{\delta} = 1-\frac{N}{K}+\frac{1}{K}{\rm tr}\left(\mathbf{I}_N+\theta(\delta/\lambda)\hat{\mathbf{R}}\right)^{-1}.
\end{equation}

Finally replacing $\delta$ with $\hat{\delta}$ in \eqref{119}, substituting \eqref{123} into \eqref{119} and after some simple algebras, we get the consistent estimate for $\gamma(\lambda)$:
	\begin{equation}\label{125}
		\widehat{\gamma(\lambda)} =  1-\frac{\left(1-\frac{N}{K}+\frac{\lambda}{K}{\rm tr}\left(\hat{\mathbf{R}}+\lambda\mathbf{I}_N\right)^{-1}\right)^2}{1-\frac{N}{K}+\frac{\lambda^2}{K}{\rm tr}\left(\hat{\mathbf{R}}+\lambda\mathbf{I}_N\right)^{-2}}.
	\end{equation}

\noindent$\bullet$ \textbf{Consistent estimator for $\mathbf{u}^H\mathbf{E}(\lambda)^2\mathbf{u}$.}

For convenience, we denote $\xi(\lambda) = \mathbf{u}^H\mathbf{E}(\lambda)^2\mathbf{u}$. Reviewing the definitions of $\mathbf{u}$ and $\mathbf{E}(\lambda)$ given in Theorem \ref{Theorem asymptotic distribution of alpha}, we can express $\xi$ by
\begin{equation}\label{142}
	\begin{aligned}
		\xi(\lambda) &= \mathbf{s}^H\left(\delta\mathbf{R}+\lambda\mathbf{I}_N\right)^{-1}\mathbf{R}\left(\delta\mathbf{R}+\lambda\mathbf{I}_N\right)^{-1}\mathbf{s}
		\\& = -\frac{1}{\lambda^2}\frac{\partial B(x)}{\partial x}\left|_{x=\frac{\delta}{\lambda}}\right.
	\end{aligned}
\end{equation}

Then we apply Girko's $G_{25}$ estimator and obtain the consistent estimate for $\xi(\lambda)$, given by
	\begin{equation}\label{127}
	\small \begin{aligned}
		\widehat{\xi(\lambda)}& = -\frac{1}{\lambda^2}\frac{\partial \hat{B}(x)}{\partial x}\left|_{x=\frac{\delta}{\lambda}}\right.
		\\& = \frac{\theta'(\delta/\lambda)}{\lambda^2}\mathbf{s}^H\left(\mathbf{I}_N+\theta(\delta/\lambda)\hat{\mathbf{R}}\right)^{-1}\hat{\mathbf{R}}\left(\mathbf{I}_N+\theta(\delta/\lambda)\hat{\mathbf{R}}\right)^{-1}\mathbf{s}.
	\end{aligned}
\end{equation}

Substituting \eqref{118} and \eqref{123} into \eqref{127} and after some manipulations, we finally get
\begin{equation}\label{128}
	\widehat{\xi(\lambda)} = \frac{\mathbf{s}^H\left(\hat{\mathbf{R}}+\lambda\mathbf{I}_N\right)^{-1}\hat{\mathbf{R}}\left(\hat{\mathbf{R}}+\lambda\mathbf{I}_N\right)^{-1}\mathbf{s}}{1-\frac{N}{K}+\frac{\lambda^2}{K}{\rm tr}\left(\hat{\mathbf{R}}+\lambda\mathbf{I}_N\right)^{-2}}.
\end{equation}

Finally using \eqref{125} and \eqref{128}, we obtain the consistent estimator for $\mu_0(\mathbf{R},\mathbf{s},\lambda) = \frac{\xi(\lambda)}{1-\gamma(\lambda)}$, given by
\begin{equation}\label{key}
	\begin{aligned}
	\widehat{\mu_0(\mathbf{R},\mathbf{s},\lambda)} &= \frac{\widehat{\xi(\lambda)}}{1-\widehat{\gamma(\lambda)}} \\&=\frac{\mathbf{s}^H\left(\hat{\mathbf{R}}+\lambda\mathbf{I}_N\right)^{-1}\hat{\mathbf{R}}\left(\hat{\mathbf{R}}+\lambda\mathbf{I}_N\right)^{-1}\mathbf{s}}{ \left(1-\frac{N}{K}+\frac{\lambda}{K}{\rm tr}(\hat{\mathbf{R}}+\lambda\mathbf{I}_N)^{-1}\right)^2}.
\end{aligned}
\end{equation}

Since $1-\frac{N}{K}+\frac{\lambda}{K}{\rm tr}\left(\hat{\mathbf{R}}+\lambda\mathbf{I}_N\right)^{-2}>1-\frac{N}{K}>0$ for all nonnegative $\lambda$, the consistency of $\widehat{\mu_0(\mathbf{R},\mathbf{s},\lambda)}$ is obtained by continuous mapping theorem. Now we complete the proof of Theorem \ref{Theorem consistent estimate for mu0}.
\section*{Appendix E. Proof of Theorem \ref{Theorem Consistent estimator of kappa}}
We define $\psi(\lambda) = \mathbf{u}^H\mathbf{E}(\lambda)\mathbf{u}$. Then we can express $\underline{\kappa}(\lambda)$ by 
\begin{equation}\label{key}
	\underline{\kappa}(\lambda) = \frac{(1-\gamma(\lambda))\psi(\lambda)^2}{\xi(\lambda)}.
\end{equation}

The consistent estimators for $\gamma(\lambda)$ and $\xi(\lambda)$ have been given in \eqref{125} and \eqref{128}, respectively.  It remains to establish the consistent estimator for $\psi(\lambda)$. We notice that
\begin{equation}\label{key}
	\psi(\lambda) = \frac{1}{\lambda}\mathbf{s}^H\left(\mathbf{I}_N+\frac{\delta}{\lambda}\mathbf{R}\right)^{-1}\mathbf{s} = \frac{1}{\lambda}B(x)\left|_{x = \frac{\delta}{\lambda}}\right..
\end{equation}

Then using Girko's $G_{25}$ estimator, we estimate $\psi(\lambda)$ by
\begin{equation}\label{key}
	\widehat{\psi(\lambda)} = \frac{1}{\lambda}\hat{B}(x)\left|_{x=\frac{\delta}{\lambda}}\right. = \frac{1}{\lambda}\mathbf{s}^H\left(\mathbf{I}_N+\theta(\delta/\lambda)\hat{\mathbf{R}}\right)^{-1}\mathbf{s}
\end{equation}
Using \eqref{123}, we get 
\begin{equation}\label{149}
	\widehat{\psi(\lambda)} = \mathbf{s}^H\left(\hat{\mathbf{R}}+\lambda\mathbf{I}_N\right)^{-1}\mathbf{s}.
\end{equation}

Finally, using \eqref{125}, \eqref{128} and \eqref{149}, we estimate $\underline{\kappa}(\lambda)$ by 
\begin{equation}\label{key}
\small	\begin{aligned}
	\widehat{\underline{\kappa} (\lambda)} &= \frac{(1-\widehat{\gamma(\lambda)})\widehat{\psi(\lambda)}^2}{\widehat{\xi(\lambda)}}
	\\& =  \frac{\left(1-\frac{N}{K}+\frac{\lambda}{K}{\rm tr}(\hat{\mathbf{R}}+\lambda\mathbf{I}_N)^{-1}\right)^2\left(\mathbf{s}^H\left(\hat{\mathbf{R}}+\lambda\mathbf{I}_N\right)^{-1}\mathbf{s}\right)^2}{\mathbf{s}^H\left(\hat{\mathbf{R}}+\lambda\mathbf{I}_N\right)^{-1}\hat{\mathbf{R}}\left(\hat{\mathbf{R}}+\lambda\mathbf{I}_N\right)^{-1}\mathbf{s}}.
	\end{aligned}
\end{equation}

Since $\widehat{\xi(\lambda)}>0$, the consistency of $	\widehat{\underline{\kappa} (\lambda)} $ is guaranteed by continuous mapping theorem. Now we complete the proof of Theorem \ref{Theorem Consistent estimator of kappa}.

\section*{Acknowledgment}
The authors would like to thank the editors and anonymous reviewers for their constructive improvements.

\bibliographystyle{IEEEtran}
\bibliography{IEEEexample}
\end{document}